\documentclass[a4paper,11pt]{article}
\pdfoutput=1 

\usepackage{jcappub}
\usepackage{graphicx}
\usepackage{dcolumn}
\usepackage{amssymb,amsmath,bm}
\usepackage{color}
\usepackage[dvipsnames]{xcolor}

\usepackage{xfrac}
\usepackage{aas_macros}
\usepackage{mathrsfs}
\usepackage{subcaption}
\usepackage{rotating}
\usepackage{chngcntr}
\newcommand{\nv}{\vec{\theta}}

\newcommand\Tstrut{\rule{0pt}{3ex}}   

\definecolor{internationalkleinblue}{rgb}{0.0, 0.18, 0.65}
\hypersetup{urlcolor=internationalkleinblue, linkcolor=internationalkleinblue, citecolor=internationalkleinblue}

\usepackage[T1]{fontenc}
\usepackage{natbib}
\title{Tomographic galaxy clustering with the Subaru Hyper Suprime-Cam first year public data release}

\author[a,1]{Andrina Nicola,}
\author[b]{David Alonso,}
\author[c,d]{Javier S\'anchez,}

\author[e]{An\v{z}e Slosar,}

\author[f]{Humna Awan,}
\author[f]{Adam Broussard,}
\author[a,g]{Jo Dunkley,}
\author[b]{Zahra Gomes,}
\author[f]{Eric Gawiser,}
\author[h]{Rachel Mandelbaum,}
\author[i,j,k,l]{Hironao Miyatake,}
\author[m]{Jeffrey A. Newman,}
\author[n]{Ignacio Sevilla,}
\author[e,o]{Sarah Skinner,}
\author[p]{Erica Wagoner}

\author{\\(The LSST Dark Energy Science Collaboration)}

\affiliation[a]{Department of Astrophysical Sciences, Princeton University, Peyton Hall, Princeton NJ 08544-0010, USA}
\affiliation[b]{Department of Physics, University of Oxford, Denys Wilkinson Building, Keble Road, Oxford OX1 3RH, United Kingdom}
\affiliation[c]{Fermi National Accelerator Laboratory, Batavia, IL, 60510, USA}
\affiliation[d]{Department of Physics and Astronomy, University of California, Irvine, CA 92697, USA}
\affiliation[e]{Brookhaven National Laboratory, Physics Department, Upton, NY 11973, USA}
\affiliation[f]{Rutgers University, Physics \& Astronomy Department, Piscataway, NJ 08854, USA}
\affiliation[g]{Department of Physics, Princeton University, Princeton, New Jersey 08544, USA}
\affiliation[h]{McWilliams Center for Cosmology, Department of Physics, Carnegie Mellon University, Pittsburgh, PA 15213, USA}
\affiliation[i]{Institute for Advanced Research, Nagoya University, Nagoya 464-8601, Japan}
\affiliation[j]{Division of Physics and Astrophysical Science, Graduate School of Science, Nagoya University, Nagoya 464-8602, Japan}
\affiliation[k]{Kavli Institute for the Physics and Mathematics of the Universe (Kavli IPMU, WPI), UTIAS, The University of Tokyo, Chiba 277- 8583, Japan}
\affiliation[l]{Jet Propulsion Laboratory, California Institute of Technology, Pasadena, CA 91109, USA}
\affiliation[m]{Department of Physics and Astronomy and PITT PACC, University of Pittsburgh, Pittsburgh, PA, 15260, USA}
\affiliation[n]{Centro de Investigaciones Energ\'eticas, Medioambientales y Tecnol\'ogicas (CIEMAT), Madrid, Spain}
\affiliation[o]{Department of Physics, Missouri University of Science and Technology, Rolla, MO 65401, USA}
\affiliation[p]{Department of Physics, University of Arizona, Tucson, AZ, 85721, USA}

\emailAdd{anicola@astro.princeton.edu}

\abstract{We analyze the clustering of galaxies in the first public data release of the Hyper Suprime-Cam Subaru Strategic Program. Despite the relatively small footprints of the observed fields, the data are an excellent proxy for the very deep photometric datasets that will be acquired by the Large Synoptic Survey Telescope, and are therefore an ideal test bed for the analysis methods being implemented by the LSST Dark Energy Science Collaboration. We select a magnitude limited sample with $i<24.5$ and analyze it in four tomographic redshift bins covering the range $0.15\lesssim z\lesssim1.5$. We carry out a Fourier-space analysis of the two-point clustering of this sample, including all auto- and cross-correlations between bins. We demonstrate the use of map-level deprojection methods to account for non-physical fluctuations in the galaxy number density caused by observational systematics. Through a halo occupation distribution analysis, we place constraints on the characteristic halo masses of this sample as a function of redshift, finding a good fit up to scales $k_{\rm max}=1\,{\rm Mpc}^{-1}$, including both auto- and cross-correlations. Our results show monotonically decreasing average halo masses with increasing redshift, which can be interpreted in terms of the drop-out of red galaxies at high redshifts for a flux-limited sample, consistent with previous analyses. In terms of photometric redshift systematics, we show that additional care is needed in order to marginalize over uncertainties in the redshift distribution in galaxy clustering, even for samples of this small size, and that these uncertainties can be significantly constrained by including cross-bin correlations. We are able to make a $\sim3\sigma$ detection of the effects of lensing magnification in the HSC data. Our results are stable to variations in the amplitude of density fluctuations $\sigma_8$ and the cold dark matter abundance $\Omega_c$ and we find constraints that agree well with measurements from Planck and low-redshift probes. Finally, we use our analysis pipeline to study the clustering of galaxies as a function of limiting flux, and provide a simple fitting function for the linear galaxy bias for magnitude limited samples as a function of limiting magnitude and redshift.}

\begin{document}
\maketitle
\flushbottom

\section{Introduction}\label{sec:intro}
  The past two decades have seen a revolution in our understanding of the Universe and its constituents. While observations of the cosmic microwave background (CMB) play a pivotal role in anchoring the standard  cosmological model \cite{Planck:2018}, they are not a direct probe of the physics of the low-redshift Universe. In order to directly characterize properties of dark energy and modified gravity, we need to measure the expansion history and growth  \cite{1903.12016}. At the moment, this is only possible with optical large scale structure surveys.

  Over a decade ago, the Dark Energy Task Force divided the evolution of optical experiments into approximate stages \cite{0609591}. The current Stage III experiments, including spectroscopic surveys such as eBOSS \cite{1707.09322,Zhu:2018,1712.08064,1801.02689,1801.02891} and VIPERS \cite{1611.07048,1612.05645,1708.00026} and photometric surveys such as DES \cite{1708.01531,1708.01530,1708.01536}, the Hyper Suprime-Cam survey \cite{1704.05858,2019PASJ...71...43H,1906.06041} and the Kilo-Degree Survey / VIKING-450 \cite{1902.11265,1812.06077,Joudaki:2019} are approaching completion and the field is preparing for the Stage IV surveys, such as the ground-based DESI \cite{1611.00036,1907.10688} and LSST \cite{0912.0201,1809.01669, Ivezic:2019}, as well as major satellite missions like Euclid \cite{1606.00180} and WFIRST \cite{1904.01174}.

  Together with transformational sensitivity increases in the Stage IV surveys, the challenges of understanding and controlling systematic biases and uncertainties are becoming considerably more difficult \cite{1808.07335}. For photometric surveys, the deep surveys sensitive to many more sources inevitably come with crowded fields where blending affects a significant fraction of all sources. The increased blending leads to new uncertainties in isolating sources, measuring their fluxes (required for photometric redshifts) and inferring their weak gravitational lensing shear estimates \citep{2016ApJ...816...11D,2018MNRAS.475.4524S,2019MNRAS.483.2487J}. These effects in turn lead to subtle sample selection effects, which are amplified by the interaction of the point spread function (PSF) with blending \cite{1708.01533,1905.01324,1907.10572}. But even with perfect measurements of fluxes, our ability to infer source redshift distributions is hampered by incompleteness in spectroscopic samples used to train and calibrate redshifts \cite{1903.09325}. Finally, at the depths of Stage IV photometric surveys, instrumental and observational effects, such as scattered light from bright objects, errors in star-galaxy separation, reddening by Milky Way dust, and varying observing conditions, will imprint spurious fluctuations on the observed galaxy density field. These fluctuations will, if not correctly accounted for, masquerade as intrinsic large-scale fluctuations in the cosmic density field. Therefore, Stage IV photometric experiments are likely to be limited by our ability to remove systematic biases and minimize systematic uncertainties rather than the intrinsic statistical power of the survey.

  The community is well aware of these challenges lying ahead. There are two main approaches to preparing for the arrival of data. On one hand, we are building sophisticated, realistic mock data sets, such as LSST Dark Energy Science Collaboration (DESC) Data Challenge 2 \cite{1909.07340,1907.06530}. On the other hand, we are (re-)analyzing existing precursor data to validate our methods and codes in a realistic environment. In this paper, we focus on the latter approach and perform an analysis of photometric galaxy clustering using the first data release (DR1) of the Hyper Suprime-Cam Subaru Strategic Program (HSC-SSP) \cite{2018PASJ...70S...8A}. This dataset bears significant similarities to the data expected from LSST, both in terms of survey depth and photometric bands as well as primary data reduction and catalog generation. Specifically, HSC covers essentially the same bands as LSST, with the exception of missing the UV $u$-band fluxes. In this work, we focus on a magnitude-limited galaxy sample with limiting magnitude $i<24.5$, which is similar to the expected 5-$\sigma$ detection limit for LSST after one year \cite{1809.01669} ($i<25.1$). Finally, the primary data reduction codes used in HSC-SSP and planned for LSST are of the same lineage, employing many of the same methods \cite{2018PASJ...70S...5B,1812.03248}. With the exception of a rather small sky area of around 90 square degrees used in this paper, the HSC-SSP data is therefore a perfect proxy for future LSST data. This analysis has the additional attraction that no other photometric galaxy clustering study has been carried out with these data.

  In this work, we present an analysis of photometric galaxy clustering using HSC DR1 data\footnote{We note that we do not blind our results in any stage of the analysis.}. We split the data into four tomographic redshift bins and measure both the auto- and cross-power spectra of all bins, correcting for observational systematics using mode deprojection. We then use the halo model coupled with a halo occupation distribution to fit the data and derive constraints on astrophysical parameters, marginalizing over photometric redshift (photo-$z$) uncertainties. To ensure the stability of our results, we perform a suite of complementary analyses and find our constraints to be robust to photometric redshift uncertainties. We additionally perform extended analyses in which we compute constraints on lensing magnification in our sample and allow for variations in the $\Lambda$CDM cosmological parameters, respectively. Finally, we use our data to derive a simple fitting function for the linear galaxy bias as a function of redshift and limiting magnitude of the sample.

  In our analysis, we make several simplifying approximations; in particular we employ a simple semi-analytical halo model to derive theoretical predictions for the data, which will most certainly not be sophisticated enough for the full sky LSST analysis. The analysis of LSST data will likely require  a combination of  bias expansion approaches \cite{0902.0991,1402.5916,1611.09787,1910.07097} and power spectrum emulators \cite{1804.05865,1705.03388}. However, we find that a simple halo model suffices to describe the data at the current level of precision.

  This paper is structured as follows. In Section \ref{sec:data}, we describe the HSC-SSP data and the generation of a set of galaxy over-density maps together with maps of potential systematics. In Section \ref{sec:methods}, we detail the methodology employed in our analysis and in Section \ref{sec:results}, we present results alongside numerous tests. We discuss and conclude in Section \ref{sec:discussion}. More detailed descriptions of the generation of systematics maps are deferred to the Appendix. 

\section{Data}\label{sec:data}
  \begin{table}
  \centering
  \begin{tabular}{|l|l|}
  \hline
   {\bf Cut} & {\bf Comment} \\
   \hline
   \texttt{detect\_is\_primary}=True & Basic quality cuts, \\
   \texttt{icmodel\_flags\_badcentroid}=False & see \cite{2018PASJ...70S..25M,2018PASJ...70S...5B}\\
   \texttt{icentroid\_sdss\_flags}=False & \\
   \texttt{iflags\_pixel\_edge}=False & \\
   \texttt{iflags\_pixel\_interpolated\_center}=False & \\
   \texttt{iflags\_pixel\_saturated\_center}=False & \\
   \texttt{iflags\_pixel\_cr\_center}=False & \\
   \texttt{iflags\_pixel\_bad}=False & \\
   \texttt{iflags\_pixel\_suspect\_center}=False & \\
   \texttt{iflags\_pixel\_clipped\_any}=False & \\
   \texttt{meas.ideblend\_skipped}=False & \\
   \texttt{iblendedness\_abs\_flux}$<10^{-0.375}$ & \\
   \hline 
   \texttt{[g,r,z,y]centroid\_sdss\_flags}=False & Strict photometry cuts\\
   \texttt{[g,r,i,z,y]cmodel\_flux\_flags}=False & \\
   \texttt{[g,r,i,z,y]flux\_psf\_flags}=False & \\
   \texttt{[g,r,z,y]flags\_pixel\_edge}=False & \\
   \texttt{[g,r,z,y]flags\_pixel\_interpolated\_center}=False & \\
   \texttt{[g,r,z,y]flags\_pixel\_saturated\_center}=False & \\
   \texttt{[g,r,z,y]flags\_pixel\_cr\_center}=False & \\
   \texttt{[g,r,z,y]flags\_pixel\_bad}=False & \\
   \hline
   \texttt{icmodel\_mag}$-$\texttt{a\_i} $<$ 24.5 & Magnitude limit\\ 
   \hline
   \texttt{icmodel\_flux}$>10\,$\texttt{icmodel\_flux\_err} & 10$\sigma$ detections\\
   \hline
   \texttt{[g,r,y,z]cmodel\_flux}$>5\,$\texttt{[g,r,y,z]cmodel\_flux\_err} & 5$\sigma$ detection  (required \\
   & only in 2 other bands)\\
   \hline
   \texttt{iclassification\_extendedness}=1 & Star-galaxy separator\\
   \hline
  \end{tabular}
  \caption{Summary of the selection cuts performed to the original dataset to retrieve the sample considered in our analysis.} \label{tab:cuts_summary}
  \end{table}  

  \begin{table}
  \centering
  \begin{tabular}{|l|r|c|l|}
   \hline
   {\bf Field name} & $N_{\rm gal}$ & {\bf Area} (deg$^2$) & $f_{\rm sky}$ \\
   \hline
   GAMA09H  & 1,697,713 & 14.5 & $3.5\times10^{-4}$ \\
   GAMA15H  & 1,695,364 & 15.1 & $3.7\times10^{-4}$ \\
   HECTOMAP &   639,970 &  5.1 & $1.2\times10^{-4}$ \\
   VVDS     & 2,340,965 & 20.6 & $5.0\times10^{-4}$ \\
   WIDE12H  & 1,220,816 & 11.6 & $2.8\times10^{-4}$ \\
   XMM-LSS   & 2,139,629 & 20.8 & $5.0\times10^{-4}$ \\
   \hline
   Total    & 9,734,457 & 87.7 & $2.1\times10^{-3}$ \\
   \hline
  \end{tabular}
  \caption{Summary of the 6 different fields used in our analysis. The second column lists the number of galaxies in the DR1 catalog passing the cuts in Table \ref{tab:cuts_summary}. The third and fourth columns show the area and corresponding sky fraction covered by each field. The area was calculated as the sum of pixel areas allowed by the sky mask described in Section \ref{ssec:methods.mask}. We note that the HECTOMAP field is not included in our analysis due to its smaller area.} \label{tab:field_summary}
  \end{table}  

  \begin{table}
  \centering
  \begin{tabular}{|l|l|l|l|}
    \hline
    $z_{\rm phot}^{\rm ini}$ & $z_{\rm phot}^{\rm end}$ & $\bar{z}$ & $N_{\rm gal}$ \\
    \hline
    0.15 & 0.5  & 0.57 & 1,750,274 \\
    0.5  & 0.75 & 0.68 & 1,766,939 \\
    0.75 & 1.0  & 0.91 & 1,702,685 \\
    1.0  & 1.5  & 1.26 & 1,752,359 \\
    \hline
  \end{tabular}
  \caption{Summary of the 4 redshift bins used in our analysis. The first two columns show the photo-$z$ bin edges. We used the {\tt photoz\_best} redshift estimator for the photo-$z$ code {\tt Ephor\_AB} to assign galaxies to different bins. The third column shows the mean redshift of each bin calculated from the fiducial redshift distributions described in Section \ref{ssec:methods.nz}. The last column shows the number of galaxies in each bin. The bin edges were chosen to roughly contain an equal number of galaxies in each of them.} \label{tab:bins_summary}
  \end{table}
  
  In this work, we use data from the first data release of the Hyper Suprime-Cam Subaru Strategic Program (HSC DR1 hereon)\footnote{\url{https://hsc-release.mtk.nao.ac.jp}.}. The release is extensively documented in \cite{2018PASJ...70S...8A}, and here we only provide the details of the galaxy sample and associated data used for our clustering analysis.

  HSC-SSP is a photometric galaxy survey that has been awarded 300 nights on the Subaru Telescope starting in 2014. DR1 includes data from 61.5 nights observed to three different depths: Wide (108 square degrees to $i\sim 26.4$), Deep (26 square degrees to $i\sim26.5$) and UltraDeep (4 square degrees to $i\sim 27$). In this work we focus on the HSC Wide field, which has been observed in five broadband filters ($grizy$) and is distributed among 6 fields  of areas varying between 5 and 20 square degrees. The image quality is impressive with a median $i$-band seeing of around 0.6 arcsec, which is considerably better than other comparable surveys (e.g. DES with around 0.9 arcsec in the $riz$-bands) and also likely better than the median seeing expected to be achieved by LSST. The data are processed with \texttt{hscPipe} \cite{2018PASJ...70S...5B} and are available to the community through a public database.
  
  We use a magnitude-limited sample constructed from the HSC Wide DR1 sample by imposing data cuts that are similar to those used to create the HSC shear catalog \cite{2018PASJ...70S..25M}. The exact cuts are shown in Table~\ref{tab:cuts_summary}, and can be summarized as follows: besides a minimal set of quality cuts (selecting only primary detections with well-measured fluxes in all bands, removing objects near bad pixels, deblender artifacts, etc.), we impose an overall apparent magnitude cut in the extinction-corrected band $i_{\rm corr}<24.5$. This choice was based on a study of the survey depth-completeness relation in order to select a homogeneous and complete sample of high-confidence ($>10\sigma$) detections (see Section \ref{ssec:methods.syst}). We also select only objects with significant detections ($>5\sigma$) in at least two of the 4 remaining bands ($g,\,r,\,z,\,y$), and remove all objects classified as stars by the data reduction pipeline, using the ``extendedness'' classifier as described in \cite{2018PASJ...70S..25M,2018PASJ...70S...5B}. The resulting sample consists of 9,734,457 objects and covers $\sim88$ square degrees distributed across the 6 HSC DR1 fields, as described in Table \ref{tab:field_summary}. The $i$-band magnitude cut is by far the most stringent one, discarding approximately half of the total sample.

  All objects have photometric redshift measurements from 6 different codes as presented in \cite{2018PASJ...70S...9T}. We use the {\tt photoz\_best} redshift estimator assigned by the {\tt Ephor\_AB} method as a marker to divide the sample into four tomographic samples containing roughly equal galaxy numbers. The {\tt photoz\_best} estimator is defined to minimize the risk that the true galaxy redshift lies outside the range $z_{\rm true}\pm 0.15(1+z_{\rm true})$, where $z_{\rm true}$ is the galaxy's true redshift. A preliminary Fisher matrix \cite{Fisher:1935} study showed that the information content saturates quickly when slicing the data into more than four samples. The redshift bins are described in Table \ref{tab:bins_summary}, and the associated redshift distributions are discussed in Section \ref{ssec:methods.nz}.

\section{Methods}\label{sec:methods}

Our basic method is to measure the two point function of all possible combinations of galaxy density fluctuation fields. We work in the Fourier domain, so our basic quantity is the angular power spectrum $C_\ell$. Given four tomographic samples, our measurement consists of four auto-power spectra and six cross-power spectra. In this section we discuss how we construct maps of galaxy density fluctuations and associated maps of potential systematics, how we use these to measure the power spectra and their covariance matrix and finally how these are modeled within the context of the halo model.

  \subsection{Pixels and maps}\label{ssec:methods.pix}
    Our $C_\ell$-based analysis requires us to make maps of different quantities. The HSC DR1 is distributed across the 6 small fields ($\lesssim20\,{\rm deg.}^2$) summarized in Table \ref{tab:field_summary}. In this case, storing maps covering the full sky down to arcminute resolution, and performing operations on them such as spherical harmonic transforms, would be computationally inefficient and unnecessary. Instead, we perform separate power spectrum measurements on each individual field, with maps defined on rectangular sky patches covering them. Furthermore, the small size of each field allows us to make use of the flat-sky approximation safely \citep{2014JCAP...10..007N}. This leads to additional gains in speed, since spherical harmonic transforms can be replaced by the far more efficient fast Fourier transforms (FFTs). Since we use a cilindrical projection with the equator at the center of each of our fields to define our flat-sky coordinates (rather than the more common gnomonic projection \citep{2002A&A...395.1077C}), the main source of curved-sky distortions occur at large latitudes, which are small ($\lesssim 1.5^\circ$) in all cases.
    
    To generate maps of all the quantities described in this section, we make use of a rectangular pixelization scheme using the Plate Carr\'ee projection (labelled {\tt CAR} in the World Coordinate System standard \cite{2002A&A...395.1077C}). In this case, pixels are simply defined by equal intervals of colatitude $\theta$ and azimuth $\phi$. To minimize the distortions caused by the flat-sky approximation, we place the projection reference point (i.e. a point in the equator $\theta=\pi/2$) at the center of each field. We use square pixels of size $\alpha_{\rm pix}=0.6$ arcmin on a side, corresponding to a Nyquist frequency $\ell_{\rm Nyquist}=18,000$. The pixel size was chosen as a compromise between the need to have several galaxies in each pixel on average and the need to study small-scale clustering (the corresponding comoving wavenumber is $k\sim5\,{\rm Mpc}^{-1}$ at redshift $z=1$), as well as to describe the variation in survey coverage accurately. The maps are defined on a rectangular patch large enough to cover all objects in each field, leaving a buffer of 10 masked pixels on all edges to avoid boundary effects when computing the FFTs.
  
  \subsection{Survey mask}\label{ssec:methods.mask}
    \begin{figure}
      \centering
      \includegraphics[width=0.49\textwidth]{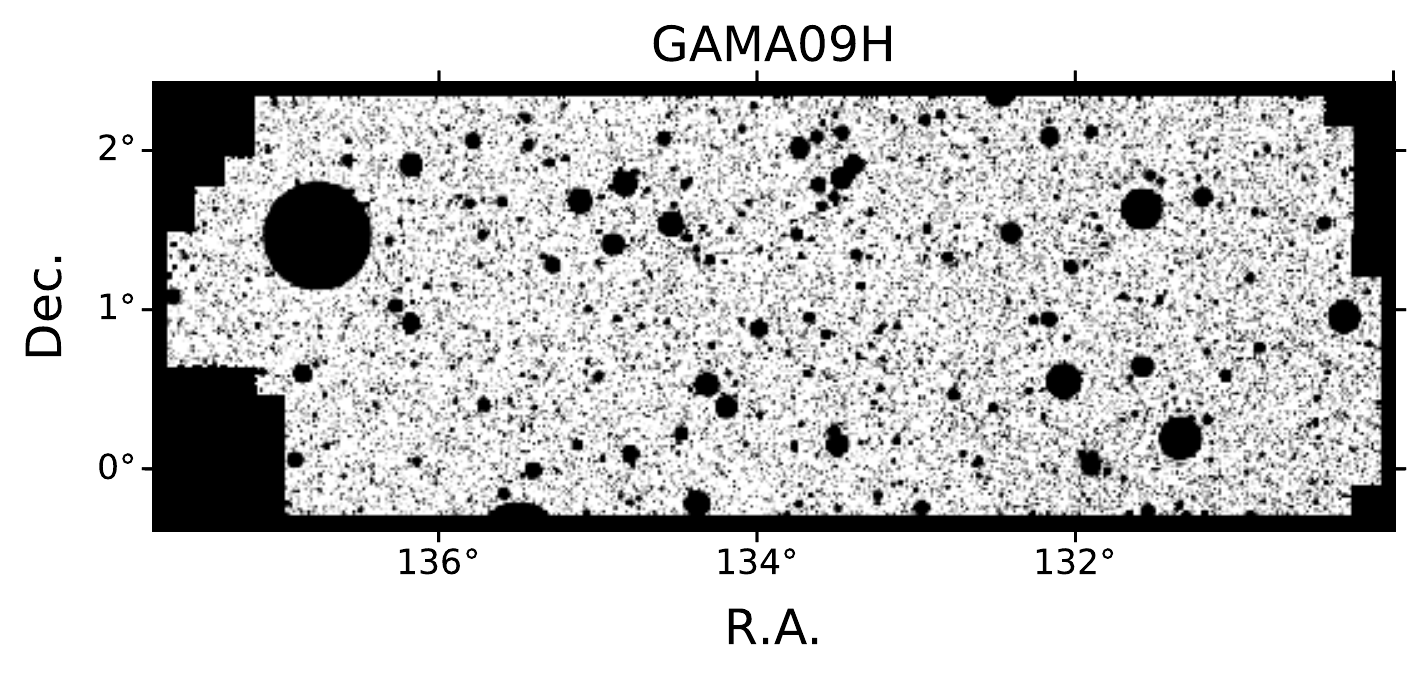}
      \includegraphics[width=0.49\textwidth]{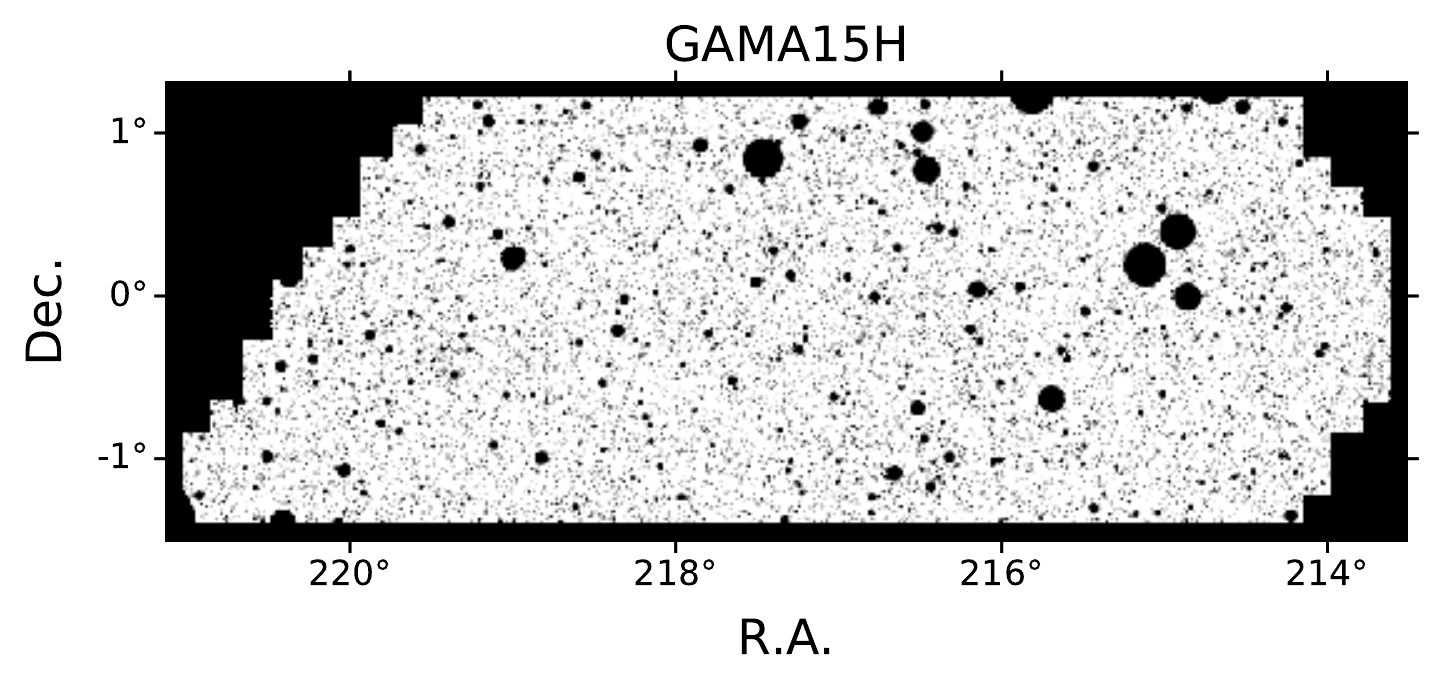}
      \includegraphics[width=0.49\textwidth]{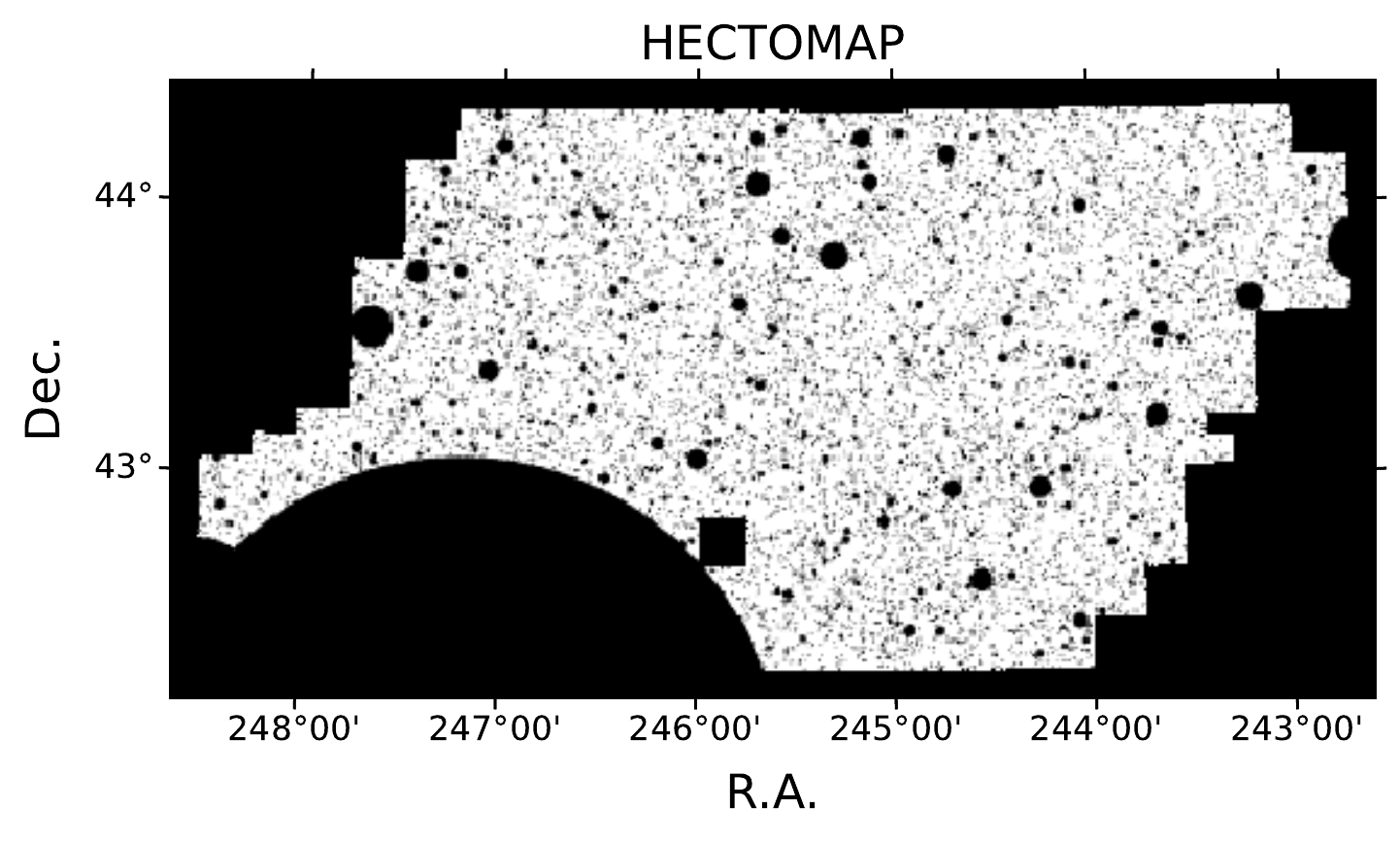}
      \includegraphics[width=0.49\textwidth]{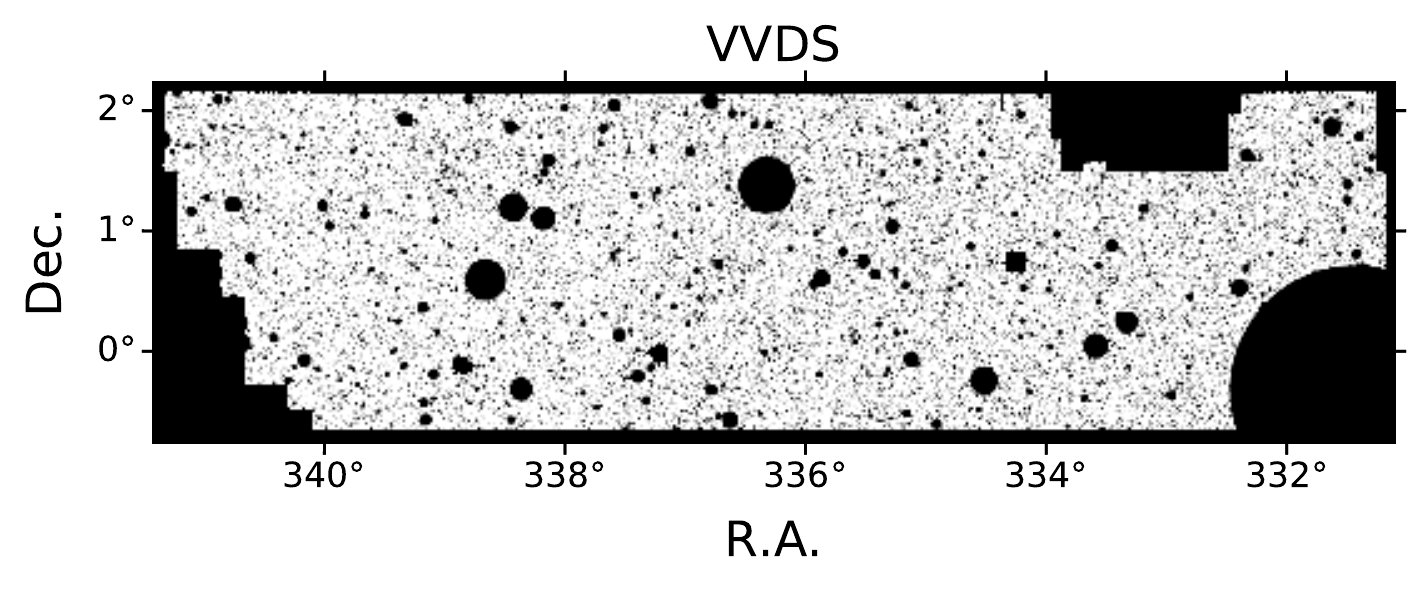}
      \includegraphics[width=0.49\textwidth]{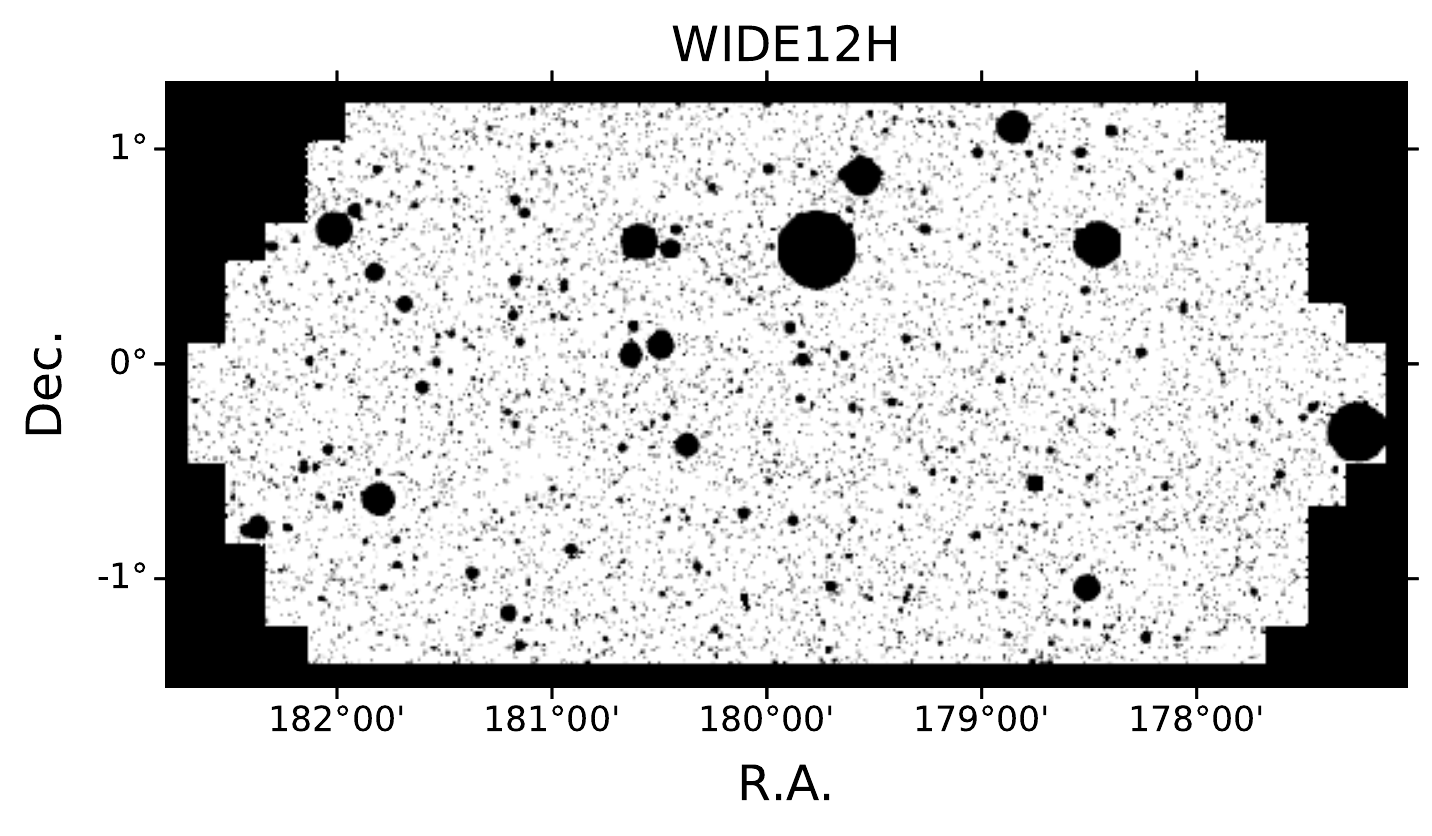}
      \includegraphics[width=0.49\textwidth]{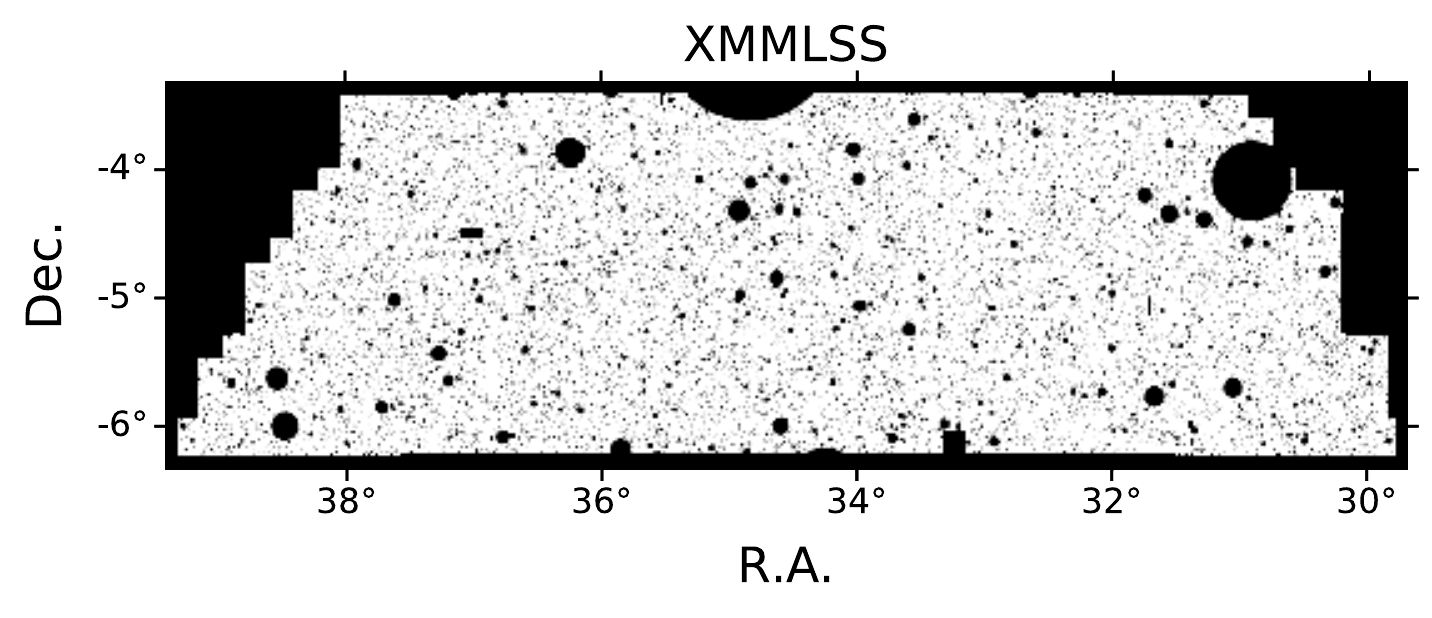}
      \caption{Sky masks for the 6 fields used in this analysis. The map pixels contain values between 0 and 1, corresponding to the fraction of the pixel's area not covered by the fiducial bright-object mask.}
      \label{fig:masks}
    \end{figure}
    The reconstruction of the survey geometry is a central step in order to obtain unbiased estimates of the angular power spectrum. This information is encoded in the so-called ``survey mask'', which minimally contains binary information about which areas of the sky should (mask $=1$) or should not (mask $=0$) be used in the analysis. The basis for our survey mask is the so-called ``bright-object mask'' \cite{2018PASJ...70S...8A}, provided with the HSC DR1, which flags sources that are close to bright stars (mag $<17.5$), with a magnitude-dependent exclusion radius (the so-called ``Sirius'' mask, see \cite{2018PASJ...70S...8A} for details). The information about the bright object mask is encoded in the HSC DR1 at the catalog level in terms of per-object flags. We transform this information into a pixelized sky map through a multi-step process:
    \begin{enumerate}
      \item We start by creating a low-resolution binary mask based on the presence of objects from the raw catalog in a given pixel. This mask has a pixel size of 0.6 arcmin. The large number density of the raw catalog  ($n_g\sim30\,{\rm arcmin}^{-2}$) is high enough that masked pixels are unlikely to correspond to intrinsically empty regions of the sky, but rather completely unobserved pixels.
      \item We upgrade the low-resolution binary mask to a higher resolution ($0.2$ arcmin), and remove all pixels containing objects flagged by the bright object mask. We then remove all disconnected, unmasked groups of pixels, corresponding to spurious islands within the exclusion radius of a bright object with no sources in the catalogs.
      \item We downgrade the resulting mask back to the original resolution through an averaging procedure, producing a map quantifying the observed fraction of each pixel.
    \end{enumerate}
    We verified that the resulting mask is robust by comparing the obtained power spectra to those found with an observed fraction map defined simply as the fraction of objects in each pixel outside the bright object mask. The resolution of our fiducial mask ($0.6$ arcminutes) defines the resolution of all maps used in this analysis. After this procedure, we further mask all pixels with a $10\sigma$ depth below our magnitude limit of $i<24.5$, where the depth is estimated as described in Section \ref{ssec:methods.syst}.

    This defines our fiducial masks, which are shown in Figure \ref{fig:masks} for each field. As part of our analysis of systematic biases (see Section \ref{sssec:results.spectra.syst}), we also study the effect of masking regions with significant contamination from the different observing conditions (described in Section \ref{ssec:methods.syst}) on our measurements. In addition it has been observed that the NOMAD star catalog \cite{2004AAS...205.4815Z}, which is one of the datasets used to construct the bright-object mask described above, is contaminated by a small fraction of bright nearby galaxies ($\sim10\%$). To study the impact of this contamination on our clustering measurements, we therefore also estimate the angular power spectra using a more recent version of the star mask (the so-called ``Arcturus'' mask described in \cite{2018PASJ...70S...7C}).

  \subsection{Systematics maps}\label{ssec:methods.syst}
    \begin{figure}
      \centering
      \includegraphics[width=0.49\textwidth]{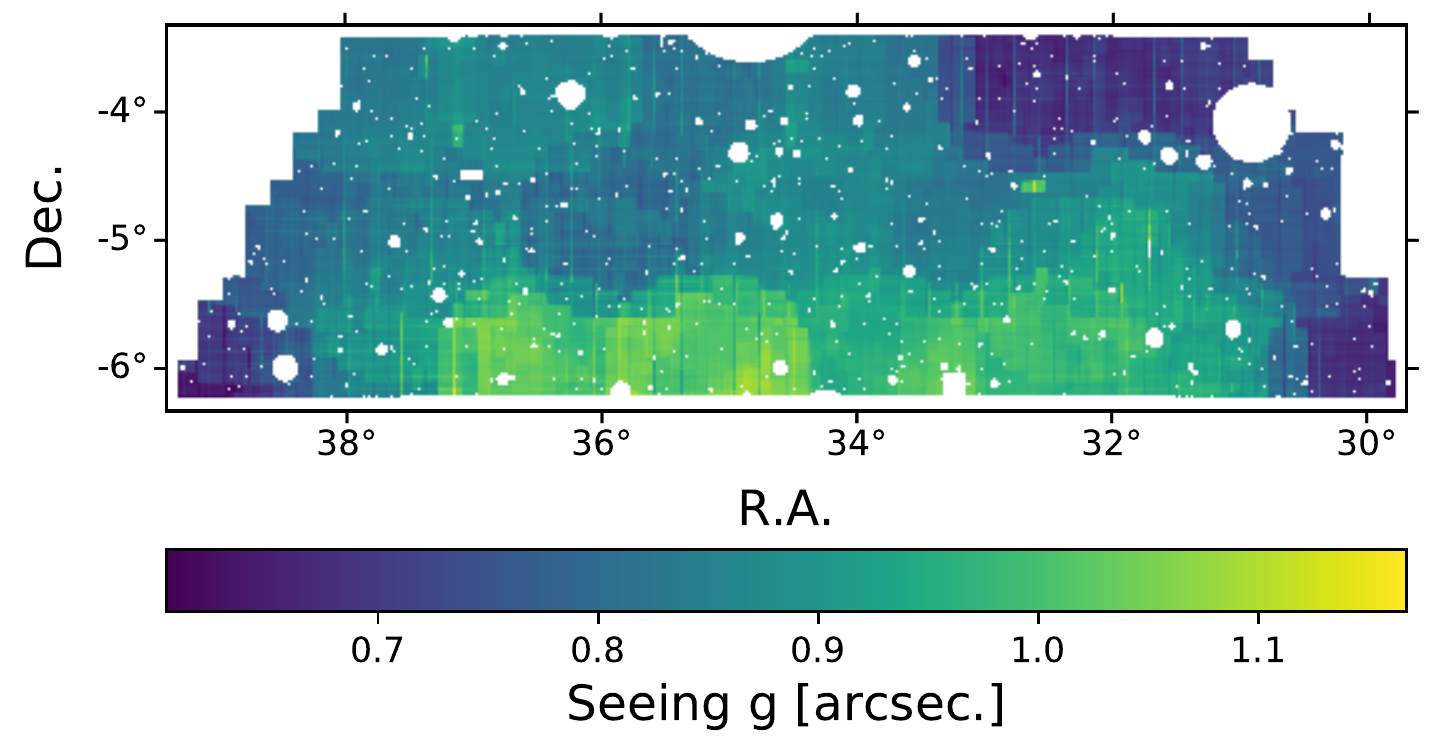}
      \includegraphics[width=0.49\textwidth]{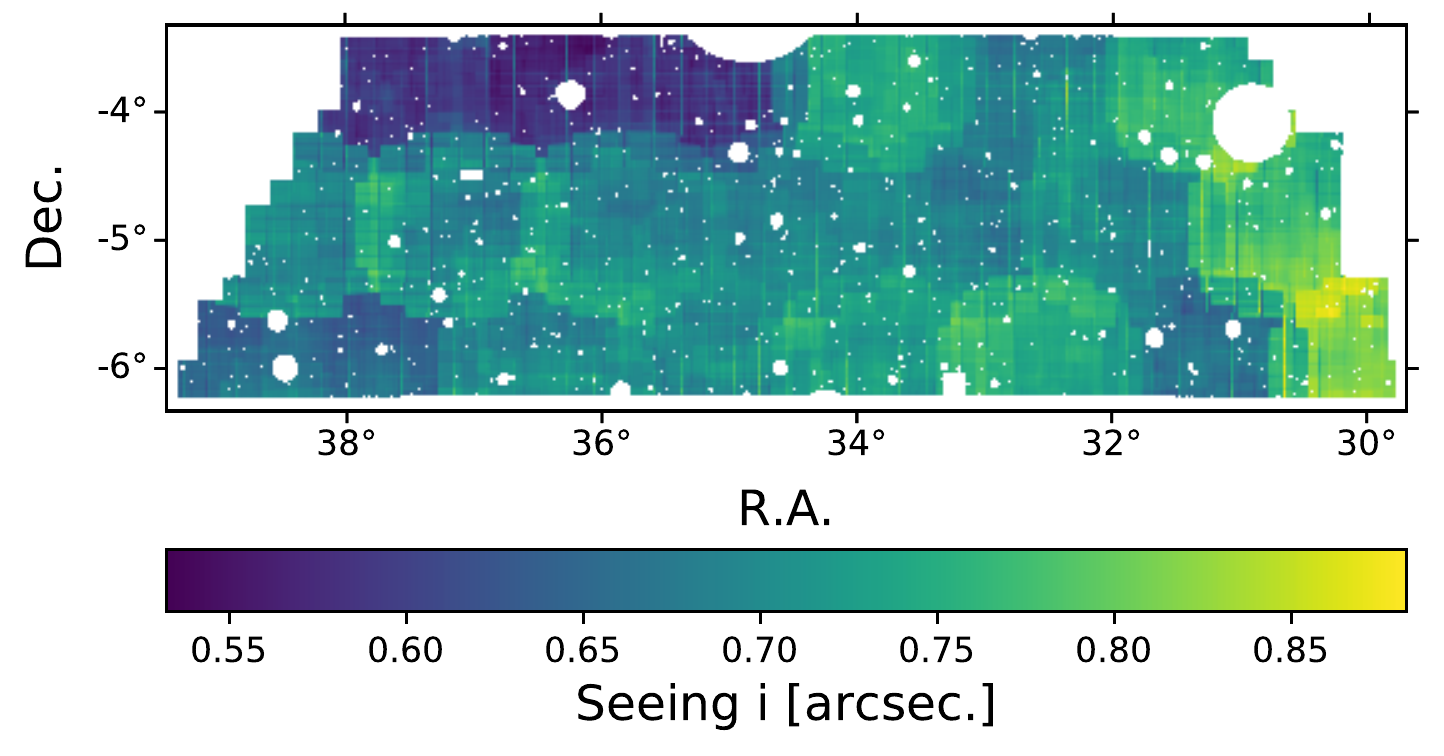}
      \includegraphics[width=0.49\textwidth]{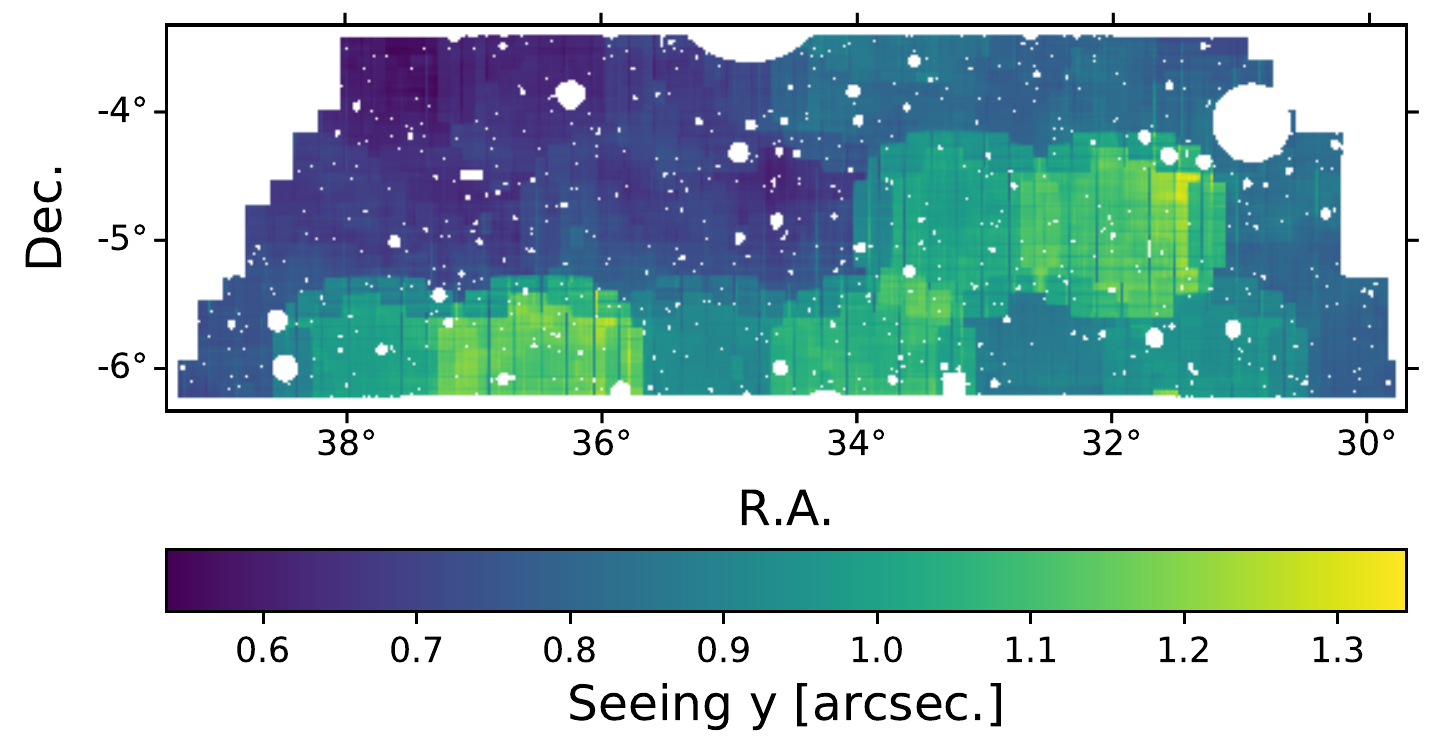}
      \includegraphics[width=0.49\textwidth]{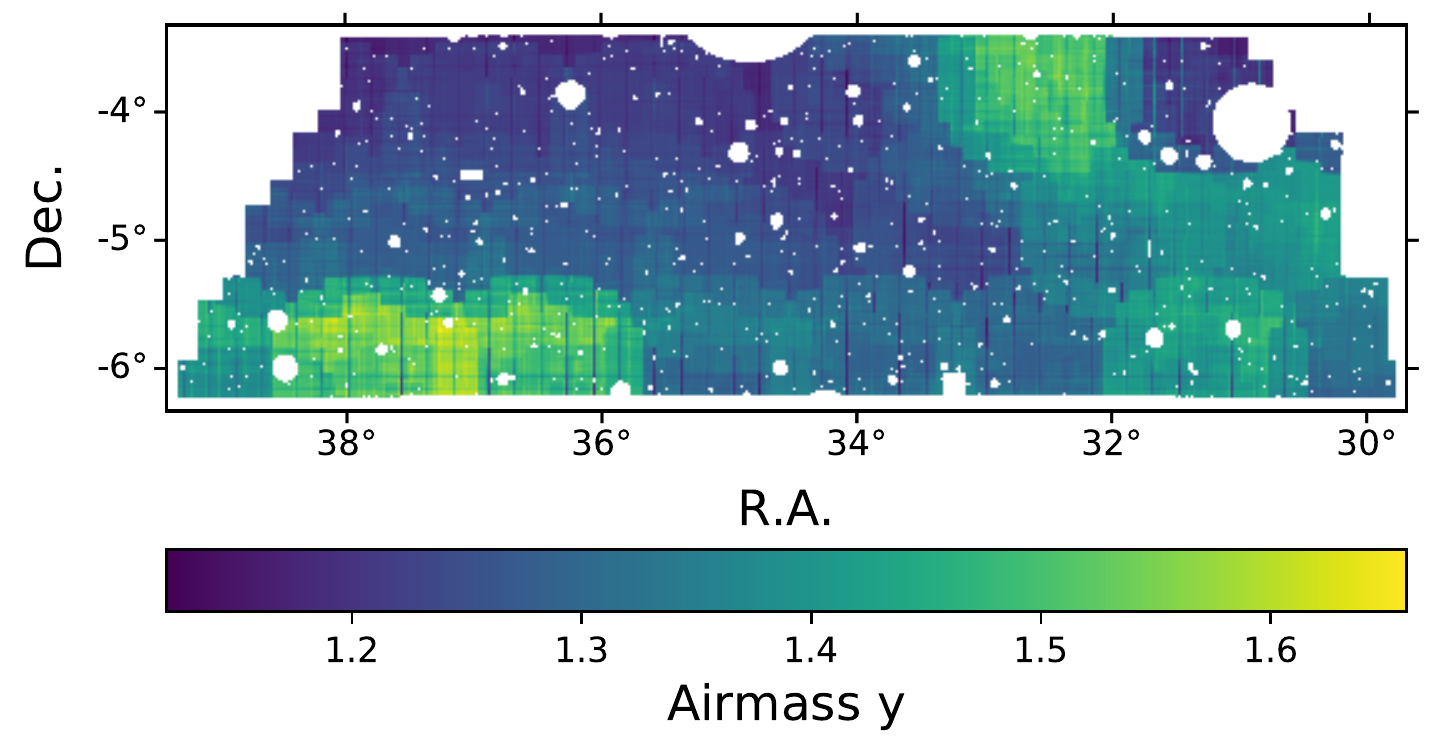}
      \includegraphics[width=0.49\textwidth]{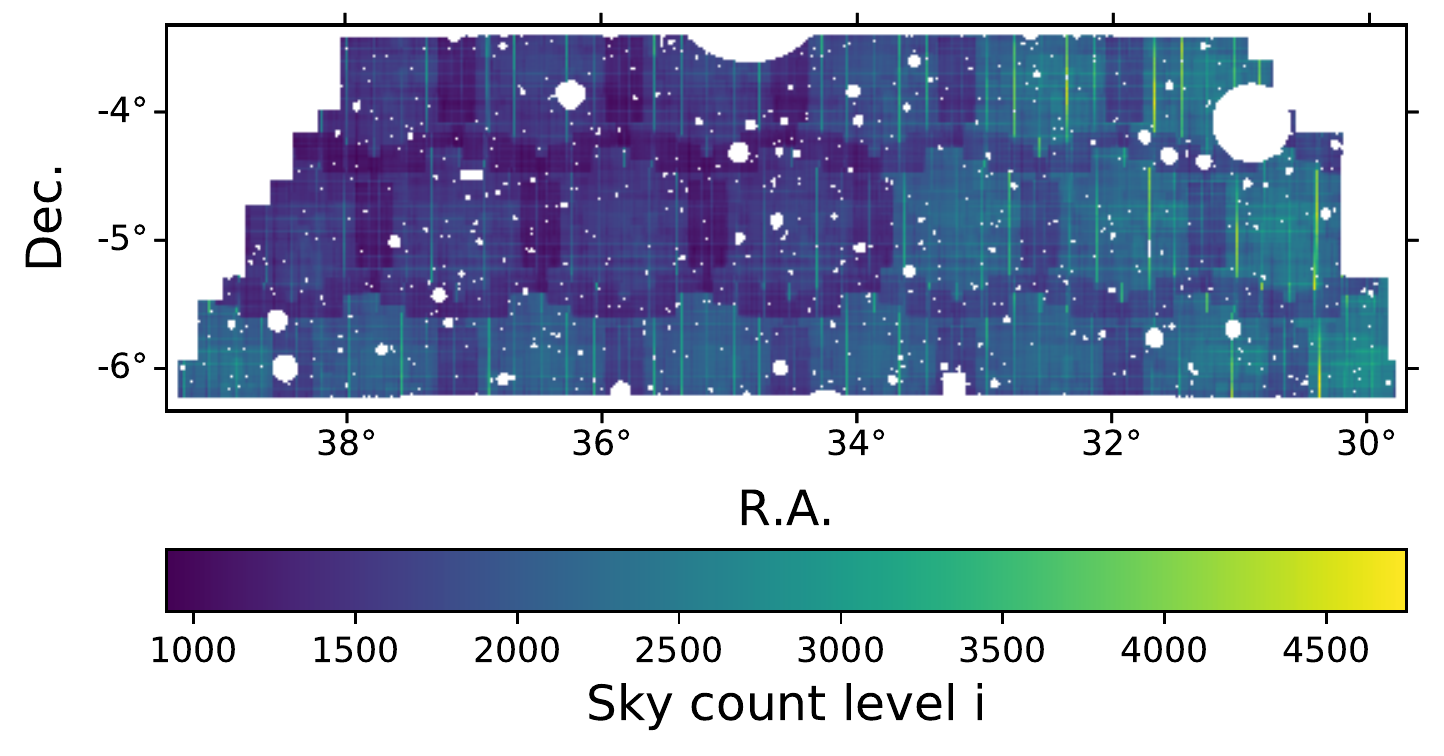}
      \includegraphics[width=0.49\textwidth]{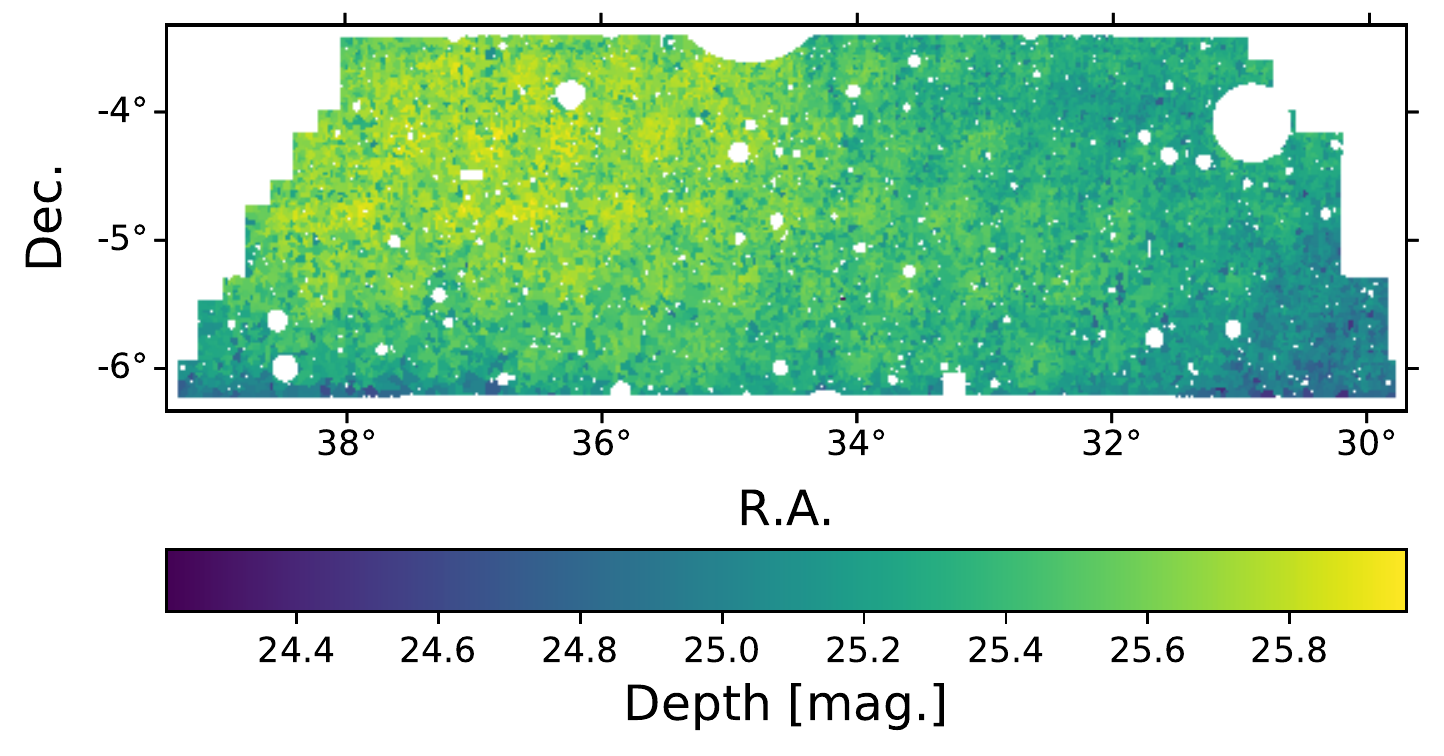}
      \includegraphics[width=0.49\textwidth]{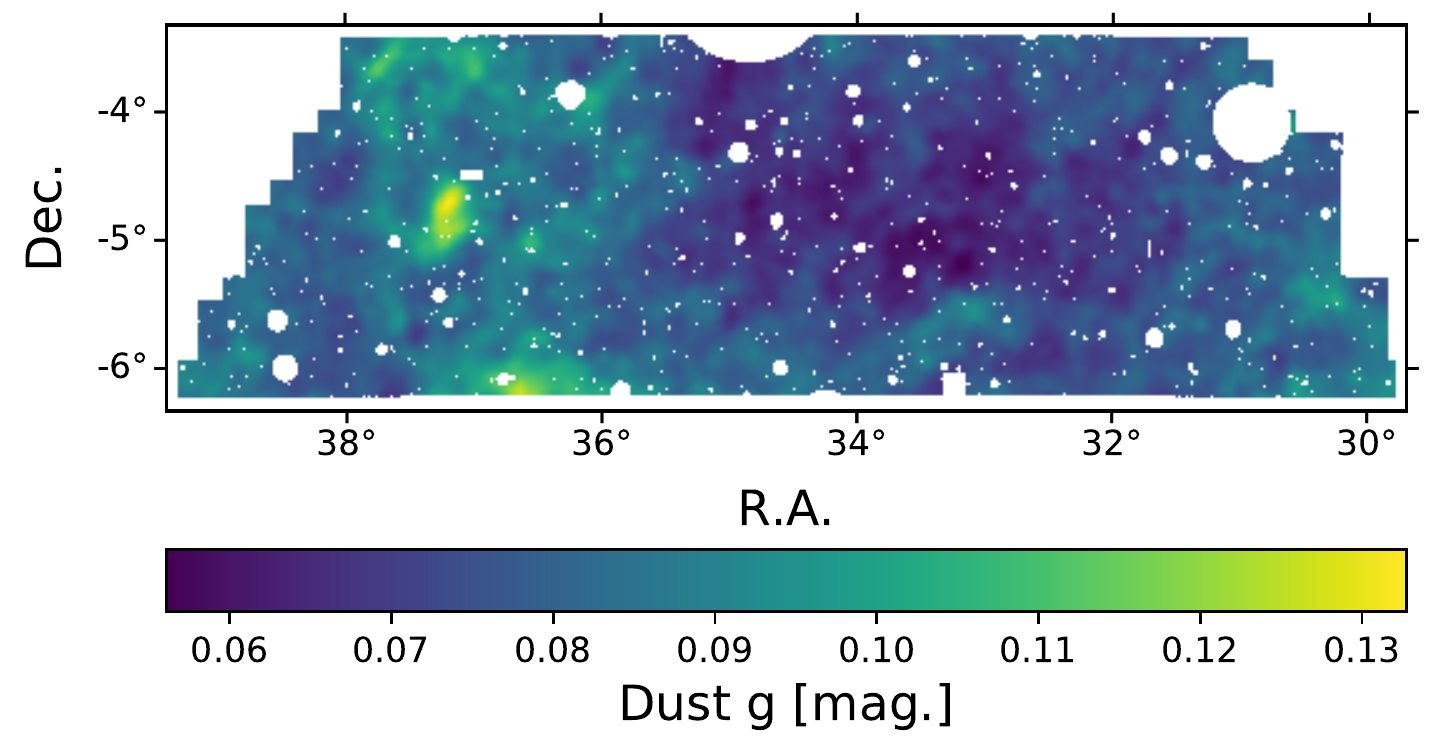}
      \includegraphics[width=0.49\textwidth]{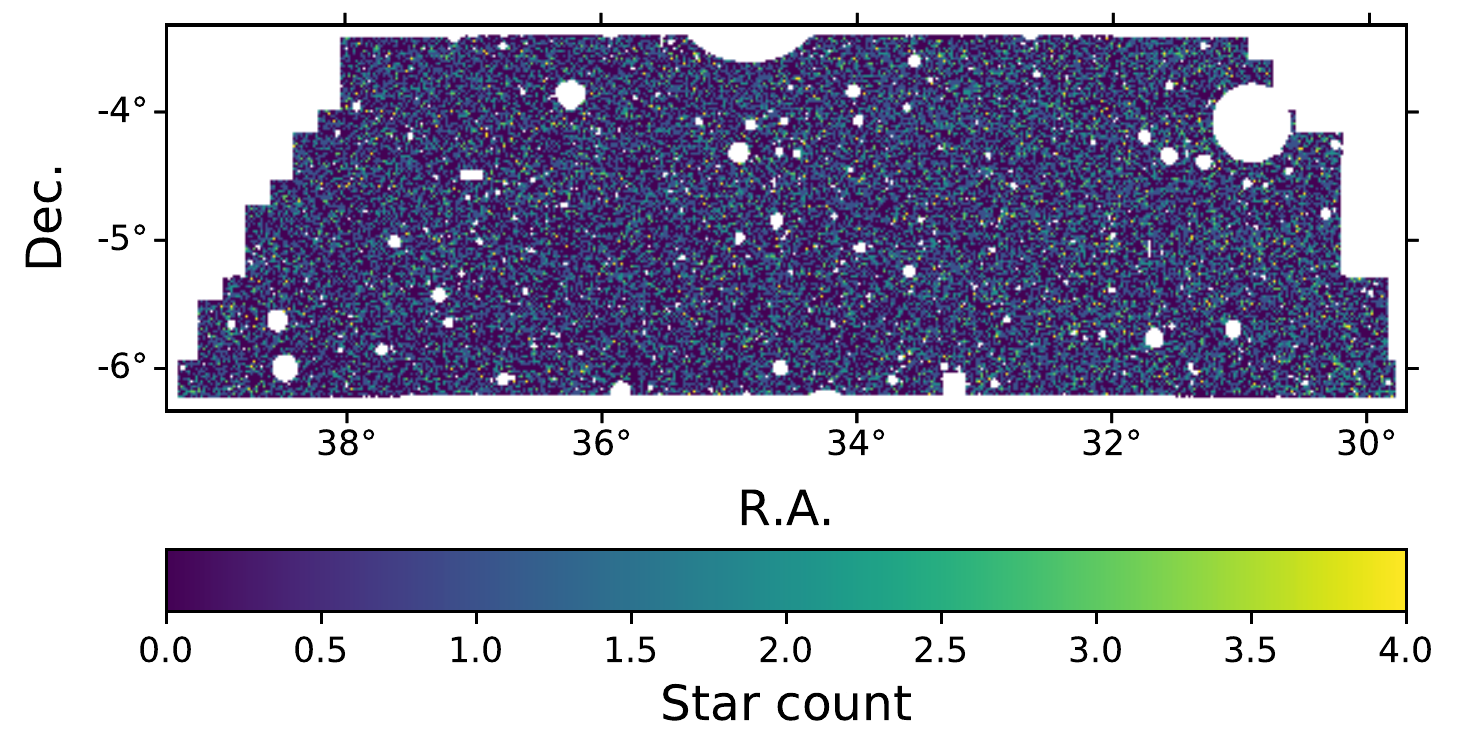}
      \caption{Maps of different observational systematics that could cause an artificial modulation in the inferred galaxy overdensity. The maps correspond to the XMM-LSS field, and were obtained as described in Section \ref{ssec:methods.syst}. From left to right and top to bottom, the different panels show maps of the seeing in the $g$, $i$ and $y$ bands, airmass in the $y$ band, sky level in the $i$ band, $10\sigma$ $i$-band depth, dust absorption and star counts per pixel. We can visually appreciate the existing correlations between different systematics (e.g. seeing and airmass), which are automatically taken into account by the deprojection method described in Section \ref{ssec:methods.cell} and in \cite{2019MNRAS.484.4127A}.}
      \label{fig:sysmap}
    \end{figure}
    A number of astrophysical and observational systematic effects can modify the observed distribution of galaxies, and therefore may bias our inference of their clustering properties. To mitigate this effect, we deproject maps of these systematics from our data, as described in Section \ref{ssec:methods.cell}. We generate maps of the most plausible sources of systematic variations in the galaxy number density:
    \begin{enumerate}
      \item {\bf Survey depth}: we estimate the $10\sigma$ survey depth as a function of angular position on the sky as follows. For each pixel, we compute the average $i$-band {\tt cmodel} flux error of all objects classified as stars that fall into it\footnote{It is worth noting that {\tt cmodel} flux errors are underestimated for galaxies. Nevertheless we use them to define our depth map due to the fact that our sample selection is based on {\tt cmodel} fluxes and magnitudes. Additionally, the corresponding depth map should still provide an appropriate estimate of the relative spatial variations in depth, which is its main role as a contaminant map for clustering measurements.}. The corresponding survey depth is then given by this mean flux error multiplied by 10 and translated into a magnitude. Not all pixels contain enough stars to estimate a reliable mean flux error. We use a nearest-neighbor interpolation to assign a depth value to those pixels that contain fewer than 4 stars. We verified that we obtain similar depth estimates using other methods, such as that outlined in \cite{2018PASJ...70S..25M}, in which 10$\sigma$ depth is defined as the mean {\tt cmodel} magnitude of all sources with $i$-band signal-to-noise ratio between 9 and 11. We emphasize that our method uses only stars, discarding all sources classified as galaxies. The use of galaxies to estimate any systematics map is dangerous in the context of template deprojection, since the stochastic noise in the map associated with the discrete sampling of these sources correlates directly with the galaxy distribution, and can give rise to spurious correlations that bias the estimated power spectrum after deprojection (see Section \ref{ssec:methods.cell}). We provide further details in Appendix \ref{app:depth}.
      \item {\bf Dust extinction}: we make maps of dust absorption in each band using the data from \cite{1998ApJ...500..525S} projected onto the 6 HSC DR1 fields.
      \item {\bf Star contamination}: we produce a map of the number density of stars using all objects passing our sample cuts but classified as stars by the star-galaxy separator. As described in Section \ref{ssec:methods.cell}, the method used to remove the impact of this systematic is slightly different from the rest. Note that, if a small number of galaxies have been mis-classified as stars, deprojecting this map could lead to an artificial underestimation of the clustering amplitude. We have however verified that the effect on the power spectra of deprojecting the star map is negligible (see Section \ref{sssec:results.spectra.syst}). Furthermore, the purity and completeness of the star and galaxy samples associated with this classifier are high for the magnitude range considered in our analysis ($\gtrsim 95 \%$ for galaxies), as can be seen from Fig. 17 in Ref.~\cite{2018PASJ...70S...5B}.
      \item {\bf Observing conditions}: the HSC DR1 provides metadata for each filter exposure, containing information about a number of observing conditions. We make coadded maps of these following a procedure similar to that described in \cite{2016ApJS..226...24L}. For each pixel in our map we gather all exposures that fully or partially overlap with it. For each quantity $Q$ in filter $f$, this allows us to build a list of values of $Q$ for each exposure\footnote{We note that we omit any exposure from CCD 9, since it was found to yield unreliable measurements and was never used in the HSC coadd images.}. We then compute the weighted mean of these values and assign the result to the pixel. The weights used to coadd different exposures consist of the product of the area overlap between pixel and exposure and an approximation of the weights used by the HSC pipeline to produce coadded images \cite{2018PASJ...70S...5B}. Since we do not have direct access to the latter, we used, as coadd weights, the inverse of the sky level of each exposure, which should be close to inverse-variance weighting. Following this procedure, we produce maps of the following quantities: airmass, CCD temperature, seeing, PSF ellipticity, exposure time, sky count level, root-mean square deviation of the sky count and number of visits.
    \end{enumerate}
    Figure \ref{fig:sysmap} shows examples of some of these maps.
    
    Note that an underlying assumption of this step is that the coadded mean of different exposures fully captures the connection between fluctuations in observing conditions and the artificial fluctuations they cause on the number counts. In general we could also consider other cumulants of the per-exposure distribution of observing condition values. As shown in Sec.~\ref{sssec:results.spectra.syst}, we do not observe an important contamination from these systematics in our results, and we therefore leave this more thorough study for future work.

  \subsection{Redshift distributions}\label{ssec:methods.nz}
    \begin{figure}
      \centering
      \includegraphics[width=0.9\textwidth]{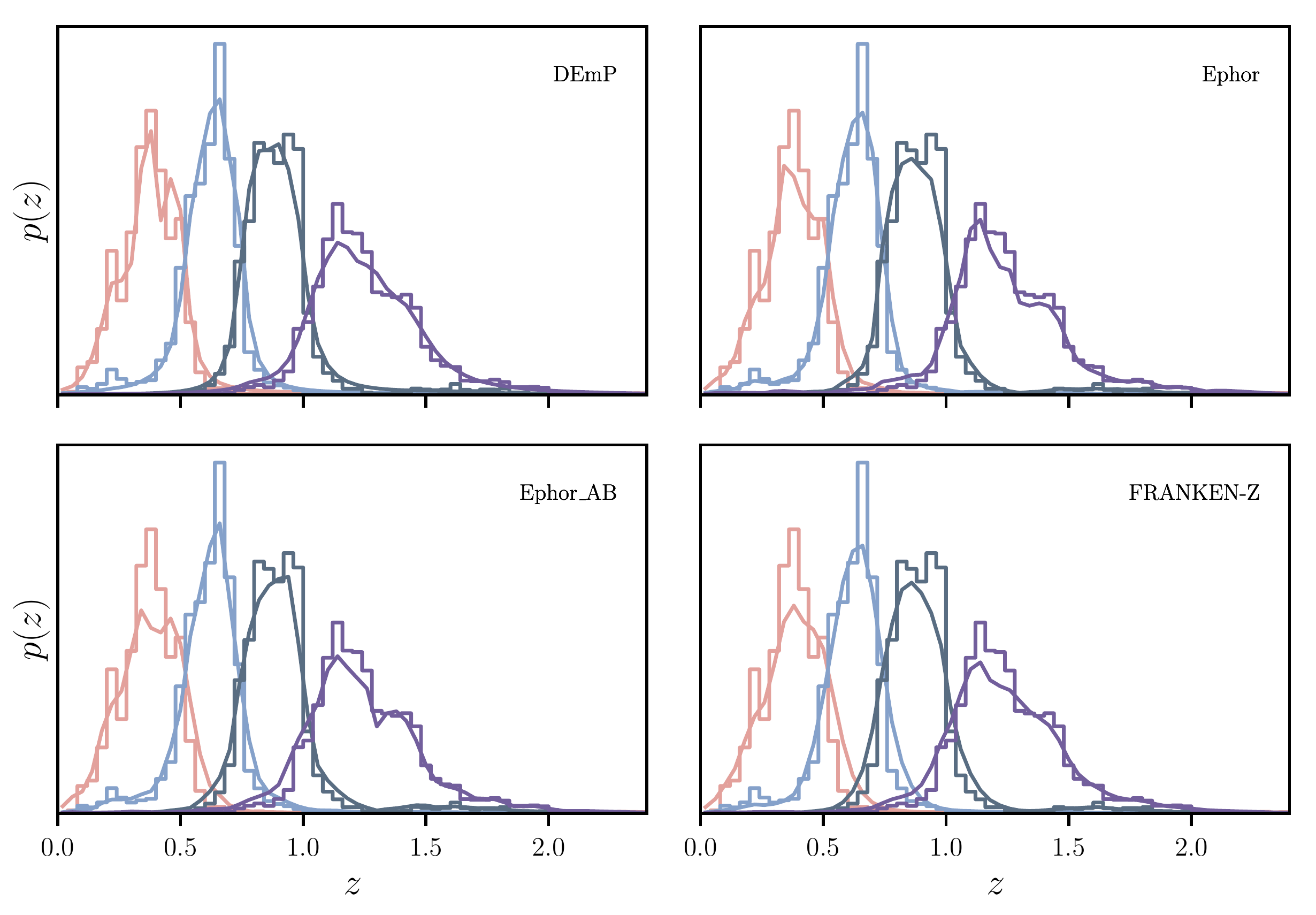}
      \caption{Redshift distributions used in this analysis. The histograms in each panel are the same, and correspond to the redshift distributions estimated from the COSMOS 30-band catalog. These are the fiducial redshift distributions used in our analysis. The solid curves in each panel show the stacked redshift distributions inferred by stacking the per-object photo-$z$ probability distributions obtained with the four alternative photo-$z$ codes explored here ({\tt DEmP}, {\tt Ephor}, {\tt Ephor\_AB} and {\tt FRANKEN-Z} from left to right and top to bottom). The redshift distributions for the four different redshift bins listed in Table \ref{tab:bins_summary} are shown in different colors. Note that all four redshift estimation codes use the same COSMOS 30-band catalog so the level of visual agreement might not provide a realistic measure of uncertainty on the redshift distributions.}
      \label{fig:nzs}
    \end{figure}
    The underlying redshift distributions of our tomographic samples ($p^i(z)$ in Eq. \ref{eq:cell_gg_limber}) are a central component of the theory model used to infer astrophysical and cosmological parameters. Estimating redshift distributions for photometric samples is a non-trivial problem that has been studied extensively in the literature \citep{2008MNRAS.390..118L,2008ApJ...684...88N,2018MNRAS.478..592H}. We use two different methods to estimate $p^i(z)$.
    
    To estimate the fiducial redshift distributions used in our main analysis, we make use of the COSMOS 30-band photometric catalog of \cite{2016ApJS..224...24L}. The idea is to use the high-fidelity photo-$z$ estimates in the COSMOS 30-band data as the truth, in which case one can estimate the redshift distribution by simply making a histogram these redshifts with appropriate weights. These weights are calculated following the procedure described in \cite{2017MNRAS.465.1454H,2019PASJ...71...43H}, which we summarize here for completeness:
    \begin{enumerate}
      \item We first cross-match all objects in the COSMOS 30-band catalog with sources in the HSC COSMOS field that satisfy our sample selection cuts (see Section \ref{sec:data}). Matches are found as pairs of objects with an angular separation smaller than 1 arcsecond. All unmatched objects in either catalog are discarded. We note that the unmatched objects represent $\sim15-20\%$ of the HSC sample lying on the COSMOS 30-band area. This is partially due to the details of the different sky masks applied to both catalogs. We do not observe the unmatched sample to populate any particular region in color space.
      \item For each object in the COSMOS 30-band data with a match in HSC, we find its $N_{\rm neigh}^{\rm COSMOS}=20$ nearest neighbors in the COSMOS 30-band sample within the 5-dimensional space of HSC apparent magnitudes $(g,r,i,z,y)$ using a Euclidean metric. We record the distance to the furthest neighbor in this space.
      \item We then compute the number of HSC sources $N_{\rm neigh}^{\rm HSC}$ found within the same radius in magnitude space. The weight applied to the COSMOS 30-band object is then given by the ratio $N^{\rm HSC}_{\rm neigh}/N^{\rm COSMOS}_{\rm neigh}$ normalized by the total number of objects in each sample. We verified that, applying these weights, we can reproduce the magnitude and color distributions of the HSC sample with the matched COSMOS 30-band sources.
    \end{enumerate}
    There are a number of caveats associated with this method to estimate redshift distributions. First, the COSMOS 30-band photometric redshifts have a lower redshift accuracy and precision than a purely spectroscopic sample. At our magnitude limit, the normalized median absolute deviation of the COSMOS 30-band photometric redshifts is $\sigma\simeq0.03$ \cite{2016ApJS..224...24L}. Second, the small area covered by COSMOS may lead to sample variance uncertainties in the inferred redshift distributions that are difficult to quantify, particularly if the color-redshift relation has an environmental dependence. Finally, the photo-$z$ codes used in HSC DR1 were trained on the COSMOS 30-band data, which could lead to circularity in the estimation of the redshift distributions.
    
    Other methods to estimate source redshift distributions have been used in the literature. For instance, the first-year cosmology analysis of the Dark Energy Survey \citep{1708.01530,2018MNRAS.478..592H} used an initial guess redshift distribution estimated by drawing from the photo-$z$ posterior of all sources in each bin. These were then calibrated by shifting them in redshift, with the best-fit value and uncertainty in the shift parameter $\Delta z$ determined directly from a matched sample of COSMOS galaxies \citep{2018MNRAS.478..592H} and through cross-correlations with redMaGiC galaxies \citep{2018MNRAS.481.2427C}. As described in Sections \ref{sssec:methods.theory.photoz_syst} and \ref{sssec:methods.theory.photoz_syst}, we follow a similar approach to propagate photo-$z$ uncertainties, but we additionally explore uncertainties in the distribution widths.

    In order to study the dependence of our results on the method used to estimate the fiducial redshift distributions, we have also produced alternative estimates through a stacking approach. In this case, for a given photo-$z$ code, we produce an estimate of the redshift distribution of each tomographic bin by adding the photo-$z$ probability distributions of all objects in that bin. This is not a mathematically consistent method to recover the ensemble redshift distributions in the absence of perfect priors. However, the resulting distributions allow us to explore the level of uncertainty in the underlying $p(z)$s. Following this procedure we generate 4 alternative estimates of $p^i(z)$ for the photo-$z$ codes {\tt DEmP}, {\tt Ephor}, {\tt Ephor\_AB} and {\tt FRANKEN-Z}. These are some of the best-performing algorithms presented in \cite{2018PASJ...70S...9T}, and constitute a fair representation of the underlying photo-$z$ uncertainties. The different redshift distributions for all tomographic bins and photo-$z$ methods are shown in Figure \ref{fig:nzs}. The different distributions are visually compatible with each other, although it must be noted that this is, to some extent, by construction, given that the different photo-$z$ codes were trained with the same data from the COSMOS 30-band catalog.

  \subsection{Angular power spectra}\label{ssec:methods.cell}
    We compute angular power spectra using a flat-sky pseudo-$C_\ell$ (PCL) algorithm \citep{2002ApJ...567....2H} as implemented in {\tt NaMaster}\footnote{\url{https://github.com/LSSTDESC/NaMaster}.}. The reader is referred to the code's paper \cite{2019MNRAS.484.4127A} for a detailed description of the estimator, but we provide a brief summary here for completeness.
    
    In the absence of a sky mask, the flat-sky auto-power spectrum could be simply estimated by Fourier transforming a given map $a_{\bf l}={\rm FT}[a(\nv),{\bf l}]$ (where ${\rm FT}$ denotes a Fourier transform operation, and ${\bf l}$ is a 2D wavenumber), and averaging its modulus squared over bins of $\ell\equiv|{\bf l}|$. In any practical situation however, it will be desirable to apply weights on the map to e.g. downweight noisy areas or altogether remove pixels that haven't been observed. In this case the observed (``masked'') map is $\tilde{a}(\nv)=w(\nv)a(\nv)$, where $w(\nv)$ is a weight map. The Fourier coefficients of the observed map are therefore given by a convolution of the true Fourier coefficients with the Fourier transform of the mask, coupling different ${\bf l}$ modes in the original map. The result of using the na\"ive estimator of the power spectrum described above on the masked map is therefore a version of the true underlying power spectrum where different $\ell$s are coupled through the so-called {\sl mode-coupling matrix} $M_{\ell\ell'}$:
    \begin{equation}
      \left\langle \tilde{C}^{ab}_\ell \right\rangle = \sum_{\ell'} M_{\ell\ell'}(w_a,w_b)\,C^{ab}_{\ell'}.
    \end{equation}
    Here, $\tilde{C}^{ab}_\ell$ is the power spectrum of two observed fields $\tilde{a}=w_a\,a$ and $\tilde{b}=w_b\,b$, and $C^{ab}_\ell$ is the true power spectrum. As explicitly written above, the mode-coupling matrix depends exclusively on the properties of the survey masks \citep{2002ApJ...567....2H}, and not on the underlying signal maps. The PCL algorithm calculates the mode-coupling matrix analytically using the properties of the weight maps, and uses it to obtain an unbiased estimate of $C^{ab}_\ell$.
    
    The overdensity maps used are constructed as $\delta_{g,p}=N_p/(\bar{N}\,w_p)-1$, where $N_p$ is the number of sources in pixel $p$, $w_p$ is the survey mask described in Section \ref{ssec:methods.mask}, quantifying the unmasked area fraction in each pixel, and $\bar{N}$ is the mean number of sources per pixel, estimated as $\bar{N}=\sum_p N_p/\sum_p w_p$.
    
    An important aspect of power spectrum estimation that is particularly relevant for galaxy clustering studies is accounting for the effect of sky contaminants on the final summary statistic. In this analysis we have done so using a technique called ``template deprojection''. We start by compiling a list of maps of quantities that can potentially cause artificial perturbations in the observed number density of sources. These include all the quantities described in Section \ref{ssec:methods.syst}.
    
    For small levels of contamination, we can start by assuming that these contaminants affect the observed galaxy overdensity at a linear level:
    \begin{equation}\label{eq:deproj1}
      \delta_g^{\rm obs}(\nv) = \delta_g^{\rm true}(\nv) + A_{\rm syst}\,\Delta_{\rm syst}(\nv),
    \end{equation}
    where $\Delta_{\rm syst}$ is a template map of the fluctuation of a given contaminant around its mean across the survey footprint, and $A_{\rm syst}$ is an unknown linear factor. Template deprojection methods avoid systematic biases by removing all modes from the observed maps that are common to any of the systematic template maps, effectively projecting the input map onto the subspace that is orthogonal to all the contaminant templates. This is equivalent to building a Gaussian model for the observed map using Equation \ref{eq:deproj1} and marginalizing over the free amplitude $A_{\rm sys}$. In the context of pseudo-$C_\ell$ estimators this is achieved in practice by obtaining the best-fit value of $A_{\rm syst}$, subtracting the corresponding best-fit contaminant contribution from the maps and analytically accounting for the associated loss of modes when computing the angular power spectrum. Further details about this method can be found in \cite{2017MNRAS.465.1847E,2019MNRAS.484.4127A}.
    
    We must note that star contamination is a special type of contaminant, since it is an additive contribution to the observed galaxy number density $n_g$, not its overdensity. In the simplest scenario, a fraction $f$ of stars contribute to the observed galaxy density: $n_g^{\rm obs}=n^{\rm true}_g+f\,n_s$, where $n_s$ is the local number density of stars. A fluctuation in the star density around the mean $\Delta_s(\nv)=n_s(\nv)/\bar{n}_s-1$, therefore produces both a multiplicative and an additive effect on $\delta_g$:
    \begin{equation}
      \delta_g^{\rm obs}(\nv) = (1-F_s)\,\delta_g^{\rm true}(\nv)+F_s\,\Delta_s(\nv),
    \end{equation}
    where $F_s$ is the fraction of the sample made out of stars (i.e. $F_s\equiv f\bar{n}_s/(\bar{n}_g+f\bar{n}_s)$ with the notation above). The linear term is taken care of by the deprojection procedure, and we correct the final map of $\delta_g$ by a factor $1/(1-F_s)$, where we estimate $F_s\simeq0.02$ independent of redshift from the HSC deep COSMOS field. This estimate is consistent with the star contamination found in the HSC DR1 shape catalog \citep{2018PASJ...70S..25M}.

    As noted above, the power spectra are computed in rings of $\ell=|{\bf l}|$, which we will call bandpowers here. We use 17 piecewise-linear, contiguous bandpowers with edges $\ell=$(100, 200, 300, 400, 600, 800, 1000, 1400, 1800, 2200, 3000, 3800, 4600, 6200, 7800, 9400, 12600, 15800)\footnote{This choice is motivated by mimicking logarithmic binning as closely as possible, while avoiding very small bin widths at low angular multipoles, which typically arise for logarithmic binning schemes.}. Due to this bandpower averaging, the estimated power spectra cannot, strictly speaking, be compared with theoretical predictions estimated at token multipoles (e.g. the midpoint of each band). The effect of this averaging can however be taken into account exactly as a linear operation of the form $\bar{C}_b=\sum_\ell F^b_{\ell} C_\ell$, where $C_\ell$ is the theoretical prediction evaluated at all integer multipoles $\ell$, $\bar{C}_b$ is the prediction for the $b$-th bandpower and the bandpower windows $F^b_{\ell}$ incorporate the effects of mode-coupling, averaging into bandpowers and the inversion of the binned mode-coupling matrix.
    
    Finally, a noise bias term must be subtracted from all auto-power spectra. In the case of galaxy clustering, and assuming this noise to be entirely due to Poissonian shot noise, this can be done analytically as described in \cite{2019MNRAS.484.4127A}. In short, the noise power spectrum before mode-decoupling (i.e. before multiplying by the inverse mode-coupling matrix), can be calculated as:
    \begin{equation}\label{eq:nell}
      \tilde{N}_\ell = \Omega_{\rm pix}\,\frac{\bar{w}}{\bar{N}},
    \end{equation}
    where $\bar{w}$ is the mean value of the survey mask across the map, $\bar{N}$ is the mean number density of sources per pixel, and $\Omega_{\rm pix}$ is the pixel area in units of steradians.

  \subsection{Modeling the signal}\label{ssec:methods.theory}
  \subsubsection{Projected quantities and power spectra}\label{sssec:methods.theory.cellpk}
    Our main observable is the projected overdensity of galaxies $\delta^i_g(\nv)$ as a function of sky position $\nv$ in a given redshift bin labelled by $i$. This is related to the 3D galaxy overdensity $\Delta_g$ through
    \begin{equation}
      \delta^i_g(\nv)=\int \mathrm{d}z\,p^i(z)\,\Delta_g\left(t(z),\chi(z)\nv\right),
    \end{equation}
where we have assumed a flat cosmological model, i.e. $\Omega_{k} = 0$ for simplicity.
    In the above equation, $t(z)$ and $\chi(z)$ are the cosmic time and radial comoving distance as a function of redshift, and $p^i(z)$ is the redshift bin window function, given by the true redshift distribution of objects in the bin normalized to unit area.
  
    Given the small size of the sky patches covered by HSC DR1, we will adopt the flat sky approximation for simplicity, in which case $\nv$ is a 2D vector. It is common to decompose $\delta^i_g(\nv)$ into its Fourier coefficients
    \begin{equation}
      a^i_{\bf l}\equiv \int\frac{\mathrm{d}\theta^2}{2\pi}e^{-i{\bf l}\cdot\nv}\delta^i_g(\nv).
    \end{equation}
    The variance of the Fourier coefficients is the so-called angular power spectrum $\langle a^i_{\bf l}a^{j*}_{{\bf l}'}\rangle = C^{ij}_\ell\,\delta^{\cal D}({\bf l}-{\bf l}')$, where $\delta^{\cal D}$ is the 2D Dirac delta function. The 3D power spectrum $P_{gg}(z,{\bf k})$ is defined analogously for the 3D Fourier coefficients of $\Delta_g$. Both quantities are related to each other through:
    \begin{equation}\label{eq:cell_gg_limber}
      C^{ij}_\ell = \int \mathrm{d}z\,\frac{H(z)}{\chi^2(z)} p^i(z)p^j(z)\,P_{gg}\left(z,k=\frac{\ell+1/2}{\chi(z)}\right),
    \end{equation}
    where $H(z)$ is the expansion rate at redshift $z$, and we have used the so-called Limber approximation\footnote{We note that we have compared the power spectra for the redshift distributions employed in this work obtained with the Limber approximation and those obtained making no approximations. We find the differences between the power spectra to be around $\sfrac{\Delta C_{\ell}}{C_{\ell}} \sim 10^{-4}-10^{-3}$, and thus negligible compared to the uncertainties. In the following, we therefore use the Limber approximation to compute projected power spectra for computational speed.} \citep{Limber:1953, Kaiser:1992, Kaiser:1998}. In this work, we compute theoretical predictions for angular power spectra using the DESC Core Cosmology Library (\texttt{CCL}\footnote{\url{https://github.com/LSSTDESC/CCL}.}) \cite{Chisari:2019}.

  \subsubsection{Halo Occupation Distribution}\label{sssec:methods.theory.hod}
    In order to model $P_{gg}(z,k)$ we use a halo occupation distribution (HOD) model \citep{2000MNRAS.318.1144P,2002PhR...372....1C,2002ApJ...575..587B,2005ApJ...633..791Z,2013MNRAS.430..725V}. In this halo model-based prescription we model the galaxy content of dark matter haloes as a function of halo mass. Details about HOD parameterizations can be found in \cite{2011ApJ...736...59Z}. In short, the galaxy power spectrum receives contributions from the so-called 1-halo and 2-halo terms:
    \begin{equation}
      P_{gg}(z,k) = P_{gg,{\rm 1h}}(z,k) + P_{gg,{\rm 2h}}(z,k),
    \end{equation}
    where
    \begin{align}
      & P_{gg,{\rm 1h}}(k)=\frac{1}{\bar{n}_g^2} \int \mathrm{d}M\,\frac{\mathrm{d}n}{\mathrm{d}M} \bar{N}_c\,\left[\bar{N}_s^2u_s^2(k)+2\bar{N}_su_s^2(k)\right],\\
      & P_{gg,{\rm 2h}}(k)=\left(\frac{1}{\bar{n}_g} \int \mathrm{d}M\,\frac{\mathrm{d}n}{\mathrm{d}M}\,b_h(M)\,\bar{N}_c\,\left[1+\bar{N}_su_s(k)\right]\right)^2\,P_{\rm lin}(k).
    \end{align}
    The quantity $M$ denotes halo mass, which we give in units of Solar mass $M_{\odot}$ throughout this work. In addition, $\sfrac{\mathrm{d}n}{\mathrm{d}M}$ is the halo mass function, $b_h(M)$ is the halo bias, $\bar{N}_c(M)$ denotes the mean number of central galaxies, $\bar{N}_s(M)$ is the mean number of satellites for halos containing a central galaxy\footnote{Note that, since $\bar{N}_s$ is the average number of satellites for halos with $N_c=1$, the mean number of satellites for all halos is just $\bar{N}_s\bar{N}_c$, since we assume that halos can only contain satellites if they have a central, and that there can be at most one central.}, and $u_s(k)$ is the Fourier transform of the normalized density profile of satellite galaxies. Finally, $P_{\rm lin}(k)$ is the linear matter power spectrum and $\bar{n}_g$ denotes the total mean galaxy density, given by
    \begin{equation}
      \bar{n}_g=\int \mathrm{d}M\,\frac{\mathrm{d}n}{\mathrm{d}M}\bar{N}_c(M)\left[1+\bar{N}_s(M)\right].
      \label{eq:ng_hod}
    \end{equation}
    A central assumption in standard HOD parameterizations is that centrals/satellites follow a Bernoulli/Poisson distribution. The model used here also assumes that halos can only contain satellites if they contain a central.

    Following \cite{2011ApJ...736...59Z}, we parametrize the number of centrals and satellites as a function of mass as:
    \begin{align}
      &\bar{N}_c(M)=\frac{1}{2}\left[1+{\rm erf}\left(\frac{\log(M/M_{\rm min})}{\sigma_{\log M}}\right)\right],\\
      &\bar{N}_s(M)=\Theta(M-M_0)\left(\frac{M-M_0}{M_1'}\right)^\alpha,
    \end{align}
    where $\Theta(x)$ is the Heavyside step function. Furthermore, we assume that $N_c$ follows a Bernoulli distribution with probability $p=\bar{N}_c$, and that the number of satellites is Poisson-distributed with mean $\bar{N}_s$. Finally, we model the distribution of satellites to follow that of the dark matter, and therefore $u_s$ is a Navarro-Frenk-White profile, given by \cite{Navarro:1996}:
    \begin{equation}
      u_s(k|M)=\frac{\sin x\left[{\rm Si}\left((1+c)\,x\right)-{\rm Si}(x)\right]+\cos x\left[{\rm Ci}\left((1+c)x\right)-{\rm Ci}(x)\right]-\frac{\sin(cx)}{(1+c)x}}{{\rm ln}(1+c)-\frac{c}{1+c}},
    \end{equation}
    where $x=\sfrac{k R_\Delta}{c}$, $R_\Delta$ is the halo radius, $c=c(M)$ is the concentration parameter, and ${\rm Si}/{\rm Ci}$ are the sine and cosine integral functions. We define $R_\Delta$ as the radius that encloses $\Delta=200$ times the background matter density. In this work, we model the halo mass function $\sfrac{\mathrm{d}n}{\mathrm{d}M}$ following \cite{Tinker:2010}\footnote{More accurate estimates of the mass function can be obtained through the use of emulators \citep{2019ApJ...872...53M}. We verified that our analysis was not too sensitive to the choice of parameterization, and therefore we leave a more thorough study of these choices for future work.}. Furthermore, we employ the concentration-mass relation $c(M)$ (which depends on the choice of $\Delta$) derived by \cite{Duffy:2008}.

    In order to include lensing magnification in the theory prediction (see Section \ref{ssec:results.magnification}), we also need to model the galaxy-matter and matter-matter power spectra (see Eq. \ref{eq:cell_gg_wmag}). Following the HOD parametrization, the 1-halo and 2-halo contributions to $P_{gm}(k)$ are given by:
    \begin{align}
      & P_{gm,{\rm 1h}}(k)=\frac{1}{\bar{n}_g\bar{\rho}_M} \int \mathrm{d}M\,\frac{\mathrm{d}n}{\mathrm{d}M} M\,\bar{N}_c\,\left[1+\bar{N}_su_s(k)\right],\\
      & P_{gm,{\rm 2h}}(k)=\left(\frac{1}{\bar{n}_g} \int \mathrm{d}M\,\frac{\mathrm{d}n}{\mathrm{d}M}\,b_h(M)\,\bar{N}_c\,\left[1+\bar{N}_su_s(k)\right]\right)\,P_{\rm lin}(k),
    \end{align}
    where $\bar{\rho}_M$ is the comoving matter density. $P_{mm}(k)$ is given by the {\tt Halofit} fitting function \cite{Smith:2003} with the revisions of \cite{Takahashi:2012}.

    It is a well-known fact that the simple halo model implementation described here is not able to accurately describe the ``quasi-linear'' scales in the transition between the 1-halo and 2-halo-dominated regimes, corresponding to $k\sim0.1\,{\rm Mpc}^{-1}$ at $z\sim0$ \citep{2015MNRAS.454.1958M}. To correct for this inaccuracy, we multiply $P_{gg}$ and $P_{gm}$ by a universal scale-dependent factor, given by the ratio between the {\tt Halofit} and the pure halo model predictions for $P_{mm}$:
    \begin{equation}
      R(z,k)=\frac{P^{\rm Halofit}_{mm}(z,k)}{P^{\rm halo\,model}_{mm}(z,k)}.
    \end{equation}
We compute the quantity $R(z,k)$ for our fiducial cosmological model given in Sec.~\ref{ssec:methods.constr}. The amplitude of this correction is close to unity at large scales but can reach values of $\sim 1.3$ for large $k$, c.f. Ref.~\cite{2015MNRAS.454.1958M}.
 
    We expect mean galaxy properties to evolve as a function of redshift $z$ for the magnitude-limited sample considered in this analysis. Instead of fitting a separate HOD model to each redshift bin, we fit redshift-dependent functions to $M_{\mathrm{min}}(z)$, $M_{0}(z)$ and $M_{1}(z)$ and choose a functional form given by:
    \begin{equation}
      \log{M_{i}(z)} = \mu_{i} + \mu_{i, p} \left(\frac{1}{1+z} - \frac{1}{1+z_{p}}\right),
    \end{equation}
    where $\log$ is the logarithm to base 10, $i \in [\mathrm{min}, 0, 1]$ and $z_{p}$ denotes a pivot redshift, which we set to $z_{p} = 0.65$. This functional form is motivated by an initial analysis in which we separately fit an HOD to each auto-power spectrum and determine a function consistent with the observed redshift evolution of the three parameters $M_{\mathrm{min}}(z)$, $M_{0}(z)$ and $M_{1}(z)$. We additionally include a pivot redshift $z_{p}$ in our parametrization to remove degeneracies between the fitted parameters.

  \subsubsection{Photometric redshift systematics modeling}\label{sssec:methods.theory.photoz_syst}
    Uncertainties in photometric redshifts represent one of the major systematic uncertainties in photometric galaxy clustering analyses. To lowest order, errors in photo-$z$'s cause shifts in the means and changes in the width of the derived redshift distribution for a population of galaxies. In this work, we therefore choose to parametrize the impact of photo-$z$ errors on the derived galaxy redshift distributions $p_{i}(z)$ using a two-parameter model given by
    \begin{equation}
      p_{i}(z) \propto \hat{p}_{i}(z_{c} + (1 + z_{w, i})(z-z_{c}) + \Delta z_{i}),
      \label{eq:photo-z-model}
    \end{equation} 
    where the index $i$ runs over the number of redshift bins considered in our analysis. In the above equation, $\hat{p}_{i}(z)$ denotes the estimated redshift distribution while $p_{i}(z)$ is the underlying true distribution. The parameters $\Delta z_{i}$ account for shifts in the means of the distributions and changes to their widths are parameterized through $z_{w, i}$. The quantity $z_{c}$ is kept constant in our analysis and is set to the redshift at which $\hat{p}_{i}(z)$ attains its maximal value. The normalization of the redshift distribution $p_{i}(z)$ depends on both $\Delta z_{i}$ and $z_{w, i}$ and we do not account for it in Eq.~\ref{eq:photo-z-model} for clarity.
    
  \subsubsection{Covariance matrices}\label{sssec:methods.theory.covar}
    We use an analytical procedure to estimate the uncertainties of our measured power spectra, inspired by the methods used by \cite{Krause:2017}. As shown in \cite{2009MNRAS.395.2065T,Takada:2013}, the covariance matrix for large-scale structure data can be decomposed into a disconnected trispectrum part, essentially equivalent to the covariance of a Gaussian random field with the same power spectrum as the data, a connected part, caused by the non-Gaussian nature of the density field, and a super-sample covariance term (labeled SSC here)\footnote{A comparison of the relative contributions of these three terms to the total covariance can be found in Fig.~\ref{fig:covariance-contributions}.}. The SSC contribution accounts for the coherent shift in the amplitude of density fluctuations within the surveyed volume caused by long wavelength modes larger than the survey.
    
    We estimate the Gaussian covariance of the pseudo-$C_\ell$ estimator as described in \cite{2004MNRAS.349..603E,2019arXiv190611765G}. The covariance of the observed pseudo-power spectra is given by
    \begin{equation}\label{eq:pcl_cov_0}
      {\rm Cov}\left(\tilde{C}^{ab}_\ell,\tilde{C}^{cd}_{\ell'}\right)=\sum_{{\bf l}_1,{\bf l}_2} C^{ac}_{\ell_1}C^{bd}_{\ell_2}W^a_{{\bf l}{\bf l}_1}W^b_{{\bf l}{\bf l}_2}W^c_{{\bf l}'{\bf l}_1}W^d_{{\bf l}'{\bf l}_2} + C^{ad}_{\ell_1}C^{bc}_{\ell_2}W^a_{{\bf l}{\bf l}_1}W^b_{{\bf l}{\bf l}_2}W^c_{{\bf l}'{\bf l}_2}W^d_{{\bf l}'{\bf l}_1},
    \end{equation}
    where the quantities $W^x_{{\bf l}{\bf l}'}$ are coupling coefficients depending only on the mask of $x$ (see \cite{2019arXiv190611765G} for further details). Without further approximations, computing the covariance matrix would therefore imply solving a 4-dimensional integral for each pair $(\ell,\ell')$. This is an $O(\ell_{\rm max}^6)$ operation which easily becomes computationally unfeasible. Since the coupling coefficients $W^x_{{\bf l}{\bf l}'}$ are usually highly peaked around ${\bf l}={\bf l}'$, we can proceed further by approximating the power spectra to be constant within the support of $W^x_{{\bf l},{\bf l}'}$. Effectively this implies approximating $C^{ac}_{\ell_1}C^{bd}_{\ell_2}$ in Eq \ref{eq:pcl_cov_0} as
    \begin{equation}
      C^{ac}_{\ell_1}C^{bd}_{\ell_2}\simeq C^{ac}_{(\ell}C^{bd}_{\ell')}\equiv\frac{1}{2}\left(C^{ac}_\ell C^{bd}_{\ell'}+C^{ac}_{\ell'} C^{bd}_\ell\right).
    \end{equation}
    This allows us to simplify the expression above significantly, leading to
    \begin{equation}
      {\rm Cov}\left(\tilde{C}^{ab}_\ell,\tilde{C}^{cd}_{\ell'}\right)=(2\ell'+1)^{-1}\left[C^{ac}_{(\ell}C^{bd}_{\ell')}M_{\ell\ell'}(w_aw_c,w_bw_d)+C^{ad}_{(\ell}C^{bc}_{\ell')}M_{\ell\ell'}(w_aw_d,w_bw_c)\right],
    \end{equation}
    where $M_{\ell\ell'}$ are the mode-coupling matrices described in the previous section, except now they are computed from the products of two masks. It has been shown by \cite{2019arXiv190611765G} that this approach yields a very good approximation for the power spectrum covariance, fully accounting for the effects of mode coupling due to survey geometry.

    The second contribution to the total covariance matrix for galaxy clustering is caused by the connected part of the trispectrum, which accounts for mode-coupling due to the non-Gaussian nature of the density field. In our work we compute this contribution using the halo model coupled with a halo occupation distribution. In general, this contribution is given by the angular projection of the three-dimensional trispectrum as (see e.g. \cite{Krause:2017})
    \begin{equation}
    \begin{aligned}
      \mathrm{Cov}_{\mathrm{NG}}(C^{ab}_{\ell}, C^{cd}_{\ell'}) = \frac{1}{4 \pi f_{\mathrm{sky}}} \int_{\vert \boldsymbol{\ell} \vert \in \ell_{1}} \int_{\vert \boldsymbol{\ell}' \vert \in \ell_{2}} \int \frac{\mathrm{d}^{2}\boldsymbol{\ell}}{A(\ell_{1})} \; \frac{\mathrm{d}^{2}\boldsymbol{\ell}'}{A(\ell_{2})} \; \mathrm{d}\chi \; \frac{q^{a}(\chi)q^{b}(\chi)q^{c}(\chi)q^{d}(\chi)}{\chi^{6}} \times \\ T^{abcd}(\sfrac{\boldsymbol{\ell}}{\chi}, \sfrac{-\boldsymbol{\ell}}{\chi}, \sfrac{\boldsymbol{\ell}'}{\chi}, \sfrac{-\boldsymbol{\ell}'}{\chi}).
    \end{aligned}
    \end{equation}
    The quantity $A(\ell_{i})$ denotes the area of an annulus of width $\Delta \ell_{i}$ around $\ell_{i}$, i.e. $A(\ell_{i}) \equiv \int_{\vert \boldsymbol{\ell} \vert \in \ell_{i}} \mathrm{d}^{2}\boldsymbol{\ell}$, which is approximately given by $A(\ell_{i}) \approx 2 \pi \Delta \ell_{i} \ell_{i}$ for $\ell_{i} \gg \Delta \ell_{i}$. Finally, $q^a(\chi(z))\equiv \sfrac{H(z)}{c} \;p^a(z)$ is the window function for the $a$-th redshift bin.
    
    Using the halo model, the connected part of the trispectrum $T^{abcd}$ can be written as (e.g. \cite{Takada:2013}):
    \begin{equation}
      T^{abcd} = T^{abcd, 1h} + (T^{abcd, 2h}_{22} + T^{abcd, 2h}_{13}) + T^{abcd, 3h} + T^{abcd, 4h},
    \end{equation}
    where
    \begin{equation}
    \begin{aligned}
      &T^{abcd, 1h}(\mathbf{k}_{a}, \mathbf{k}_{b}, \mathbf{k}_{c}, \mathbf{k}_{d}) = I^{0}_{abcd}(k_{a}, k_{b}, k_{c}, k_{d}), \\
      &T^{abcd, 2h}_{22}(\mathbf{k}_{a}, \mathbf{k}_{b}, \mathbf{k}_{c}, \mathbf{k}_{d}) = P_{\mathrm{lin}}(k_{ab})I^{1}_{ab}(k_{a}, k_{b})I^{1}_{cd}(k_{c}, k_{d}) + 2 \; \mathrm{perm.}, \\
      &T^{abcd, 2h}_{13}(\mathbf{k}_{a}, \mathbf{k}_{b}, \mathbf{k}_{c}, \mathbf{k}_{d}) = P_{\mathrm{lin}}(k_{a})I^{1}_{a}(k_{a})I^{1}_{bcd}(k_{b}, k_{b}, k_{c}) + 3 \; \mathrm{perm.}, \\
      &T^{abcd, 3h}(\mathbf{k}_{a}, \mathbf{k}_{b}, \mathbf{k}_{c}, \mathbf{k}_{d}) = B^{\mathrm{PT}}(\mathbf{k}_{a}, \mathbf{k}_{b}, \mathbf{k}_{cd})I^{1}_{a}(k_{a})I^{1}_{b}(k_{b})I^{1}_{cd}(k_{c}, k_{d}) + 5 \; \mathrm{perm.},\\
      &T^{abcd, 4h}(\mathbf{k}_{a}, \mathbf{k}_{b}, \mathbf{k}_{c}, \mathbf{k}_{d}) = T^{\mathrm{PT}}(\mathbf{k}_{a}, \mathbf{k}_{b}, \mathbf{k}_{c}, \mathbf{k}_{d})I^{1}_{a}(k_{a})I^{1}_{b}(k_{b})I^{1}_{c}(k_{c})I^{1}_{d}(k_{d}).
    \label{eq:halo-mod-trisp}
    \end{aligned}
    \end{equation}
    Here, ${\bf k}_{ab}\equiv {\bf k}_a+{\bf k}_b$, and the quantities $B^{\mathrm{PT}}$ and $T^{\mathrm{PT}}$ denote the matter bi- and trispectrum respectively, as estimated using tree-level perturbation theory. The full expressions for these terms can be found in \cite{Takada:2013}. Finally, $I^{n}_{a_1...b_m}$ denotes the generic halo model integral, defined as (e.g. \cite{Krause:2017}):
    \begin{equation}
      I^{n}_{a_1...a_m}(k_1,...,k_m) = \int \mathrm{d}M \frac{\mathrm{d}n}{\mathrm{d}M}b_{h, n}(M)  \left\langle\prod_{i=1}^m \left[\tilde{u}_{a_i}(k_i, M) \right]\right\rangle, 
      \label{eq:halo-mod-intg}
    \end{equation}
    where $b_{h,1}(M)\equiv b_h(M)$ is the halo bias, $b_{h,0}\equiv1$, and $\tilde{u}_{a_i}$ is the halo profile for the $i$-th field being correlated (e.g. the HOD profile described in Section \ref{sssec:methods.theory.hod} in the case of galaxy clustering). For simplicity, we follow \cite{Krause:2017} and approximate the 2- to 4-halo trispectrum as the linearly biased matter trispectrum and only include a probe-specific 1-halo trispectrum contribution. Specifically, we set 
    \begin{equation}
      T^{abcd} = T^{abcd, 1h} + b_{a}b_{b}b_{c}b_{d}T^{m, 2h+3h+4h},
    \end{equation}
    where $T^{abcd, 1h}$ and $T^{m, 2h+3h+4h}$ are computed following Equations \ref{eq:halo-mod-trisp}. For $T^{abcd, 1h}$, we evaluate Eq.~\ref{eq:halo-mod-intg} for the galaxy distribution, while for $T^{m, 2h+3h+4h}$, we use the corresponding expressions for the matter distribution. Finally, $b_{i}$ denotes the linear bias predicted using halo occupation distribution modeling, given by
    \begin{equation}
      b_{a}=\frac{1}{\bar{n}_{g, a}}\int \mathrm{d}M\,\frac{\mathrm{d}n}{\mathrm{d}M}b_{h}(M) \bar{N}_{c, a}(1+\bar{N}_{s, a}).
    \end{equation}
    The 1-halo trispectrum $T^{abcd, 1h}$ for galaxies also receives contributions due to shot noise \cite{Lacasa:2018}. However, these are expected to be small \cite{Lacasa:2018} and we thus neglect them in this work.

    Finally, we compute the super-sample covariance contribution following the treatment of \cite{Krause:2017}, i.e.:
     \begin{equation}
    \begin{aligned}
       \mathrm{Cov}_{\mathrm{SSC}}(C^{ab}_{\ell}, C^{cd}_{\ell'}) &= \int \mathrm{d}\chi \;\frac{q^{a}(\chi)q^{b}(\chi)q^{c}(\chi)q^{d}(\chi)}{\chi^{4}} \times \\ &\frac{\partial P_{ab}(\sfrac{\ell}{\chi}, z(\chi))}{\partial \delta_{\rm LS}}\frac{\partial P_{cd}(\sfrac{\ell'}{\chi}, z(\chi))}{\partial \delta_{\rm LS}}\sigma^{2}_{b}(z(\chi)).
    \end{aligned}
    \end{equation}
        The quantity $\sigma_b^2(z)$ is the variance of the long wavelength mode $\delta_{\rm LS}$ over the survey footprint, given by
    \begin{equation}
      \sigma_b^2(z) = \int \frac{\mathrm{d}k_\perp^2}{(2\pi)^2}P_{\rm lin}(k_\perp,z)\left|W(k_\perp,z)\right|^2.
    \end{equation}
    Furthermore $W(k_\perp,z)$ denotes the Fourier transform of the survey footprint, which we approximate as a compact circle with an area matched to our data set:
    \begin{equation}
      W(k_\perp,z)=\frac{2 J_1(k_\perp\chi(z)\theta_s)}{k_\perp \chi(z)\theta_s},\hspace{12pt} \theta_s={\rm arccos}(1-2f_{\rm sky}),
    \end{equation}
    where $J_1(x)$ is the cylindrical Bessel function of order 1.
    Finally, the quantity $\partial P_{ab}(k, z)/\partial \delta_{\rm LS}$ is the response of the power spectrum $P_{ab}$ to a large-scale density fluctuation, which we estimate using the halo model and results from perturbation theory as (e.g. \cite{Krause:2017}):
     \begin{equation}
    \begin{aligned}
      \frac{\partial P_{ab}(k, z)}{\partial \delta_{\rm LS}} =& \left( \frac{68}{21} - \frac{1}{3}\frac{\mathrm{d}\log{k^{3} P_{\mathrm{lin}}}(k, z)}{\mathrm{d}\log k} \right) I_{a}^{1}(k)I_{b}^{1}(k)P_{\mathrm{lin}}(k, z) + I_{ab}^{1}(k, k) \\ &- (b_{a} + b_{b})P_{ab}(k, z).
    \label{eq:ps-resp}  
    \end{aligned}
     \end{equation}
    The last term in Eq.~\ref{eq:ps-resp} accounts for the fact that the observed galaxy overdensity is computed using the mean galaxy density estimated inside the survey volume.
    
    For consistency with our implementation of the connected trispectrum, we compute the response function $\sfrac{\partial P_{ab}(k, z)}{\partial \delta_{\rm LS}} $ for a given probe as the linearly biased response of the matter field\footnote{In order to test the robustness of our results to this approximation, we also compute the SSC contribution to the covariance using the probe-specific halo model quantities in Eq.~\ref{eq:ps-resp}. We find our parameter constraints to be unaffected by this change and therefore resort to the approach described above for consistency.}. 

  \subsection{Parameter constraints}\label{ssec:methods.constr}
    In order to derive constraints on HOD, cosmological and systematics parameters, we assume the joint likelihood of all auto- and cross-power spectra to be Gaussian 
    \begin{align}
      \mathscr{L}(D \vert \theta) = \frac{1}{[(2\pi)^{d}\det{\mathsf{C}}]^{\sfrac{1}{2}}} e^{-\frac{1}{2}(\mathbf{C}^{\mathrm{obs}}_{\ell}-\mathbf{C}^{\mathrm{theor}}_{\ell})^{\mathrm{T}}\mathsf{C}^{-1}(\mathbf{C}^{\mathrm{obs}}_{\ell}-\mathbf{C}^{\mathrm{theor}}_{\ell})}.
      \label{eq:likelihood}
    \end{align}
    The quantity $\mathsf{C}$ denotes the joint covariance matrix, which we estimate analytically as described in Sec.~\ref{sssec:methods.theory.covar} and which we keep constant during parameter estimation \citep{2019OJAp....2E...3K}.
    We sample the likelihood in a Monte Carlo Markov Chain (MCMC) using the publicly available code \texttt{CosmoHammer} \cite{Akeret:2013}, which is based on \texttt{emcee} \cite{Foreman-Mackey2013}\footnote{We note that we have compared the MCMC results obtained using \texttt{CosmoHammer} to those obtained using a Metropolis-Hastings algorithm as implemented in \texttt{april} (\url{https://github.com/slosar/april}), finding consistent results for our test case.}. In our fiducial analysis we sample the parameter set $\boldsymbol{\theta} = \{\mu_{\mathrm{min}}, \allowbreak \, \mu_{\mathrm{min}, p}, \allowbreak \, \mu_{0}, \allowbreak \, \mu_{0, p}, \allowbreak \, \mu_{1}, \allowbreak \, \mu_{1, p}, \allowbreak \, \Delta z_{i}, \allowbreak \, z_{w, i}\}$, $i = 0, \dots ,3$, where the first six parameters describe the HOD of galaxies as outlined in Sec.~\ref{sssec:methods.theory.hod}. The remaining parameters account for photometric redshift uncertainties as described in Sec.~\ref{sssec:methods.theory.photoz_syst}. We perform additional analyses in which we separately allow for variations in the amplitude $A_{\mu}$ of the magnification bias kernel $W_{\mu}$ and the cosmological parameters $\Omega_{c}$ and $\sigma_{8}$, where $\Omega_{c}$ is the fractional cold matter density today and $\sigma_{8}$ denotes the r.m.s. of linear matter fluctuations in spheres of comoving radius 8 $h^{-1}$ Mpc. For all sampled parameters, we assume flat, uniform priors. The sampled parameters are shown alongside their priors in Tab.~\ref{tab:params}\footnote{The values of these parameters derived from our analysis will be discussed in Sec.~\ref{ssec:results.hod-constraints}.}. The remaining HOD parameters are set to $\alpha = 1.0$ and $\sigma_{\log M} = 0.4$, consistent with the simulation results of Ref.~\cite{2005ApJ...633..791Z}. Unless stated otherwise, we further fix all cosmological parameters to the best-fit values derived by the Planck Collaboration in 2018 using temperature, polarization and CMB lensing data, i.e. $\Omega_{b}=0.0493$, $\Omega_{c}=0.264$, $h=0.6736$, $n_{s}=0.9649$ and $\sigma_{8}=0.8111$ (see the fourth column in Tab.~2 in \cite{Planck:2018}).
    
    We fit all power spectra up to a maximal angular multipole $\ell_{\mathrm{max}}$ approximately corresponding to $k_{\mathrm{max}} = 1$ Mpc$^{-1}$, as determined through the Limber relation $k_{\mathrm{max}} = \sfrac{\ell_{\mathrm{max}}}{\chi(z)}$. For the auto-power spectra, we determine $\ell_{\mathrm{max}}$ at the effective redshift $z_{\mathrm{eff}}$ of the bin using our fiducial Planck 2018 cosmological model\footnote{We define the effective redshift for each tomographic bin as the mean redshift of the galaxy distribution.}. For the cross-correlations, we set $\ell_{\mathrm{max}}$ to the minimum value derived for the two redshift bins\footnote{This leads to $\{\ell_{\mathrm{max}, 00}, \allowbreak \, \ell_{\mathrm{max}, 01}, \allowbreak \, \ell_{\mathrm{max}, 02}, \allowbreak \, \ell_{\mathrm{max}, 03}, \allowbreak \, \ell_{\mathrm{max}, 11}, \allowbreak \, \ell_{\mathrm{max}, 12}, \allowbreak \, \ell_{\mathrm{max}, 13}, \allowbreak \, \ell_{\mathrm{max}, 22}, \allowbreak \, \ell_{\mathrm{max}, 23}, \allowbreak \, \ell_{\mathrm{max}, 33}\} = \{2000, \allowbreak \, 2000, \allowbreak \, 2000, \allowbreak \, 2000, \allowbreak \, 2000, \allowbreak \, 2000, \allowbreak \, 2000, \allowbreak \, 2600, \allowbreak \, 2600, \allowbreak \, 3400\}$. The maximal angular multipoles are the same for the first and second redshift bin due to our choice of bandpowers.}.
    
    The analytical covariance matrix $\mathsf{C}$ depends on cosmological and HOD parameters. In order to determine a covariance matrix that closely resembles the data, we resort to a two-step process: in a first step, we fit the data using a Gaussian covariance matrix derived from the observed data power spectra. We then use the best-fit parameters determined in this analysis to compute the full covariance matrix as described in Sec.~\ref{sssec:methods.theory.covar} and use this updated covariance matrix in all our subsequent analyses.
    
We note that the priors assumed on the photo-$z$ parameters (shifts and widths), are significantly broader than those used in the HSC cosmic shear analysis \cite{2019PASJ...71...43H}, where the prior on the shift parameters is of the order of $0.01-0.04$. This prior was estimated as the scatter between the best-fit shift parameters that recover the same shear power spectrum for different estimates of the redshift distribution and different photo-$z$ codes. We carried out an additional analysis where we quantified the shift and width parameters allowed by the sample variance uncertainties in our fiducial estimate of the redshift distribution from COSMOS. To do so, we estimated the covariance matrix of the redshift distribution amplitudes in each narrow histogram bin shown in Fig. \ref{fig:nzs}, assuming a simple cylindrical survey geometry and that volume-to-volume number density correlations follow a normal distribution given a biased linear power spectrum. From this covariance matrix, we then draw Gaussian realizations of the redshift distributions (with our fiducial estimate as the mean). For each realization, we compute the mean and the width of the corresponding $N(z)$ distributions. Estimating the standard deviation of the means and widths from all realizations, we find that sample variance uncertainty in COSMOS leads to 1-$\sigma$ shifts of $\Delta z\sim0.006$ and fractional widths $z_w\sim 0.04$. We therefore conclude that the priors used here are conservative, and encompass the range of shift and width values allowed by our uncertainties. It is interesting to note that, having access to the covariance matrix of the redshift distribution uncertainties  allows us to perform a more general characterization of the $N(z)$ uncertainties associated with sample variance beyond the simple shift-width parameterizations (e.g. through principal component analysis). We leave this type of study for future work.

Finally we note that using the galaxy number density $\bar{n}$ as an additional data point in Eq.~\ref{eq:likelihood} has the potential to improve the constraining power of the data on HOD parameters, as it effectively fixes one of the parameters (see e.g. \cite{Zhai:2017}). However, we have chosen to not include this additional constraint in our analysis, as including it would require the modeling of the effects of photometric redshift errors and observational systematics on the estimated number density $\hat{\bar{n}}$, which is beyond the scope of this paper.

    \begin{table*}
      \caption{Summary of parameters varied in the MCMC with their respective priors. The posterior means and best-fits are given for our fiducial analysis. The best-fit is defined as the maximum likelihood value of each parameter and the uncertainties denote the $68 \%$ c.l. corresponding to the equal-probability values encompassing a total probability of 0.68.} \label{tab:params}
      \begin{center}
        \begin{tabular}{cccc}
          \hline\hline 
          Parameter & Prior & Posterior mean & Best-fit \\ \hline \Tstrut                             
          $\mu_{\mathrm{min}}$ & flat $\in [0., \,15.]$ & $11.88\substack{+0.22 \\ -0.23}$ & $12.02$ \\ 
          $\mu_{\mathrm{min}, p}$ & flat $\in [-10., \,10.]$ & $-0.5\substack{+2.1 \\ -2.0}$ & $-1.3$ \\
          $\mu_{0}$ & flat $\in [0., \,15.]$ & $5.7 \pm 4.0$ & $6.6$ \\
          $\mu_{0, p}$ & flat $\in [-5., \,10.]$ & $2.5\substack{+5.1 \\ -5.0}$ & $-1.4$ \\
          $\mu_{1}$ & flat $\in [0., \,17.]$ & $13.08\substack{+0.27 \\ -0.28}$ & $13.27$ \\ 
          $\mu_{1, p}$ & flat $\in [-12., \,15.]$ & $0.9\substack{+2.7 \\ -2.6}$ & $-0.3$ \\
          $\Delta z_{0}$ & flat $\in [-0.2, \,0.2]$ & $0.000\substack{+0.096 \\ -0.091}$ & $-0.064$ \\
          $\Delta z_{1}$ & flat $\in [-0.2, \,0.2]$ & $-0.016\substack{+0.096 \\ -0.086}$ & $-0.087$ \\
          $\Delta z_{2}$ & flat $\in [-0.2, \,0.2]$ & $-0.01\substack{+0.11 \\ -0.09}$ & $-0.07$ \\
          $\Delta z_{3}$ & flat $\in [-0.2, \,0.2]$ & $0.01\substack{+0.11 \\ -0.10}$ & $-0.03$ \\
          $z_{w, 0}$ & flat $\in [-0.2, \,0.2]$ & $-0.05\substack{+0.12 \\ -0.11}$ & $-0.10$ \\
          $z_{w, 1}$ & flat $\in [-0.2, \,0.2]$ & $-0.010\substack{+0.078 \\ -0.079}$ & $-0.010$ \\
          $z_{w, 2}$ & flat $\in [-0.2, \,0.2]$ & $0.035\substack{+0.079 \\ -0.078}$ & $0.082$ \\
          $z_{w, 3}$ & flat $\in [-0.2, \,0.2]$ & $0.05\substack{+0.11 \\ -0.13}$ & $0.18$ \\ \\
          $A_{\mu}$ & flat $\in [-5., \,5.]$ & $ - $ & $ - $ \\
          $\Omega_{c}$ & flat $\in [0.1, \,0.9]$ & $ - $ & $ - $ \\
          $\sigma_{8}$ & flat $\in [0.2, \,1.5]$ & $ - $ & $ - $ \\
          \hline\hline 
        \end{tabular}
      \end{center}
    \end{table*} 

\section{Results}\label{sec:results}
  \subsection{Power spectra}\label{ssec:results.spectra}
    \subsubsection{Fiducial measurements}\label{sssec:results.spectra.fid}
      \begin{figure}
        \centering
        \includegraphics[width=0.99\textwidth]{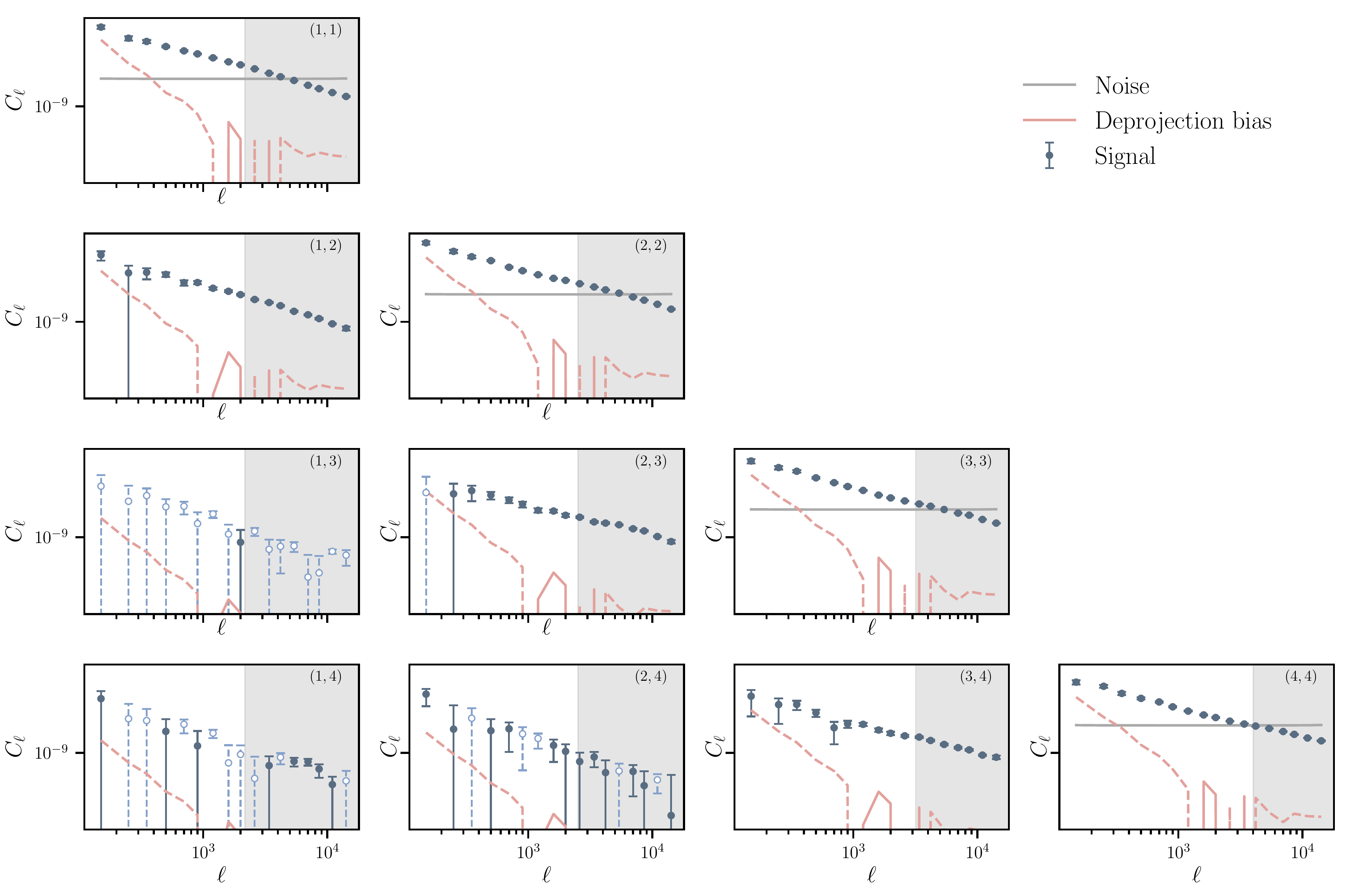}
        \caption{Summary plot showing all measured auto- and cross-power spectra. The dark green data points with error bars show the coadded power spectra (see Section \ref{sssec:results.spectra.fid}), with the hollow blue circles showing the absolute value of negative data. The Poisson noise contribution and deprojection biases subtracted from the raw power spectra are shown as gray and pink lines respectively, with the negative parts of the deprojection bias shown as dashed lines. The semi-transparent gray bands cover the range of scales excluded from our analysis. Finally, the pair of numbers in the upper right corner of each panel corresponds to the indices of the bins being cross-correlated.}
        \label{fig:cls_summary}
      \end{figure}
      We compute all auto- and cross-power spectra between the four different redshift bins (listed in Table \ref{tab:bins_summary}) in each of the 6 HSC DR1 fields (listed in Table \ref{tab:field_summary}) as described in Section \ref{ssec:methods.cell}, including the deprojection of 48 different contaminant templates (9 observing condition maps in each of the 5 HSC filters, a dust map, a star density map and a depth map). In order to use these measurements to constrain model parameters, we first coadd them into a single set of spectra. We perform this coaddition simply as a weighted average, weighting the spectra in each field by their area:
      \begin{equation}
        {\bf C}_{\rm coadd}=\frac{\sum_f A_f\,{\bf C}_f}{\sum_f A_f}, \hspace{12pt}{\sf Cov}_{\rm coadd}=\frac{\sum_f A_f^2\,{\sf Cov}_f}{\left(\sum_f A_f\right)^2}
      \end{equation}
      where ${\bf C}$ is a vector containing all power spectra, ${\sf Cov}$ is its covariance matrix, and $f$ runs through the 6 DR1 fields with area $A_f$. This procedure should be close to inverse-variance weighting assuming that the power spectrum covariance has a similar structure in all fields. The coadded spectra should therefore be close to optimally weighted, with the added advantage that the coaddition does not introduce additional scale dependence due to mode-coupling in the covariance matrix.
      
      The resulting power spectrum measurements are shown in Figure \ref{fig:cls_summary} as dark green circles with error bars (hollow blue circles show the the absolute value of negative data). In all cases, we have subtracted the shot noise bias as described in Section \ref{ssec:methods.cell}, which is also shown as a gray solid line in the auto-correlations. As described in Section \ref{ssec:methods.cell}, after deprojecting a set of contaminant templates, the power spectra estimated from the projected maps must be corrected for a bias caused by the loss of modes due to deprojection. This bias is also shown as a pink solid line in Figure \ref{fig:cls_summary}, where the dashed parts show the absolute value of the bias when it is negative. This bias is always at least a factor $\sim5$ smaller than the measured power spectra, and is most relevant on large scales. The semi-transparent gray bands in the figure cover the range of scales excluded from our analysis (described in Section \ref{ssec:methods.constr}). Unless otherwise stated, in what follows all our results will not include any of these data.
      
      The rest of this sub-section describes the different tests we have carried out to quantify the robustness of these measurements.

    \subsubsection{Consistency across fields}\label{sssec:results.spectra.consistent}
      The fact that the HSC DR1 sample is distributed across 6 different disconnected fields allows us to carry out consistency tests of the measurements in individual fields. For example, it is reasonable to expect that, if a given undetected systematic is biasing our measurements significantly, its impact would vary across different fields, and would therefore lead to inconsistent power spectrum measurements between them. We carry out two basic consistency tests, involving the number density of objects in each field (i.e. the one-point function) and the measured power spectra.
      
\begin{figure}
        \centering
        \includegraphics[width=0.7\textwidth]{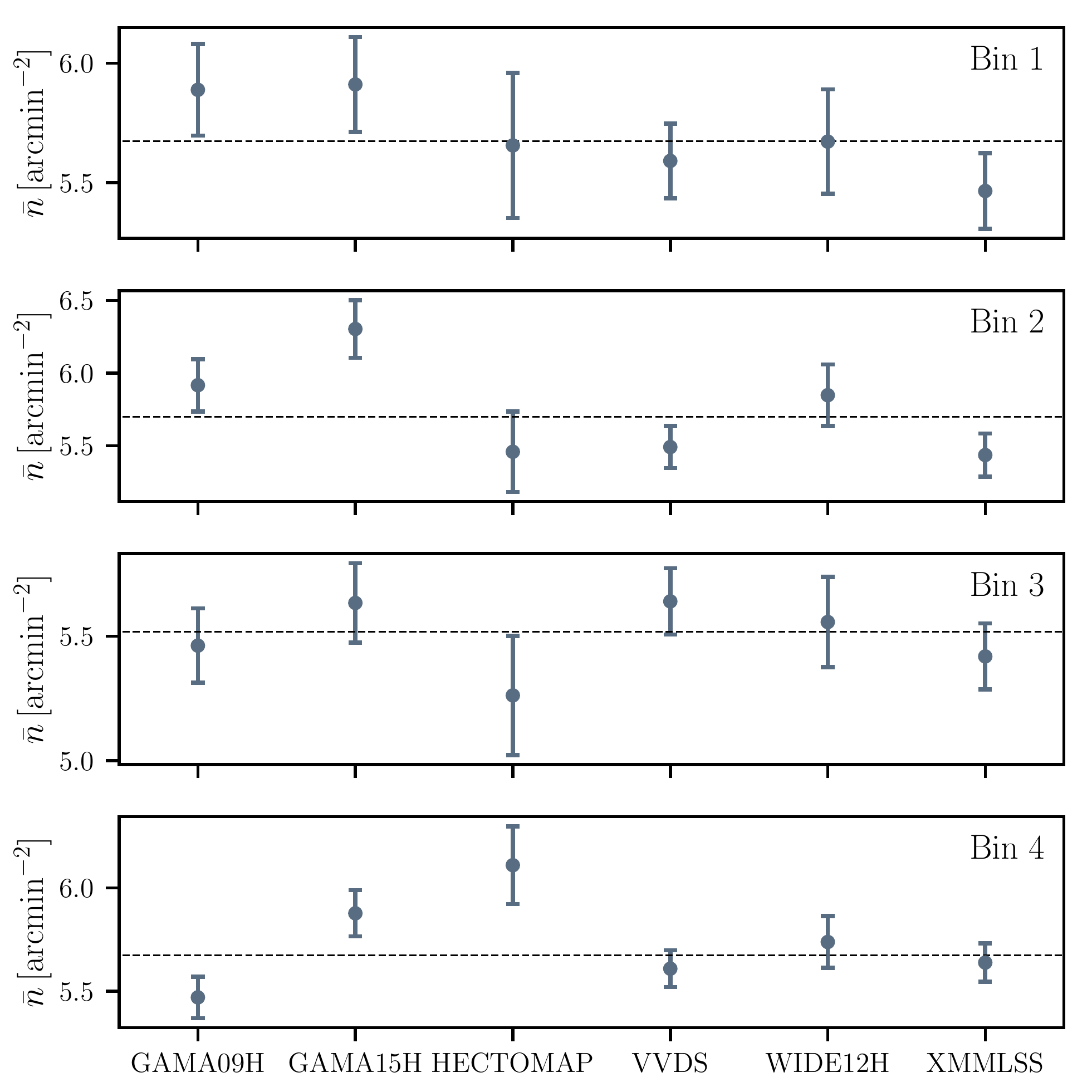}
        \caption{Source number densities estimated in the 6 different fields ($x$-axis) and in the 4 different redshift bins (in each panel, from top to bottom). We find no significant evidence of inconsistency between the number density of sources found in each field.}
        \label{fig:ndens_consistency}
      \end{figure}

      \paragraph{Number densities.} The number density of galaxies estimated in a given field is simply given by the ratio of the number of galaxies and the area of the field:
      \begin{equation}
        \hat{\bar{n}}=\frac{\sum_p N_p}{\sum_p \Omega_p}= \bar{n} \frac{\int \mathrm{d}\nv^2\,W(\nv)\left[1+\delta_g(\nv)\right]}{\int \mathrm{d}\nv^2\,W(\nv)},
      \end{equation}
      where $p$ runs over all pixels in the map, and $N_p$ and $\Omega_p$ are the number of objects and area of pixel $p$. In the second equality we have taken the continuum limit, $\bar{n}$ is the true number density, $\delta_g$ is the galaxy overdensity and $W$ is the field's mask. Using the statistics of $\delta_g$ it is straightforward to calculate the variance of the estimated $\hat{\bar n}$:
      \begin{equation}
       {\rm Var}(\hat{\bar n})=\int \frac{\mathrm{d}{\bf l}^2}{(2\pi)^2}\left|\frac{W_{\bf l}}{\int \mathrm{d}\nv^2\,W(\nv)}\right|^2\,C_\ell.
      \end{equation}
      Here ${\bf l}$ is a 2D Fourier-space wave vector, $W_{\bf l}$ is the Fourier transform of the mask and $C_\ell$ is the power spectrum of $\delta_g$\footnote{We verified the validity of this calculation by replacing $C_\ell$ by its shot-noise contribution and recovering the Poisson limit (${\rm Var}_{\rm Poisson}(\hat{\bar n})=\bar{n}/A$, where $A$ is the footprint area). We find that the error on $\hat{\bar n}$ is dominated by cosmic variance (as opposed to shot noise) by more than a factor of $\sim5$ in all cases.}. Following this procedure, we estimate the number density in each field and redshift bin as well as its uncertainty (where we use the mask described in Section \ref{ssec:methods.mask} and the best-fit theory power spectra to compute the latter). The results are shown in Figure \ref{fig:ndens_consistency}. In all cases we find no significant deviations in the number density found in each field with respect to the mean, with only one estimate out of the 24 (GAMA15H in bin 2) deviating by more than 2 $\sigma$. We therefore conclude that there is no evidence of inconsistency between fields on the basis of their number densities.

\begin{figure}
        \centering
        \includegraphics[width=0.99\textwidth]{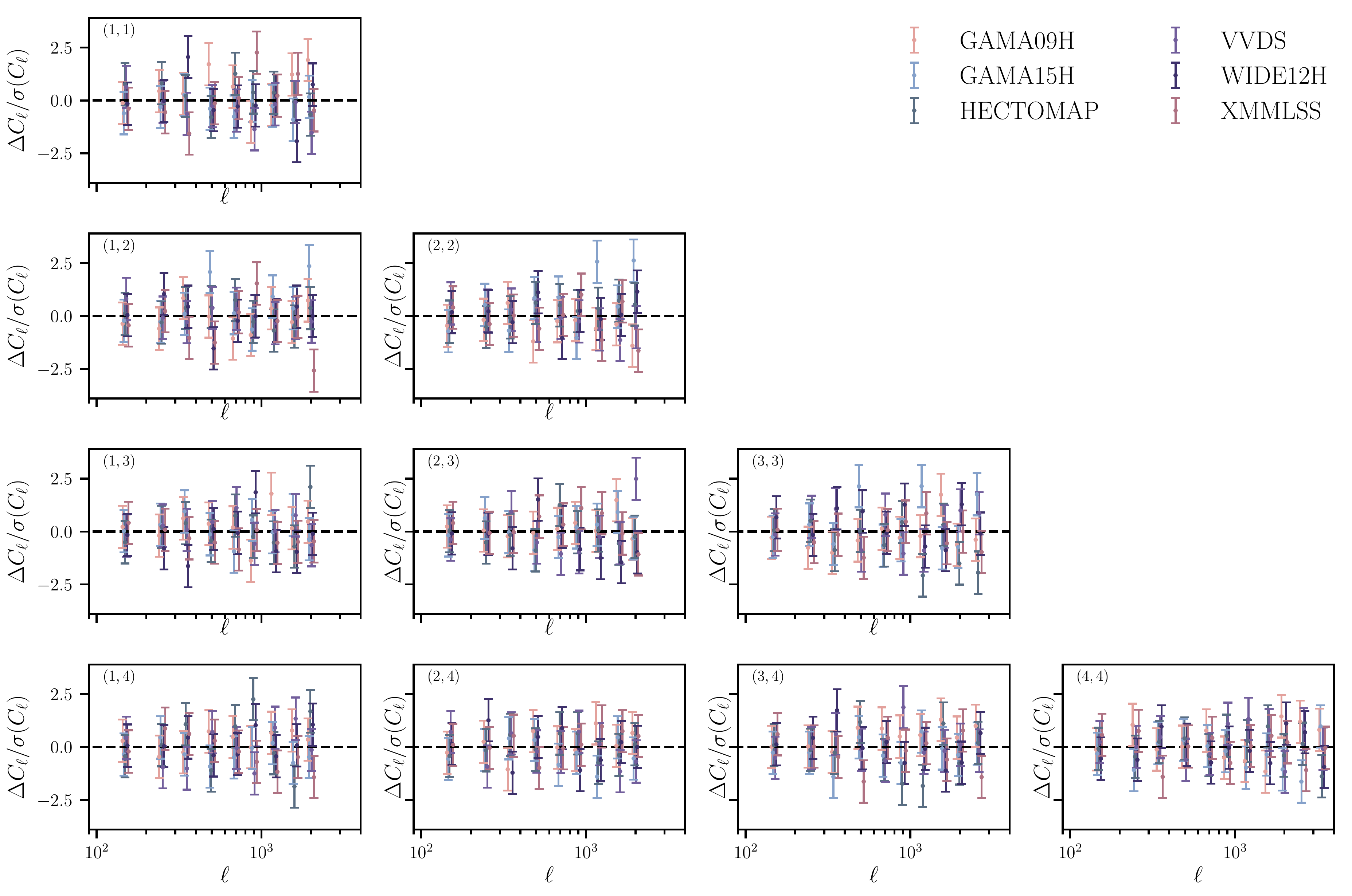}
        \caption{Difference between the power spectra measured in each field and the mean coadded spectra normalized by the 1$\sigma$ errors. The pair of numbers in the upper left corner of each panel corresponds to the indices of the bins being cross-correlated. We find consistency between the different measurements.}
        \label{fig:cls_consistency}
      \end{figure}

      \paragraph{Power spectra.} Figure \ref{fig:cls_consistency} shows the difference between the power spectra estimated in each field and the coadded power spectra normalized by the power spectrum errors for all auto- and cross-correlations. We observe a reasonable scatter with respect to the coadded spectra of up to $\sim3\sigma$, which does not immediately indicate any evidence for inconsistency between fields. As a more quantitative check for inconsistencies we carry out a $\chi^2$ analysis of this scatter. Let  $\Delta {\bf C}_f={\bf C}_f-{\bf C}_{\rm coadd}$ be the difference between the power spectra measured in field $f$ and the coadded ones. Using the notation of Section \ref{sssec:results.spectra.fid}, the covariance of $\Delta {\bf C}_f$ is given by:
      \begin{equation}
        {\sf Cov}_{\Delta_f}=\left(1-2\frac{A_f}{\sum_{f'}A_{f'}}\right){\sf Cov}_f+{\sf Cov}_{\rm coadd}.
      \end{equation}
      We can therefore quantify the significance of the power spectrum differences by computing the $\chi^2$:
      \begin{equation}
        \chi^2\equiv\Delta{\bf C}_f^T\cdot{\sf Cov}_{\Delta_f}^{-1}\cdot\Delta{\bf C}_f,
      \end{equation}
      and its probability to exceed (PTE) under the assumption that $\chi^2$ follows a ``chi-squared'' distribution with a number of degrees of freedom given by the size of $\Delta{\bf C}_f$. Doing so for all the individual auto- and cross-correlations, as well as for the combined data vector containing all of them simultaneously, we find no quantitative evidence of inconsistency between fields. All PTEs are larger than 8\%, with the vast majority of them lying above 30\%. We conclude that there is no evidence for systematic biases from the power spectrum measurements in different fields, and therefore it is safe to coadd them and use the coadded spectra to obtain model constraints. We observe that the HECTOMAP field exhibits some of the lowest PTEs in the analysis described above. Although none of these are cause for concern, we have omitted the measurements from HECTOMAP when obtaining the coadded power spectra for safety. Since this is by far the smallest field ($\sim5\,{\rm deg}^2$), the associated loss of sensitivity is negligible.
      
    \subsubsection{Robustness to contaminants} \label{sssec:results.spectra.syst}
      \begin{figure}
        \centering
        \includegraphics[width=0.99\textwidth]{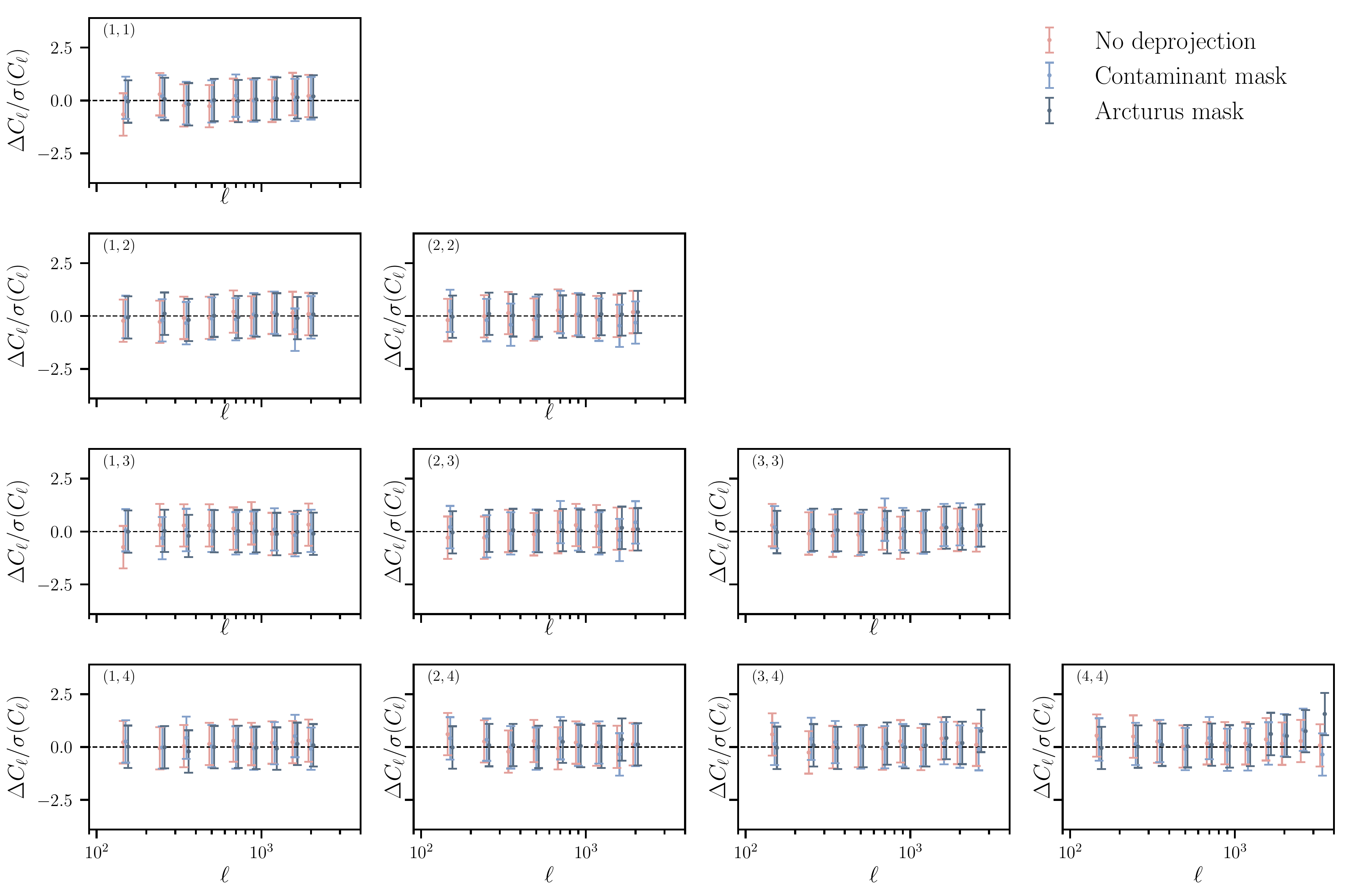}
        \caption{Difference with respect to our fiducial power spectra of alternative estimates of the power spectra normalized by their $1\sigma$ errors. The pink points show the spectra estimated without contaminant deprojection. The light blue points show spectra estimated using a conservative sky mask that removes highly contaminated regions (see Section \ref{sssec:results.spectra.syst} for details). The dark blue points show the spectra calculated using the Arcturus bright star mask.}
        \label{fig:cls_systematics}
      \end{figure}
      As described in Section \ref{ssec:methods.cell}, our main strategy to address possible contamination of the measured power spectra by systematics causing artificial density fluctuations is to project out the systematics templates discussed above from the data at the map level. This procedure can also be understood as building a linear model for the contamination (see Eq. \ref{eq:deproj1}), finding the best-fit linear coefficients for each contaminant and subtracting the corresponding contribution from the observed map. Finally, the estimated power spectra must be corrected for the loss of modes incurred (pink line in Figure \ref{fig:cls_summary}). This method is therefore able to account for any systematic contamination that is well described by a linear contribution. Since the impact of any contaminant can always be Taylor-expanded, this treatment is appropriate as long as the level of contamination is sufficiently small. In order to verify this, we have carried out two different tests.
      
      First, we have compared our fiducial power spectra, computed using contaminant deprojection, with power spectra computed without accounting for any type of contamination (i.e. estimated directly from the observed galaxy overdensity maps). This allows us to test the worst-case scenario where any source of contamination is completely ignored. The result, shown as the differences between both power spectra normalized by their 1$\sigma$ uncertainty, is shown in pink in Figure \ref{fig:cls_systematics} (note that the different curves shown there are strongly correlated with each other). In all cases we observe very small differences (smaller than $\sim0.3\sigma$) between both spectra. This suggests that the level of contamination in the raw galaxy overdensity maps is small, and the linear model implemented through template deprojection is likely accurate enough to account for it.
      \begin{figure}
        \centering
        \includegraphics[width=0.49\textwidth]{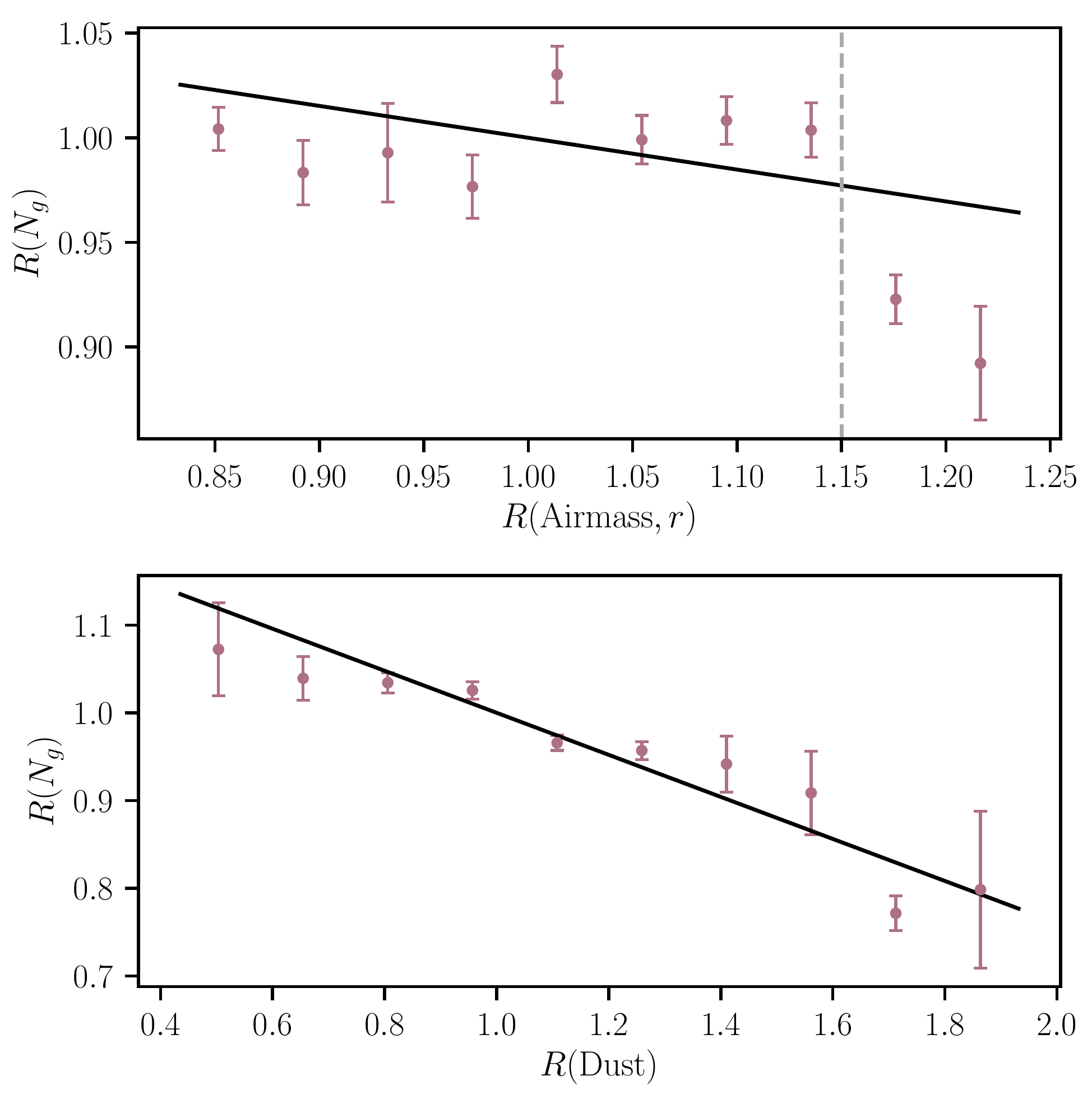}
        \includegraphics[width=0.49\textwidth]{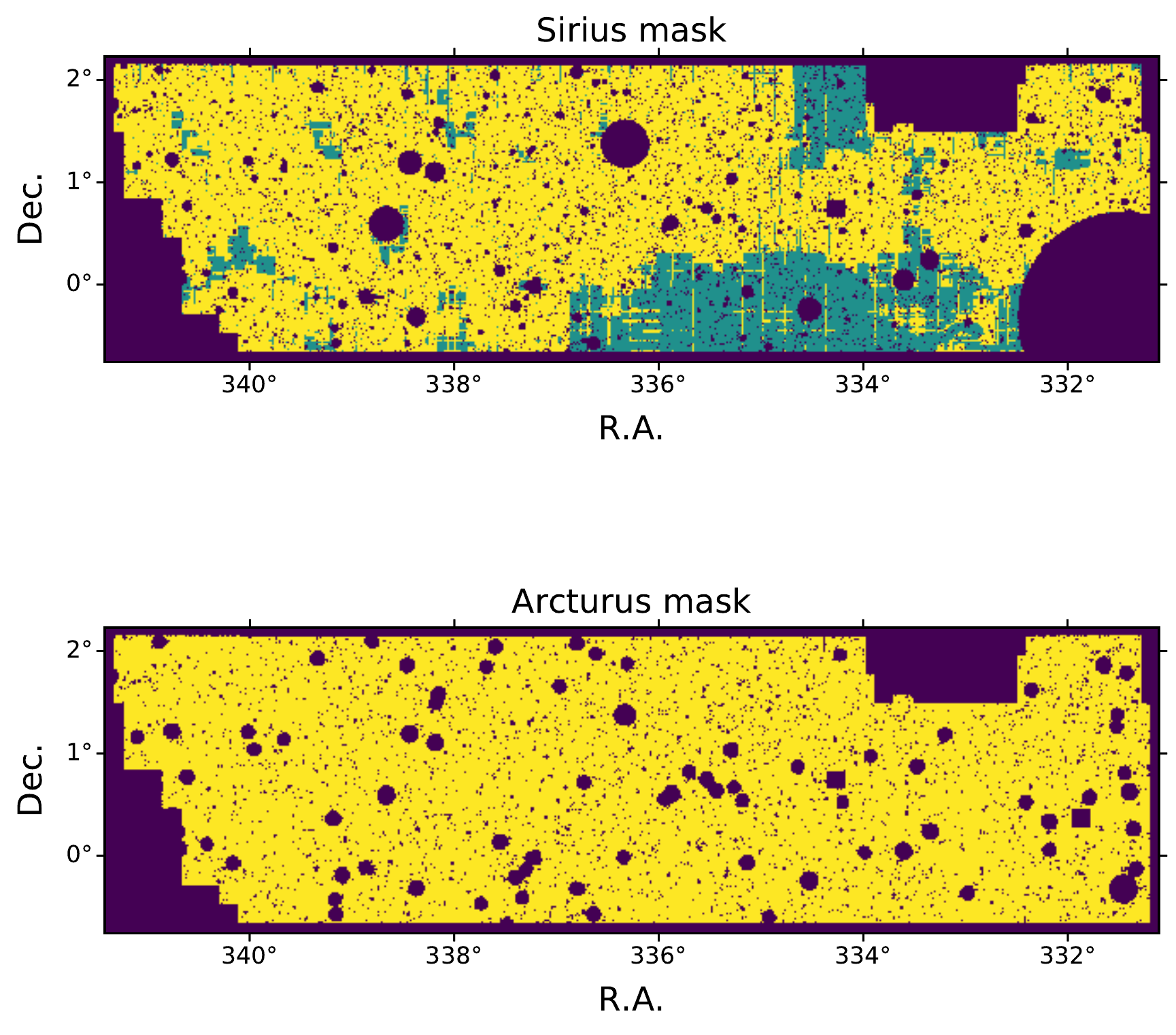}
        \caption{{\sl Left panels:} relation between the galaxy density fluctuation ($y$ axis) and the fluctuation in different systematics ($x$-axis). In all cases $R(x)\equiv \sfrac{x}{\bar{x}}$. Results are shown for $r$-band airmass (top) and dust absorption (bottom) for the third redshift bin in the VVDS field. The data are shown in burgundy, while the solid black line shows the best-fit linear relation between both quantities. In cases where a linear relation is not appropriate (as shown here for $r$-band airmass), we mask all regions where the associated contaminant is above/below a given threshold (shown as a vertical dashed line here). {\sl Right panels:} the top panel shows our fiducial ``Sirius'' mask, with the additionally masked regions associated with high levels of contamination shown in turquoise. The bottom panel shows the alternative ``Arcturus'' bright-star mask.}
        \label{fig:ndens_syst}
      \end{figure}
      
      Second, in order to explore the breakdown of the linear model used in deprojection, we have conducted a direct study of the relation between galaxy overdensity and the different systematics as follows: for each field and redshift bin, we produce a map of the relative galaxy density $R(N_g)\equiv N_g(\nv)/\bar{N}_g$, where $N_g(\nv)$ is the number of galaxies in the pixel with coordinates $\nv$, and $\bar{N}_g$ is the mean number of galaxies per pixel across the map. Then, for each of the 48 systematic templates $S$, we create a similar map $R(S)\equiv S(\nv)/\bar{S}$. We then use both maps to calculate the mean value of $R(N_g)$ in bins of $R(S)$, estimating the error on this mean via bootstrap. Finally, we produce plots of this relation for all fields, redshift bins and systematic maps, finding results such as those displayed in the left panels of Figure \ref{fig:ndens_syst}, which show the relation between the galaxy density fluctuation and the fluctuations in dust absorption and $r$-band airmass for the third redshift bin of the VVDS field (bottom and top panels respectively). In most cases, we find that the relation between galaxy density and systematic fluctuation is either flat or well approximated by a linear relation, as is the case for dust absorption in the figure. In a few cases, however, we find that a linear relation is only appropriate in parts of the range of contaminant values, and that the observed galaxy overdensity grows or decreases much faster for large or small values of $R(S)$. In these cases, fitting a linear relation over the whole range of $S$ will lead to some level of contaminant residuals that could induce a significant bias on the estimated power spectra. To verify whether this is the case, we list all cases where we find that a linear relation is not appropriate, determine the value of $R(S)$ beyond which we observe a significant increase/decrease in $R(N_g)$ (shown as a vertical dashed line in the top left panel of Figure \ref{fig:ndens_syst} for $r$-band airmass), and mask out the corresponding regions of the map. The masked regions correspond to $\sim20\%$ of the available footprint on average, and are shown in the top right panel of Figure \ref{fig:ndens_syst} in turquoise for the VVDS field. The light blue data points in Figure \ref{fig:cls_systematics} show the difference of the power spectra estimated using these more restrictive masks with respect to our fiducial power spectra, normalized by their $1\sigma$ errors. In the vast majority of cases we see only $<1\sigma$ differences between both spectra. Since small differences are to be expected when masking a significant fraction of the observed footprint, we conclude that there is no evidence of contamination in our fiducial power spectra beyond that accounted for by the template deprojection procedure.

      One final possible source of systematic bias is the effect of bright sources, which cause a depletion in the number of observed galaxies around them as described in e.g. \cite{2018PASJ...70S...7C}. To mitigate this effect we make use of the bright object mask provided with the HSC DR1 (the so-called ``Sirius'' mask) as described in Section \ref{ssec:methods.mask}. One possible problem associated with this mask is the fact that it removes regions around both bright stars as well as a small fraction of bright extra-Galactic objects. Since the latter will be correlated at some level with some of the sources used in our clustering analysis, it is important to check for a possible bias associated with masking them. To do so, and to test our fiducial power spectra against the exact procedure used to create the bright object mask, we have repeated our measurements making use of the bright star mask published by \cite{2018PASJ...70S...7C} (the so-called ``Arcturus'' mask). The top-right and bottom-right panels of Figure \ref{fig:ndens_syst} show both masks for the VVDS field. We can see that, while the masked regions are mostly centered around the same sources, the prescriptions used to define the masking radii are different (see \cite{2018PASJ...70S...7C} for further details). The dark blue data points in Figure \ref{fig:cls_systematics} show the difference with respect to our fiducial power spectra of the spectra computed using the Arcturus mask, normalized by their 1$\sigma$ errors. As before, we do not observe any statistically significant deviation between these spectra. We therefore conclude that bright sources do not impact our fiducial power spectra significantly.
      
      The scaling of the galaxy correlation function with magnitude limit is also a sensitive test to detect the presence of systematic contaminants in imaging data \citep{BookPeeblesLSS}, which has been used since the early times of galaxy surveys \citep{1996MNRAS.283.1227M}. Its use on deep samples is complicated by the unknown evolution of the galaxy luminosity function, and therefore we have not carried out this test here.
      
    \subsubsection{Shot noise subtraction} \label{sssec:results.spectra.shotnoise}
      \begin{figure}
        \centering
        \includegraphics[width=0.6\textwidth]{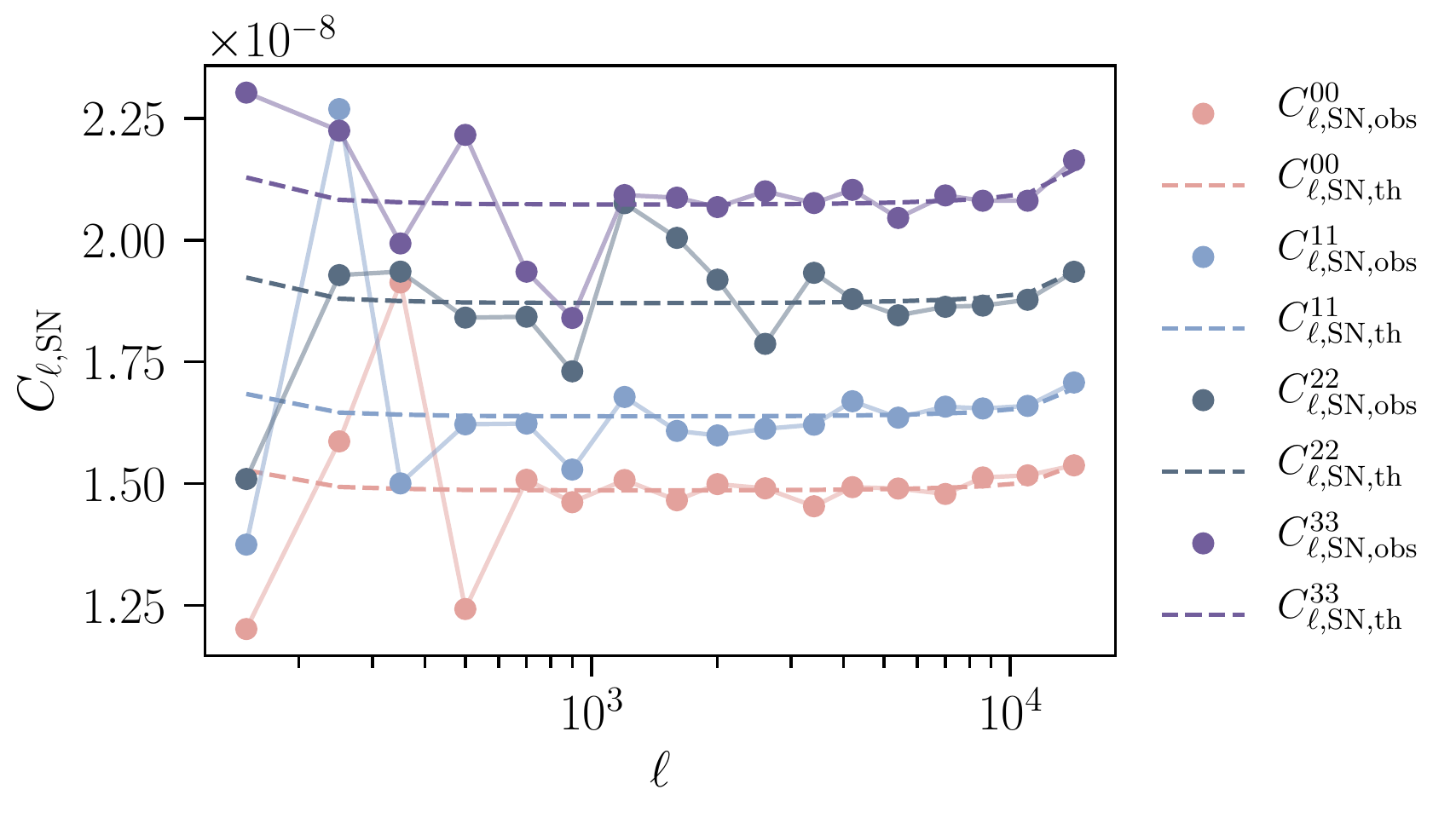}
        \caption{Comparison between the shot noise power spectra estimated analytically using Eq.~\ref{eq:nell} and estimated from the difference-map power spectra. Results are shown for the four different redshift bins. Our analytic estimate agrees well with this alternative method. The shot noise power spectra for the different bins have been offset for clarity.}
        \label{fig:shotnoise_avg}
      \end{figure}
      As described in Section \ref{ssec:methods.cell}, we subtract the shot-noise contribution to the auto-correlation power spectra using an analytical estimate, given by Eq. \ref{eq:nell}. Since it has been argued \citep{2013PhRvD..88h3507B} that non-linearities may produce deviations from this simple relation, we verify the validity of our calculation as follows.
      
      We start by splitting the galaxy sample in each field into two random subsamples with the same number of objects.  We then construct overdensity maps for each of the galaxy subsamples, which we call $\delta_1$ and $\delta_2$.  Each of these subsamples can be thought of as an independent Poisson processes that samples the same underlying smooth overdensity field $\delta$, i.e. $\delta_i = \delta + n_i$, where $n_i$ is the shot-noise contribution in $\delta_i$. Therefore, we can estimate the shot-noise power spectrum from the power spectrum of the difference between the two split maps:
      \begin{align}
      \left\langle |\delta_1 - \delta_2|^2 \right\rangle & = \langle|n_1|^2\rangle + \langle|n_2|^2\rangle,
      \end{align}
      where we have assumed that $n_1$ and $n_2$ are uncorrelated (since our two sub-catalogs are disjoint). Since the number density in each of the subsamples is half of the full sample, we can recover an estimate of the latter's shot-noise spectrum by simply dividing the power spectrum of the difference map by $4$. Figure \ref{fig:shotnoise_avg} shows the comparison between the analytic estimate in Eq. \ref{eq:nell} (transparent lines) and the estimate from the difference map power spectra (solid lines) for the four different redshift bins. Within the statistical noise of the split-map estimate, both methods agree well, particularly at high $\ell$, validating our procedure to subtract the shot-noise contribution.

  \subsection{Covariance matrix}\label{ssec:results.covariance}
\begin{figure}
      \begin{center}
        \includegraphics[width=0.5\textwidth]{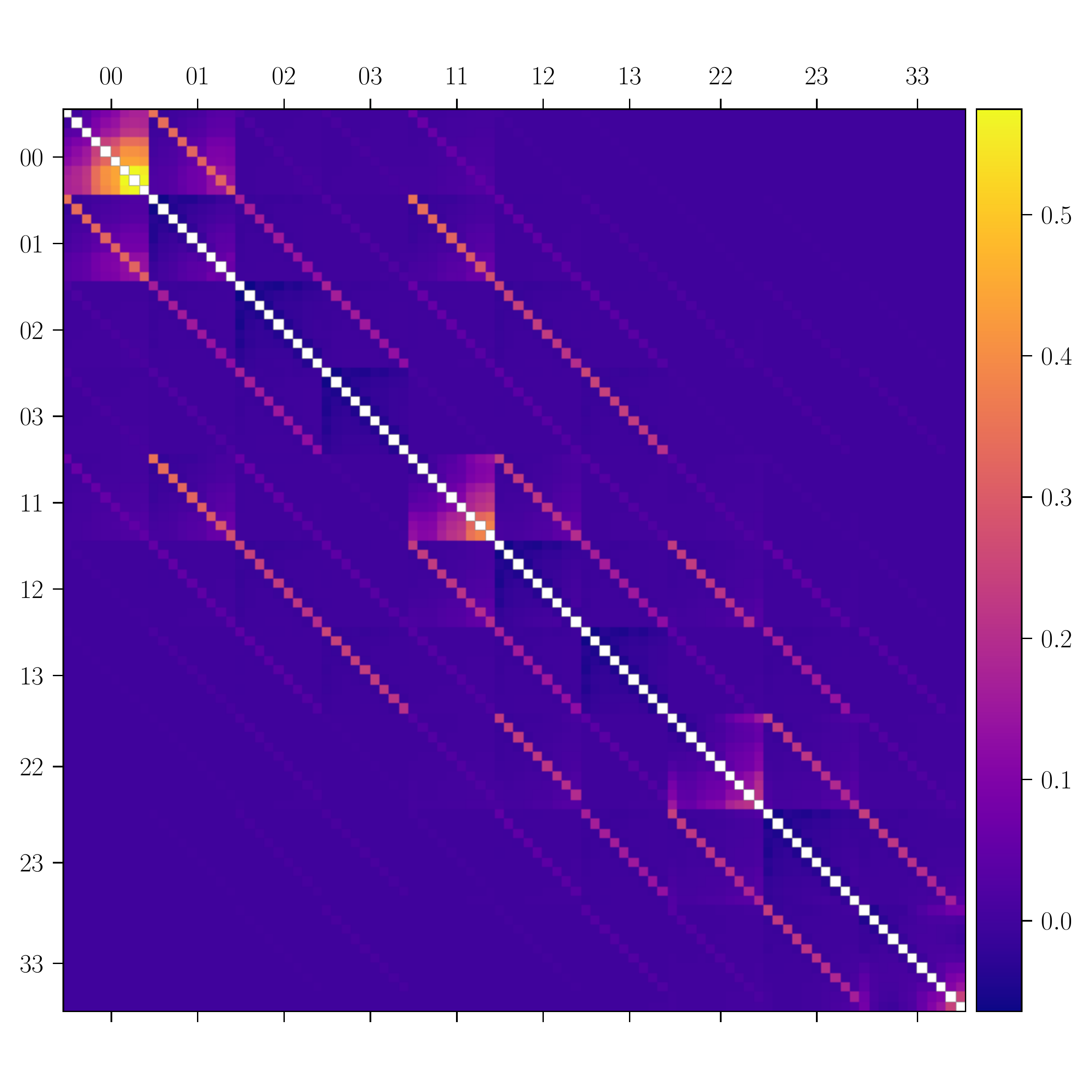}
        \caption{Correlation matrix for all auto- and cross-power spectra considered in our analysis, computed analytically as described in Sec.~\ref{sssec:methods.theory.covar}.}\label{fig:covmat}
      \end{center}
    \end{figure}

\begin{figure}
      \begin{center}
        \begin{subfigure}{0.49\textwidth}
        \includegraphics[width=\textwidth]{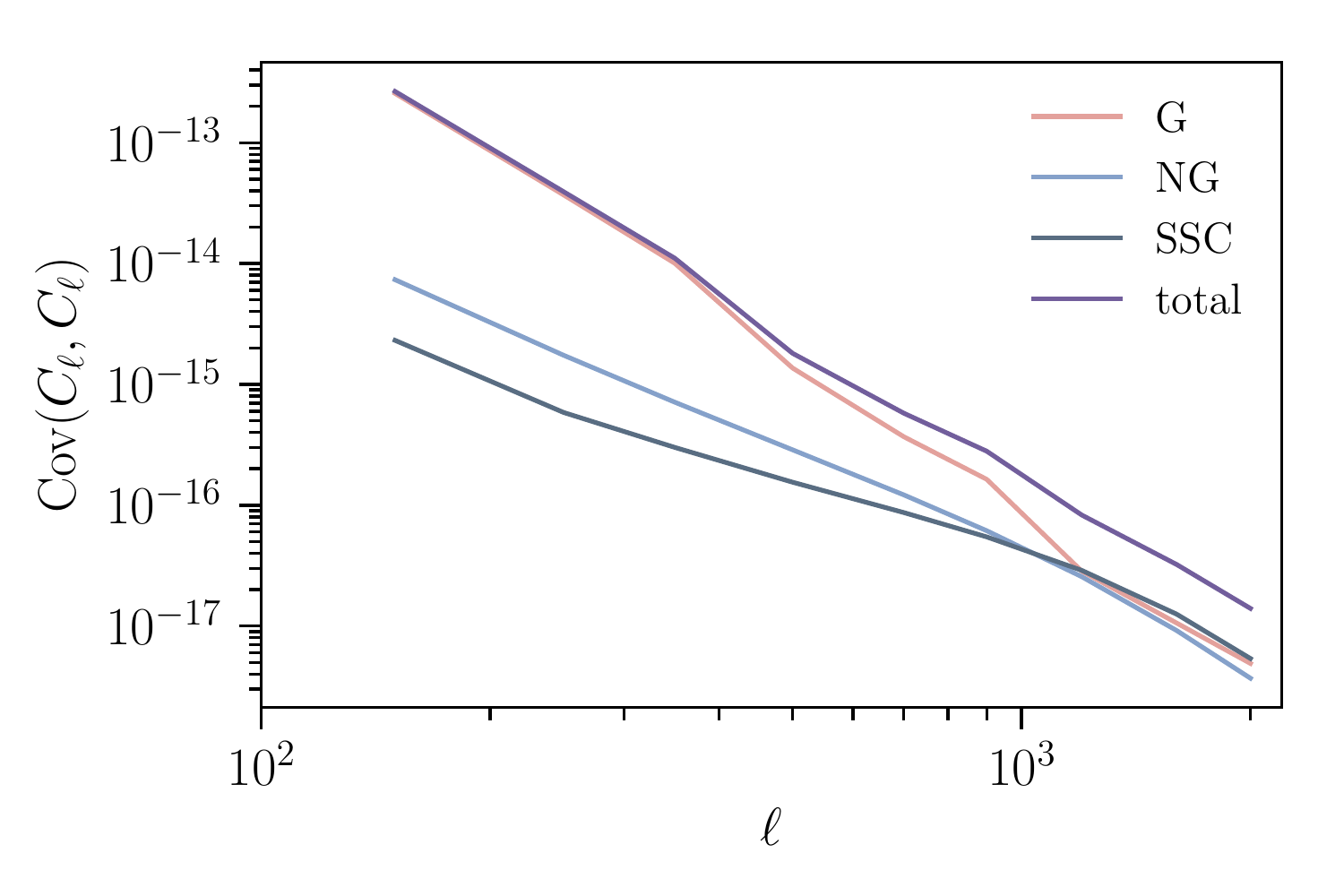}
        \caption{$C_{\ell}^{00}$}
        \end{subfigure}
        \begin{subfigure}{0.49\textwidth}
        \includegraphics[width=\textwidth]{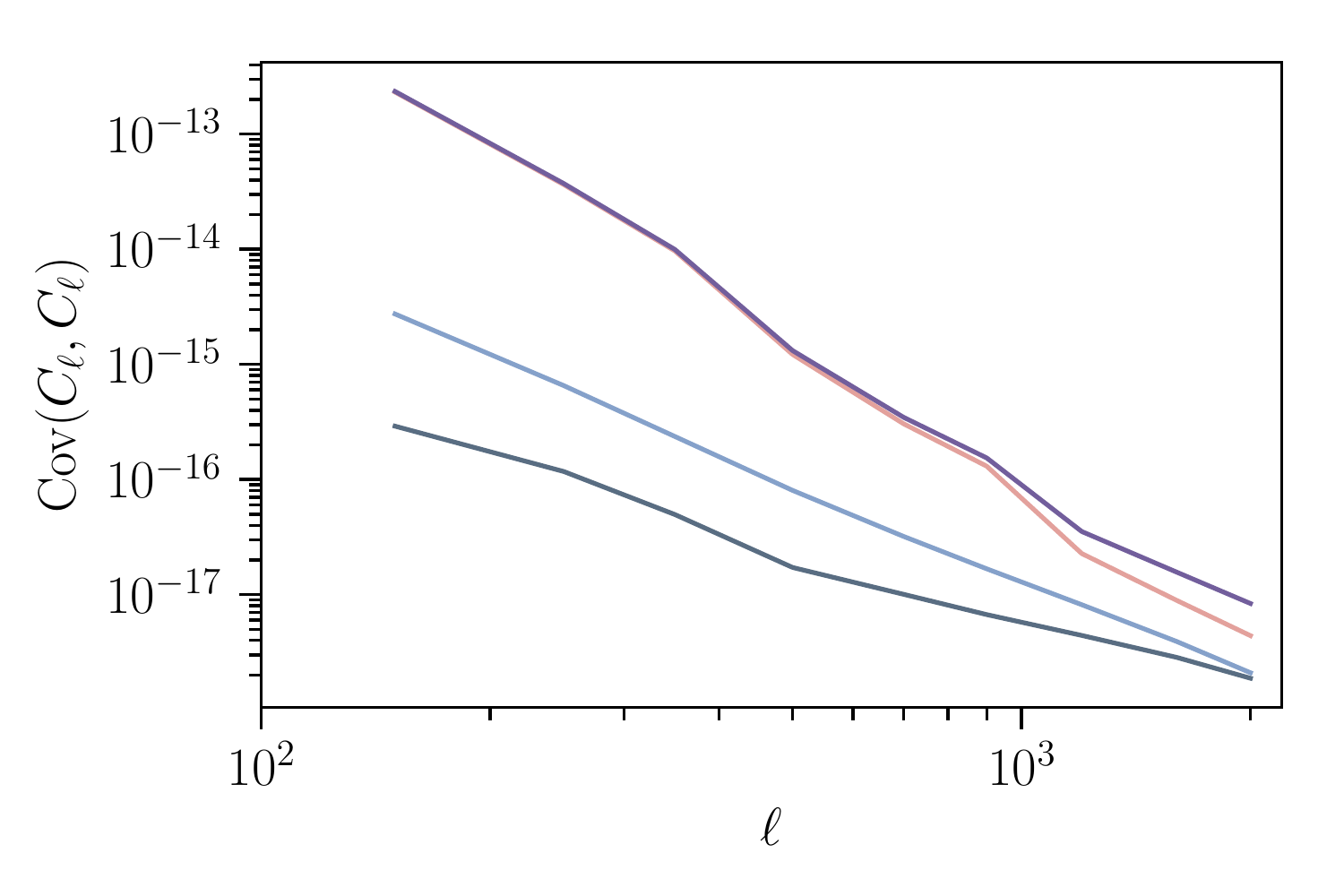} \\
        \caption{$C_{\ell}^{11}$}
        \end{subfigure}
        \begin{subfigure}{0.49\textwidth}
        \includegraphics[width=\textwidth]{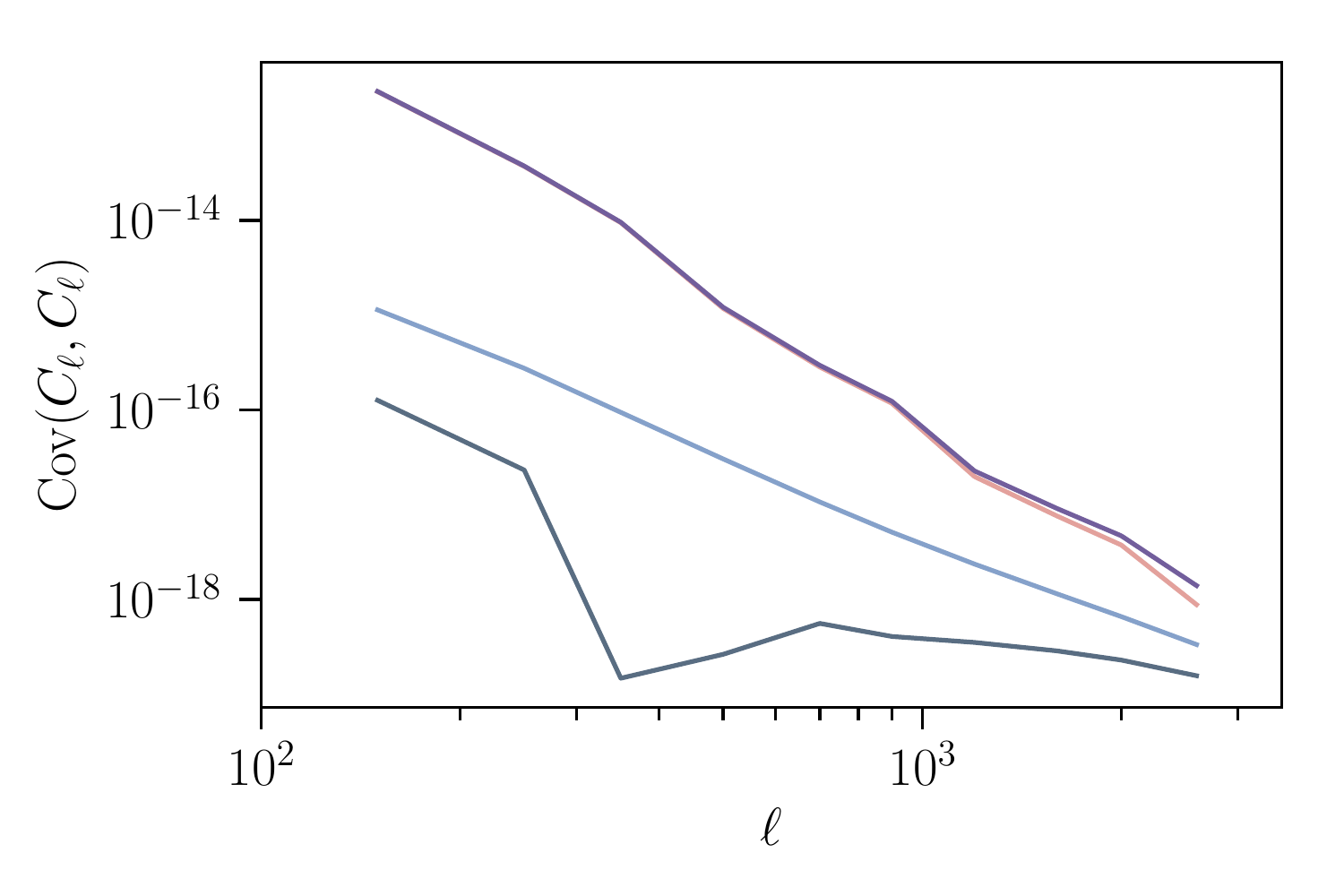}
        \caption{$C_{\ell}^{22}$}
        \end{subfigure}
        \begin{subfigure}{0.49\textwidth}
        \includegraphics[width=\textwidth]{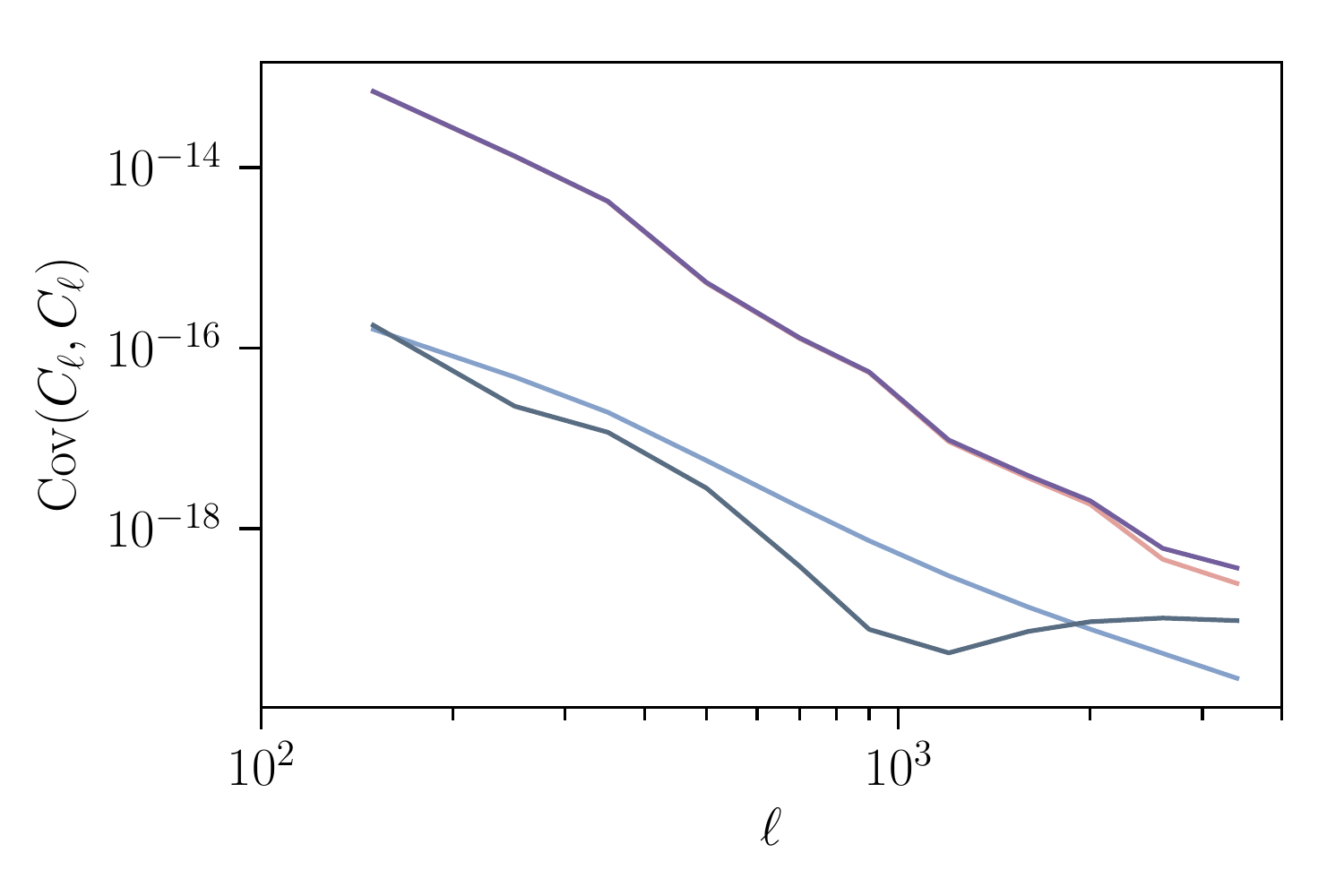}
        \caption{$C_{\ell}^{33}$}
        \end{subfigure}
        \caption{Comparison of the different contributions to the covariance matrix for the four auto-power spectra considered in our analysis.} \label{fig:covariance-contributions}
      \end{center}
    \end{figure}

    We compute the covariance matrix of all measured power spectra analytically as described in Sec.~\ref{sssec:methods.theory.covar} and the resulting correlation matrix is shown in Fig.~\ref{fig:covmat}\footnote{The correlation matrix $\mathsf{Corr}$ is obtained from the covariance matrix $\mathsf{C}$ as $\mathsf{Corr}_{ij} = \sfrac{\mathsf{C}_{ij}}{\sqrt{\mathsf{C}_{ii} \mathsf{C}_{jj}}}$.}. In Fig.~\ref{fig:covariance-contributions}, we split the auto-covariances for the four redshift bins into the Gaussian, non-Gaussian and SSC contributions. As can be seen, the Gaussian contribution is dominant on large scales while the non-Gaussian part becomes important at small scales and low redshift. The high redshift bins are mostly insensitive to non-Gaussian contributions to the covariance. Finally, the SSC is subdominant in all cases. This is mainly due to the number density correction in Eq.~\ref{eq:ps-resp}, which significantly suppresses any SSC contributions to the total covariance.

  \subsection{Constraints on HOD parameters}\label{ssec:results.hod-constraints}
    \subsubsection{Fiducial constraints}\label{sssec:results.hod-constraints.fiducial}
      Both the auto- and cross-power spectra shown in Fig.~\ref{fig:cls_summary} carry information on astrophysical, systematics and cosmological parameters. We especially expect the cross-correlations to help constrain photo-$z$ systematics parameters, as they probe the relative clustering strength in different redshift bins and thus help break degeneracies between the clustering amplitude and changes in the photometric redshift distributions. In this work, we therefore compute fiducial constraints from a joint fit to both auto- and cross-power spectra.

\begin{figure}
        \begin{center}
          \includegraphics[width=0.95\textwidth]{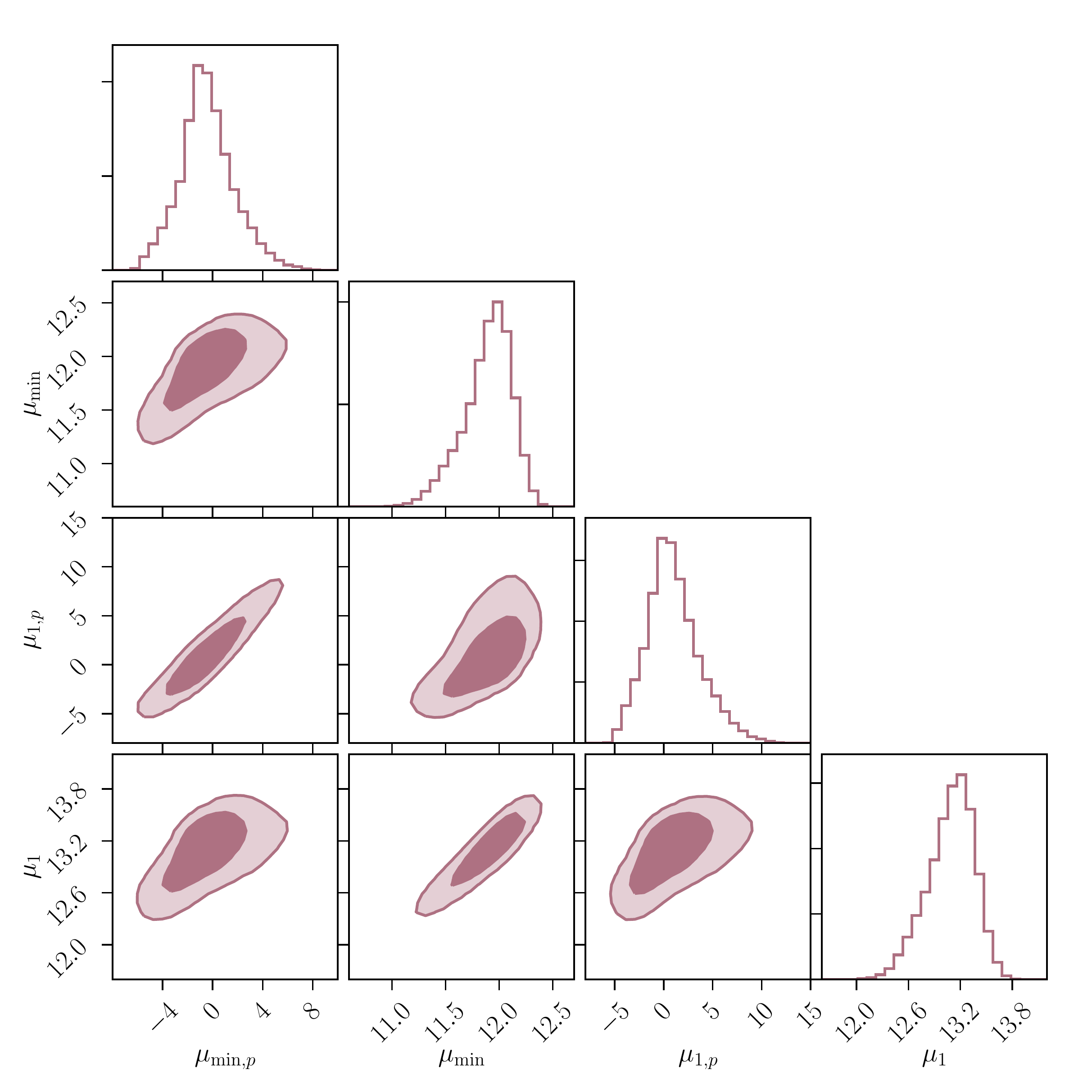}
          \caption{Fiducial constraints on HOD parameters obtained in this work. The inner (outer) contour shows the $68 \%$ c.l. ($95 \%$ c.l.).}
          \label{fig:constraints-fid-hod}
        \end{center}
      \end{figure}

\begin{figure}
        \begin{center}
          \includegraphics[width=0.95\textwidth]{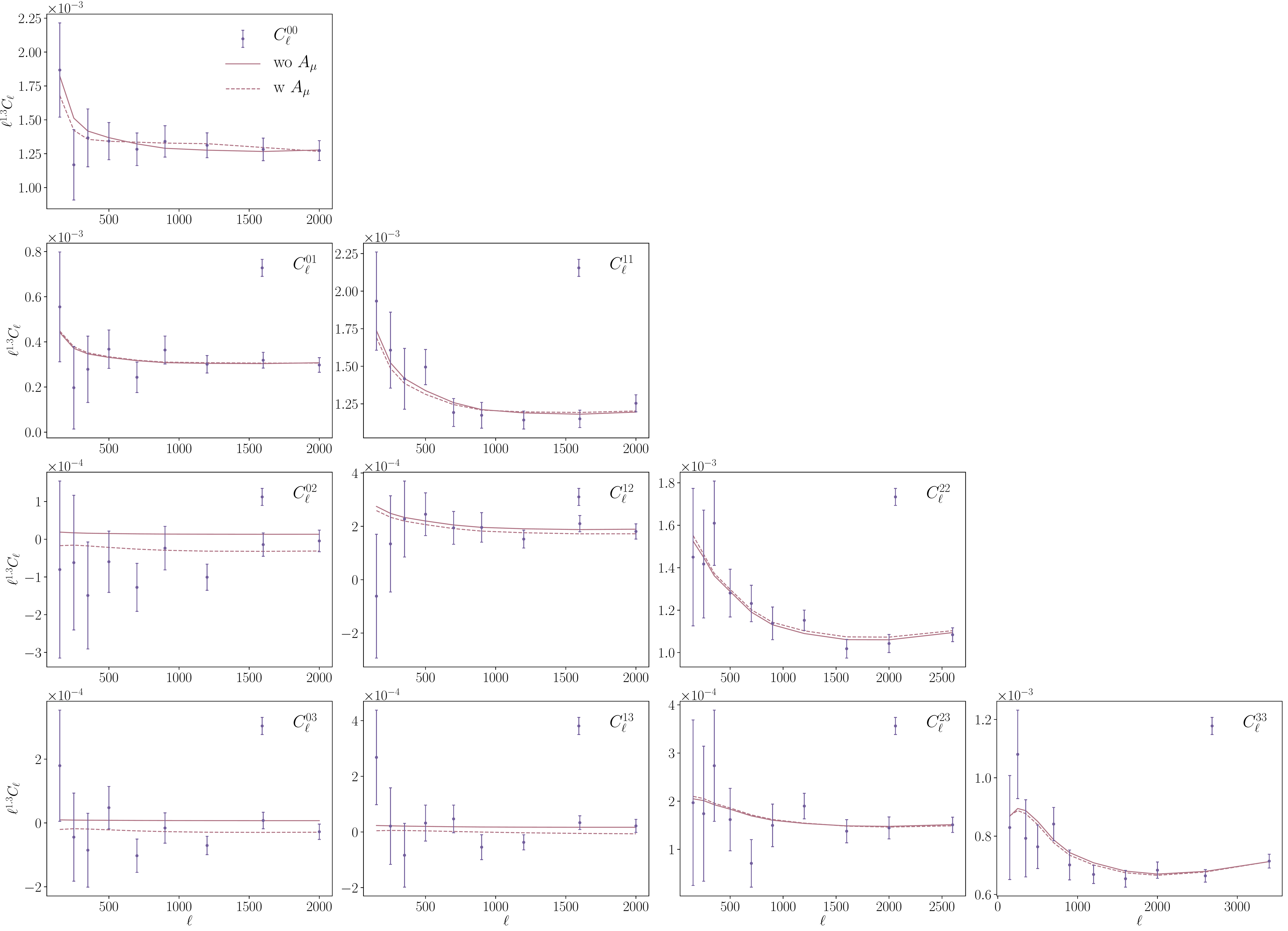}
          \caption{Measured auto- and cross-power spectra obtained in our analysis. The solid lines show the theoretical predictions derived from the best-fit model parameters not allowing for magnification, while the dashed lines show the theoretical predictions obtained allowing for a free magnification amplitude $A_{\mu}$.}
          \label{fig:cls-best-fit}
        \end{center}
      \end{figure}

      Fig.~\ref{fig:constraints-fid-hod} shows the constraints on the HOD parameters $\mu_{\mathrm{min}}, \allowbreak \, \mu_{\mathrm{min}, p}, \allowbreak \, \mu_{1}, \allowbreak \, \mu_{1, p}$ for our fiducial model described in Sec.~\ref{ssec:methods.constr}. The constraints on all fitted parameters are shown in Fig.~\ref{fig:constraints-fid-full} and the corresponding best-fit values and means are shown alongside their $68 \%$ confidence limits in Tab.~\ref{tab:params}. As can be seen from Figures \ref{fig:constraints-fid-hod} and \ref{fig:constraints-fid-full}, the data allow us to constrain $M_{\mathrm{min}}(z)$ and $M_{1}(z)$ whereas $M_{0}(z)$ is unconstrained\footnote{In the following, we therefore only show constraints on the HOD parameters $\mu_{\mathrm{min}}, \, \mu_{\mathrm{min}, p}, \, \mu_{1}, \, \mu_{1, p}$.}. Fig.~\ref{fig:cls-best-fit} shows the theoretical predictions derived from maximum likelihood parameters alongside the measured power spectra. The corresponding minimum $\chi^{2}$ is $\chi^{2} = 86.2$. Computing the degrees of freedom as $\nu = N_{\mathrm{data}} - N_{\mathrm{param}} = 94 - 14 = 80$, we obtain $\chi^{2}_{\mathrm{red}} = \sfrac{\chi^{2}}{\nu} = 1.08$ ($p$-value $= 0.30$), which shows that the data are consistent with the best-fit theoretical model\footnote{We note that this estimate of the degrees of freedom is only valid for linear models and independent basis functions (see e.g. \cite{Andrae:2010}). However, we will use it throughout this work, as it allows us to obtain a rough estimate of the goodness of fit of the different models considered in our analysis.}. In Fig.~\ref{fig:cls-best-fit-1h-2h}, we additionally show our fiducial theoretical power spectra split into their 1- and 2-halo contributions. As can be seen, the contribution of the 1-halo term to the total power spectrum is most pronounced at low redshift. At higher redshift, the importance of the 1-halo term decreases and the transition from the 1- to 2-halo-dominated regime moves to smaller angular scales.

\begin{figure}
        \begin{center}
          \begin{subfigure}[t]{0.49\textwidth}
          \includegraphics[width=\textwidth]{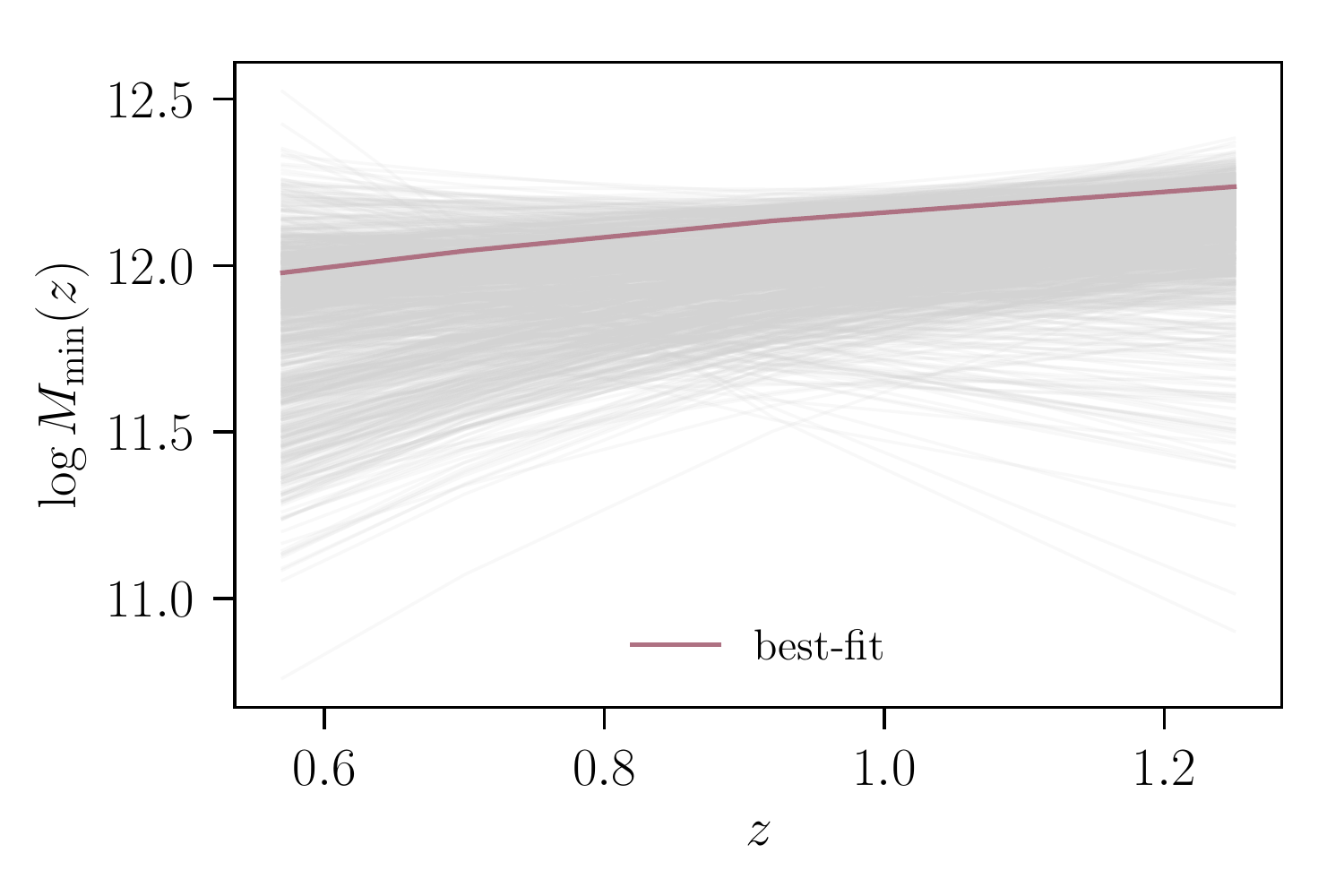}
          \end{subfigure}
          \begin{subfigure}[t]{0.49\textwidth}
          \includegraphics[width=\textwidth]{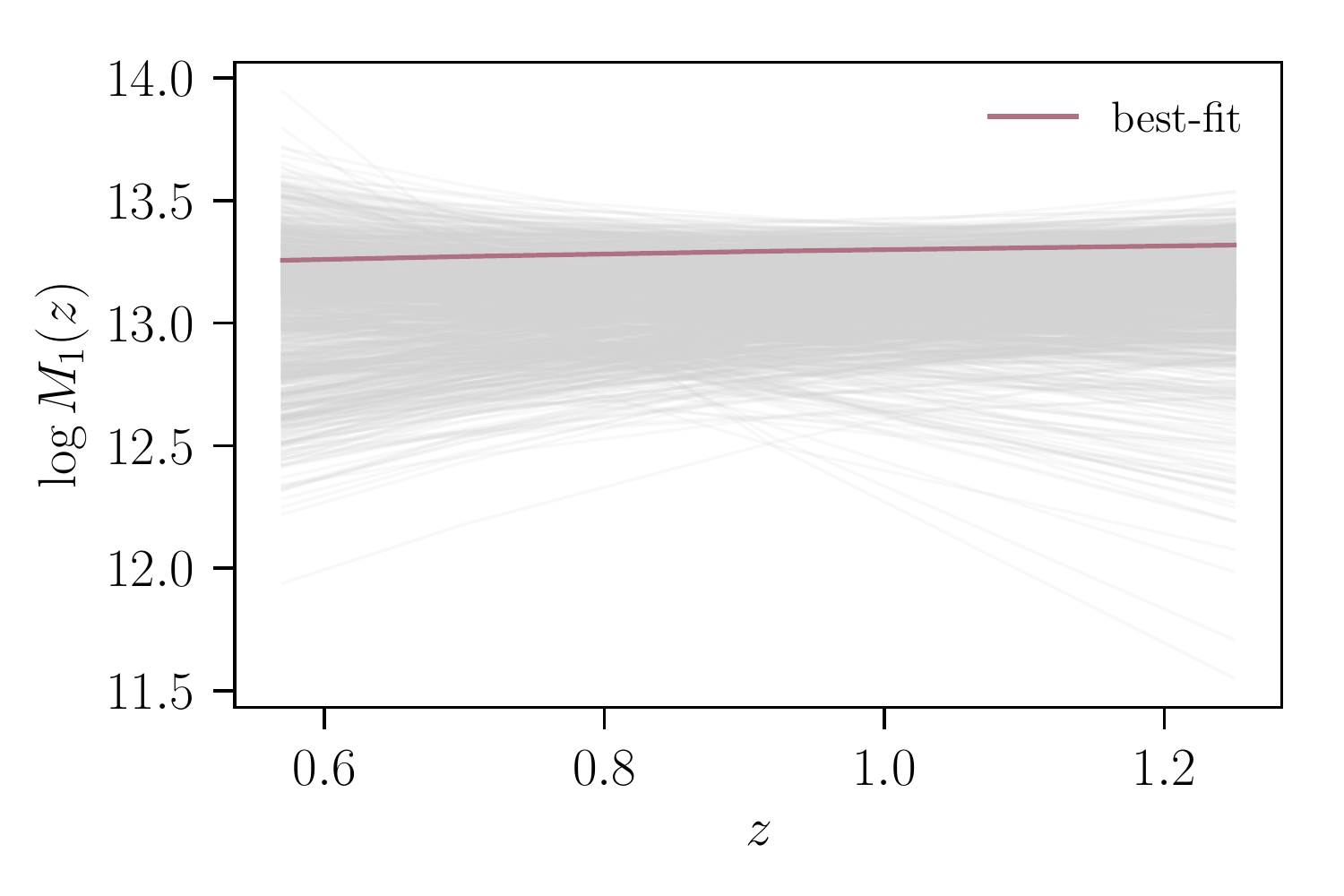}
          \end{subfigure}
          \caption{Functional forms for the logarithm of $M_{\mathrm{min}}(z)$ and $M_{1}(z)$ derived from our best-fit theoretical model alongside the functions derived from 1000 random realizations of the MCMCs to illustrate the uncertainty.} 
          \label{fig:hod-params-z-dep}
        \end{center}
      \end{figure}

In Fig.~\ref{fig:hod-params-z-dep} we show the redshift dependence of the logarithm of the minimal mass to host a central galaxy $M_{\mathrm{min}}(z)$ and the mass scale for satellites $M_{1}(z)$ obtained in our analysis. To illustrate the uncertainty on this relation, we also show a sample of curves derived from 1000 random realizations from the MCMCs. As can be seen, we find that, within our uncertainties, the data prefer both $M_{\mathrm{min}}(z)$ and $M_{1}(z)$ to be approximately redshift-independent. This is somewhat counterintuitive, as we would naively expect at least the minimal mass to host a central galaxy to increase with redshift, since galaxies will only form in the most massive halos at early times. We will discuss these findings in detail in Sec.~\ref{sssec:results.hod-constraints.constraints-interpretation}. Finally we see from Tab.~\ref{tab:params} that the minimal mass to host a satellite galaxy, $M_{0}(z)$, obtained in our analysis is quite low. This is a feature of our HOD model enforcing that no halo can host satellites if it does not contain a central galaxy, regardless of $M_{0}(z)$ (see Eq.~\ref{eq:ng_hod}). Therefore we find that the constraints on $M_{0}(z)$ are unbounded at low masses, which means that the data constrain the maximal $M_{0}(z)$ to be of the order of $M_{\mathrm{min}}(z)$, but our HOD model does not allow us to distinguish between masses smaller than $M_{\mathrm{min}}(z)$. Physically, this means that the data show preference for a model in which all halos massive enough to host a central galaxy are likely to host satellites too.

    \subsubsection{Robustness to modeling and data choices}
      In order to test the robustness of our fiducial HOD constraints to implementation and data choices, we compare the constraints from several analysis variants as follows. A summary of all robustness tests performed, and the respective constraints on HOD parameters and extended models can be found in Tab.~\ref{tab:constraints_robustness}.

      \begin{sidewaystable}
        \caption{Parameter constraints and best-fit $\chi^{2}$ values obtained for all considered analysis variants. The number of degrees of freedom $\nu$ is estimated according to $\nu = N_{\mathrm{data}}-N_{\mathrm{param}}$. In the second column of the table, the values in brackets denote the $p$-values corresponding to the observed best-fit $\chi^{2}$. The uncertainties denote the $68 \%$ c.l. corresponding to the equal-probability values encompassing a total probability of 0.68. In the cases in which they are not varied, the parameters $A_{\mu}, \Omega_{c}$ and $\sigma_{8}$ are fixed to $A_{\mu}=0$, $\Omega_{c}=0.264$ and $\sigma_{8} = 0.8111$.} \label{tab:constraints_robustness}
        \begin{center}
          \begin{tabular}{ccccccccc}
            \hline\hline 
            Analysis variant & $\sfrac{\chi^{2}}{\nu}$ & $\mu_{\mathrm{min}, p}$ & $\mu_{\mathrm{min}}$ & $\mu_{1, p}$ & $\mu_{1}$ & $A_{\mu}$ & $\Omega_{c}$ & $\sigma_{8}$ \\ \hline \Tstrut       
            fiducial & $\sfrac{86.2}{80}$ $(0.30)$ & $-0.5\substack{+2.1 \\ -2.0}$ & $11.88\substack{+0.22 \\ -0.23}$ & $0.9\substack{+2.7 \\ -2.6}$ & $13.08\substack{+0.27 \\ -0.28}$ & - & - & - \\
            auto & $\sfrac{19.2}{25}$ $(0.79)$ & $-0.9\pm 2.2$ & $11.82 \pm 0.28$ & $0.4\substack{+2.7 \\ -2.6}$ & $12.99 \pm 0.33$  & - & - & - \\
            G cov & $\sfrac{87.2}{80}$ $(0.27)$ & $-0.7\pm 2.1$ & $11.88\substack{+0.23 \\ -0.24}$ & $0.7\substack{+2.7 \\ -2.6}$ & $13.08\substack{+0.28 \\ -0.29}$ & - & - & - \\
            G+SSC cov & $\sfrac{86.2}{80}$ $(0.30)$ & $-0.4\substack{+2.0 \\ -1.9}$ & $11.89 \pm 0.20$ & $1.0\substack{+2.6 \\ -2.4}$ & $13.09\substack{+0.24 \\ -0.25}$ & - & - & - \\
            no $z_{w, i}$ & $\sfrac{88.0}{84}$ $(0.36)$ & $-0.86 \pm 0.65$ & $11.87 \pm 0.11$ & $0.33\substack{+0.89 \\ -0.90}$ & $13.07 \pm 0.15$ & - & - & - \\
            no $z_{w, i}, \Delta z_{i}$ & $\sfrac{95.2}{88}$ $(0.28)$ & $-1.09\substack{+0.62 \\ -0.77}$ & $11.78 \pm 0.13$ & $-0.11\substack{+0.73 \\ -0.94}$ & $12.93 \pm 0.16$ & - & - & - \\
            bins = 0, 1, 2 & $\sfrac{44.4}{43}$ $(0.41)$ & $-0.4\pm 2.3$ & $11.88\substack{+0.22 \\ -0.23}$ & $0.6\substack{+2.9 \\ -2.8}$ & $13.09\substack{+0.27 \\ -0.28}$ & - & - & - \\
            bins = 1, 2, 3 & $\sfrac{44.4}{46}$ $(0.54)$ & $1.2\pm 2.9$ & $11.97\substack{+0.31 \\ -0.36}$ & $3.5\substack{+3.7 \\ -3.6}$ & $13.23\substack{+0.39 \\ -0.43}$ & - & - & - \\
            pz = \texttt{Ephor\_AB} & $\sfrac{93.6}{80}$ $(0.14)$ & $0.3\substack{+2.1 \\ -1.9}$ & $12.14\substack{+0.19 \\ -0.17}$ & $1.8\substack{+2.9 \\ -2.6}$ & $13.39\substack{+0.24 \\ -0.23}$ & - & - & - \\
            pz = \texttt{Ephor} & $\sfrac{107.2}{80}$ $(0.023)$ & $0.9\pm 2.0$ & $12.15 \pm 0.17$ & $2.6\substack{+2.7 \\ -2.8}$ & $13.40\substack{+0.23 \\ -0.22}$ & - & - & - \\
            pz = \texttt{DEmP} & $\sfrac{105.4}{80}$ $(0.031)$ & $0.6\pm 1.9$ & $12.07\substack{+0.17 \\ -0.16}$ & $2.3\pm 2.6$ & $13.30\substack{+0.22 \\ -0.21}$ & - & - & - \\
            pz = \texttt{FRANKEN-Z} & $\sfrac{90.8}{80}$ $(0.19)$ & $0.0\substack{+2.0 \\ -1.8}$ & $12.12\substack{+0.18 \\ -0.16}$ & $1.4\substack{+2.8 \\ -2.4}$ & $13.38\substack{+0.23 \\ -0.22}$ & - & - & - \\
            fiducial magn. & $\sfrac{72.8}{80}$ $(0.70)$ & $-0.4\substack{+2.6 \\ -2.3}$ & $11.94\substack{+0.21 \\ -0.22}$ & $1.0\substack{+3.3 \\ -3.0}$ & $13.16\substack{+0.26 \\ -0.27}$  & - & - & - \\
            fit magn., auto+cross & $\sfrac{69.0}{79}$ $(0.78)$ & $-1.8\substack{+2.1 \\ -2.4}$ & $11.79\substack{+0.26 \\ -0.27}$ & $-0.7\substack{+2.6 \\ -2.7}$ & $12.98\substack{+0.30 \\ -0.31}$  & $2.18 \pm 0.74$ & - & - \\
            fit magn., auto & $\sfrac{19.4}{24}$ $(0.73)$ & $-0.8\substack{+2.3 \\ -2.2}$ & $11.81 \pm 0.26$ & $0.4\substack{+2.8 \\ -2.7}$ & $12.98 \pm 0.31$  & $0.6\substack{+2.7 \\ -2.6}$ & - & - \\
            fit cosmo & $\sfrac{84.4}{78}$ $(0.29)$ & $0.0\substack{+2.7 \\ -2.5}$ & $11.79\substack{+0.27 \\ -0.24}$ & $1.6\substack{+3.4 \\ -3.2}$ & $12.96\substack{+0.36 \\ -0.35}$  & - & $0.237 \pm 0.025$ & $0.81\substack{+0.15 \\ -0.14}$ \\
            \hline \hline
          \end{tabular}
        \end{center}
      \end{sidewaystable}
      
\begin{figure}
        \begin{center}
          \includegraphics[width=0.95\textwidth]{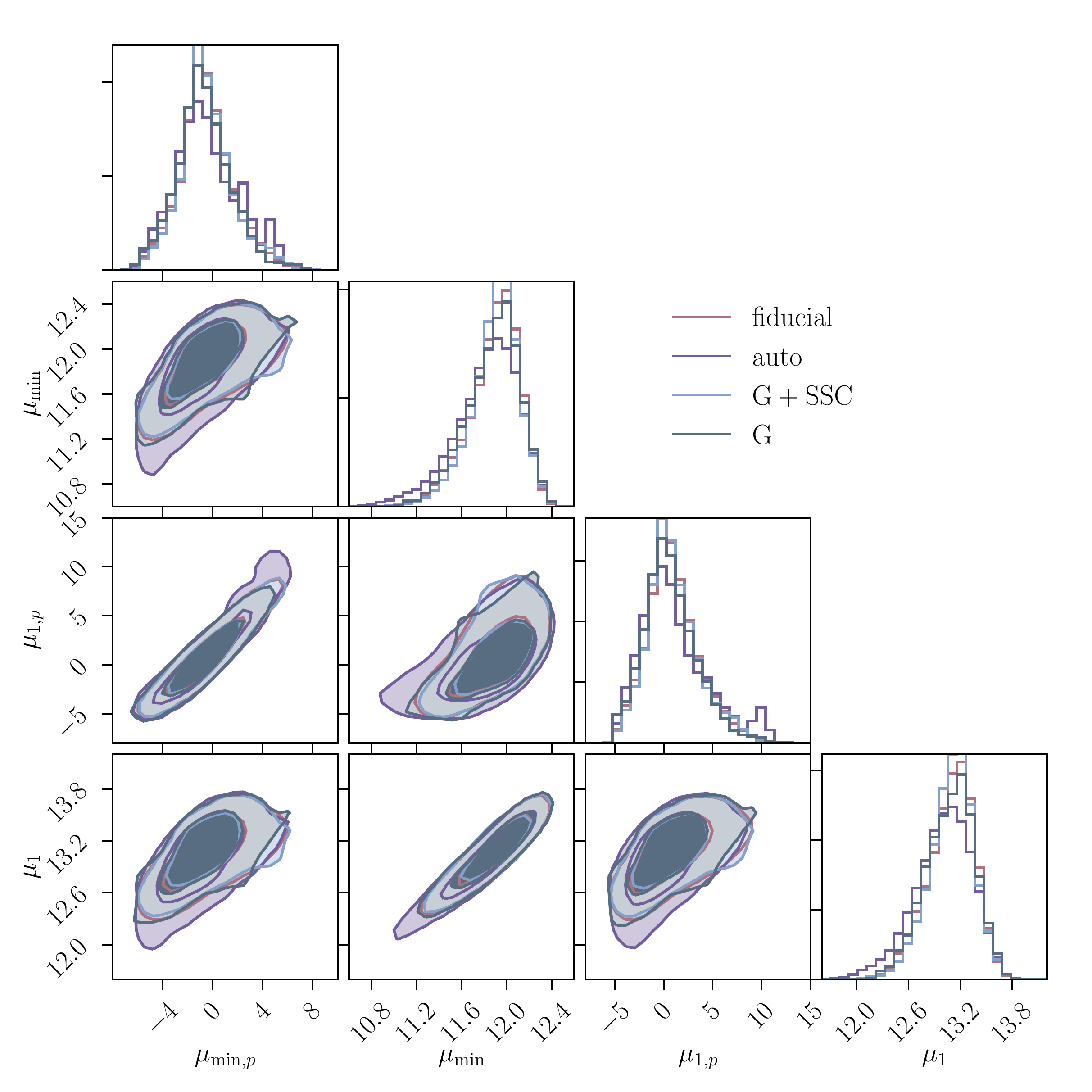}
          \caption{Comparison of our fiducial constraints on HOD parameters to those obtained from auto-power spectra alone, accounting only for Gaussian or Gaussian and SSC contributions to the covariance matrix. The inner (outer) contour shows the $68 \%$ c.l. ($95 \%$ c.l.).} \label{fig:constraints-cov=G+NG+SSC-vs-cov=G-vs-cov=G+SSC-vs-fit=auto}
        \end{center}
      \end{figure}

      As discussed above, our fiducial constraints are derived from a joint fit to both auto- and cross-power spectra. However, cross-correlations between redshift bins are especially sensitive to photometric redshift errors, such as outliers. In order to test for systematic biases affecting the cross-correlations, we therefore test the consistency of the auto- and cross-power spectra by comparing the constraints obtained using only auto-power spectra to those obtained when jointly fitting auto- and cross-power spectra. Fig.~\ref{fig:constraints-cov=G+NG+SSC-vs-cov=G-vs-cov=G+SSC-vs-fit=auto} shows the comparison between our fiducial HOD constraints and those obtained from auto-power spectra alone. As can be seen, the constraints from auto-spectra and from both auto- and cross-power spectra agree very well with each other, suggesting that uncertainties in photometric redshifts do not significantly affect the cross-correlations measured in our analysis. Furthermore, we see that the constraining power on HOD parameters is mostly unaffected by including the cross-power spectra. As an additional consistency check between auto- and cross-power spectra, we test how well the auto-spectra predict the cross-spectra. To this end we compute the $\chi^{2}$ between the observed auto- and cross-power spectra and the theoretical predictions derived from auto-power spectra only. As we will discuss below, the auto-power spectra cannot constrain the photo-$z$ systematics parameters and we therefore fix $z_{w, i} = 0, \Delta z_{i} = 0$ for this test. We find $\chi^{2} = 95.3$\footnote{We note that the number of degrees of freedom $\nu$ in this case is not well-defined as we compare the theoretical predictions obtained from auto-spectra only to the observed auto- and cross-power spectra. As a rough estimate, we can compute $\nu$ as the difference between the number of elements of the full data vector and the number of model parameters, leading to $\nu = 88$, which is equal to the number quoted in Tab.~\ref{tab:constraints_robustness}.}, which is practically equivalent to the best-fit $\chi^{2}$ obtained when fitting both auto- and cross-power spectra without accounting for photo-$z$ systematics (see Tab.~\ref{tab:constraints_robustness}). This shows that the auto-spectra are able to predict the cross-spectra and provides further confirmation of their consistency. 

      An important part of our data model is the analytical covariance matrix described in Sec.~\ref{sssec:methods.theory.covar}. In order to test the impact of the non-Gaussianity of the covariance on our results, we compare the HOD constraints obtained accounting for Gaussian (G) and Gaussian and SSC contributions (G+SSC) to our fiducial constraints, which include Gaussian, SSC and non-Gaussian (connected) contributions. As can be seen from Fig.~\ref{fig:constraints-cov=G+NG+SSC-vs-cov=G-vs-cov=G+SSC-vs-fit=auto}, these constraints agree very well with each other. The constraints from the G and the G+SSC case yield almost identical constraints, which is expected due to the suppression of the SSC contribution by the number density correction (c.f. Sec.~\ref{ssec:results.covariance}). Accounting for all non-Gaussian contributions to the covariance results in slightly broadened but consistent constraints. Comparing to Fig.~\ref{fig:covariance-contributions}, this is probably due to the fact that these corrections only affect the lowest redshift bins at small angular scales, and therefore do not have a significant impact when computing constraints from all power spectra. Finally, from Tab.~\ref{tab:constraints_robustness} we see that the reduced $\chi^{2}$ values of the data slightly increase as we remove the non-Gaussian contributions, as expected. However, the differences are not significant and we find almost identical goodness of fits, even when only accounting for the Gaussian covariance.

    \subsubsection{Robustness to photometric redshift uncertainties}
      One of the most important potential systematics in photometric galaxy clustering analyses are photometric redshift uncertainties and we therefore test the stability of our results to photo-$z$s in several different ways, as described in detail below. 
     
\begin{figure}
        \begin{center}
          \includegraphics[width=0.95\textwidth]{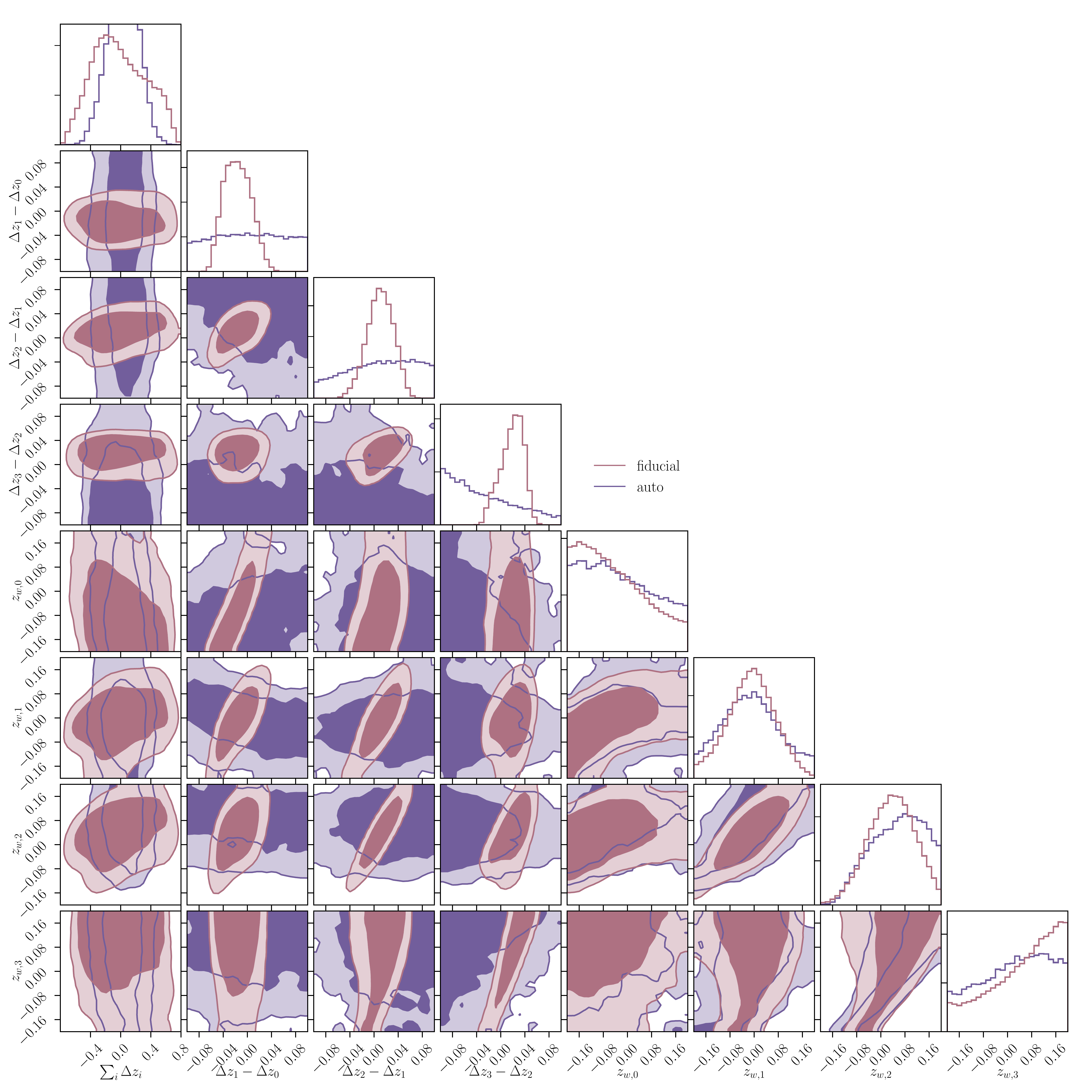}
          \caption{Comparison of our fiducial constraints on photo-$z$ systematics parameters to those obtained from auto-power spectra alone. The inner (outer) contour shows the $68 \%$ c.l. ($95 \%$ c.l.).}
          \label{fig:constraints-pz-syst-auto+cross-vs-auto}
        \end{center}
      \end{figure}

      In Fig.~\ref{fig:constraints-pz-syst-auto+cross-vs-auto}, we show the constraints on photometric redshift systematics parameters derived in our fiducial analysis. We find that the data cannot separately constrain the mean shift parameters $\Delta z_{i}$, as opposed to pairwise differences between those. We therefore show the constraints in terms of the reparameterized variables $\Sigma_{i=0}^{3} \Delta z_{i}, \Delta z_{1} - \Delta z_{0}, \Delta z_{2} - \Delta z_{1}, \Delta z_{3} - \Delta z_{2}$, which denote the sum of the $\Delta z_{i}$ and their pairwise differences respectively. While $\Sigma_{i=0}^{3} \Delta z_{i}$ and the width parameters $z_{w, i}$ are largely unconstrained, we find that we can constrain the three pairwise differences $\Delta z_{i} - \Delta z_{j}$ to within $\sigma_{\Delta z_{i} - \Delta z_{j}} \approx 0.023$ ($68 \%$ c.l.). This means that the data do not constrain the absolute position of each redshift bin, but are quite sensitive to their relative positions.

      It is instructive to investigate which part of the data drives the constraints on photometric redshift systematics. To this end we compare the constraints obtained using auto-power spectra only to those obtained from auto- and cross-power spectra (our fiducial case). As can be seen from Fig.~\ref{fig:constraints-pz-syst-auto+cross-vs-auto}, we find that the auto-spectra do not constrain the photo-$z$ systematics parameters $\Delta z_{i} - \Delta z_{j}$, as opposed to the combination of auto- and cross-power spectra. This shows that the cross-correlations drive the constraints on photometric redshift systematics and are thus essential for jointly constraining redshift systematics and astrophysical/cosmological parameters from galaxy clustering data.

\begin{figure}
        \begin{center}
          \includegraphics[width=0.95\textwidth]{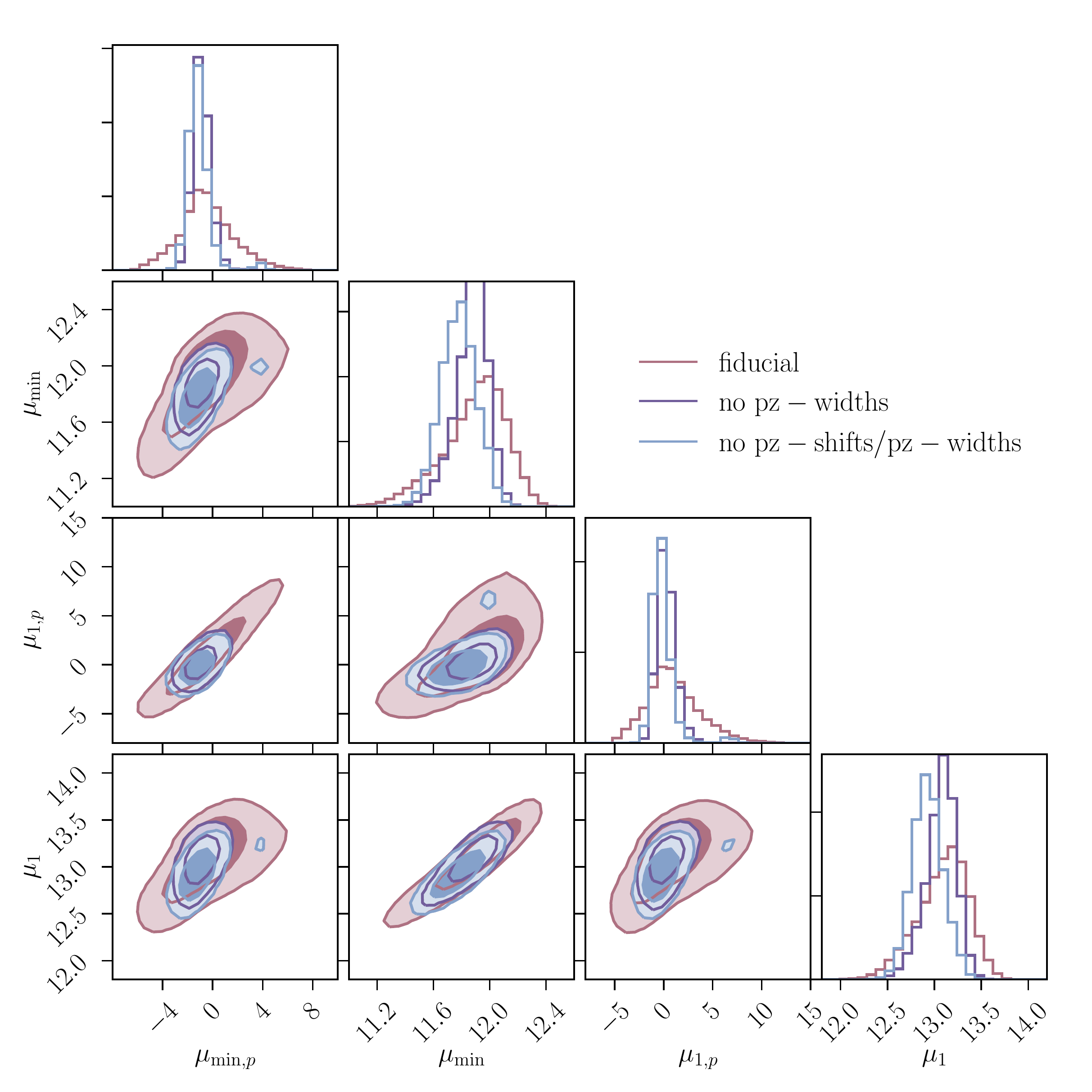}
          \caption{Comparison of our fiducial HOD constraints to those obtained setting $z_{w, i} = 0$ and $z_{w, i} = 0, \Delta z_{i} = 0$. The inner (outer) contour shows the $68 \%$ c.l. ($95 \%$ c.l.).}
          \label{fig:constraints-hod-pz-shifts-pz-widths-vs-pz-shifts-vs-no-pz-shifts}
        \end{center}
      \end{figure}

      In order to test the impact of photometric redshift uncertainties on our fiducial HOD constraints, we compare them to those obtained when separately fixing $z_{w, i} = 0$ and $z_{w, i} = 0, \Delta z_{i} = 0$. The results are shown in Fig.~\ref{fig:constraints-hod-pz-shifts-pz-widths-vs-pz-shifts-vs-no-pz-shifts}. As expected, we find that the constraints on HOD parameters weaken as we include more freedom in the photo-$z$ error model. However, as can be seen both from Fig.~\ref{fig:constraints-hod-pz-shifts-pz-widths-vs-pz-shifts-vs-no-pz-shifts} and Tab.~\ref{tab:constraints_robustness}, the constraints obtained in the three cases agree very well. In addition, we find acceptable $\chi^{2}$ values both for the model with $z_{w, i} = 0$ and $z_{w, i} = 0, \Delta z_{i} = 0$. This suggests that our fiducial HOD constraints are robust to photo-$z$ uncertainties in the COSMOS 30-band catalog, as parametrized through Eq.~\ref{eq:photo-z-model}. 

\begin{figure}
        \begin{center}
          \includegraphics[width=0.95\textwidth]{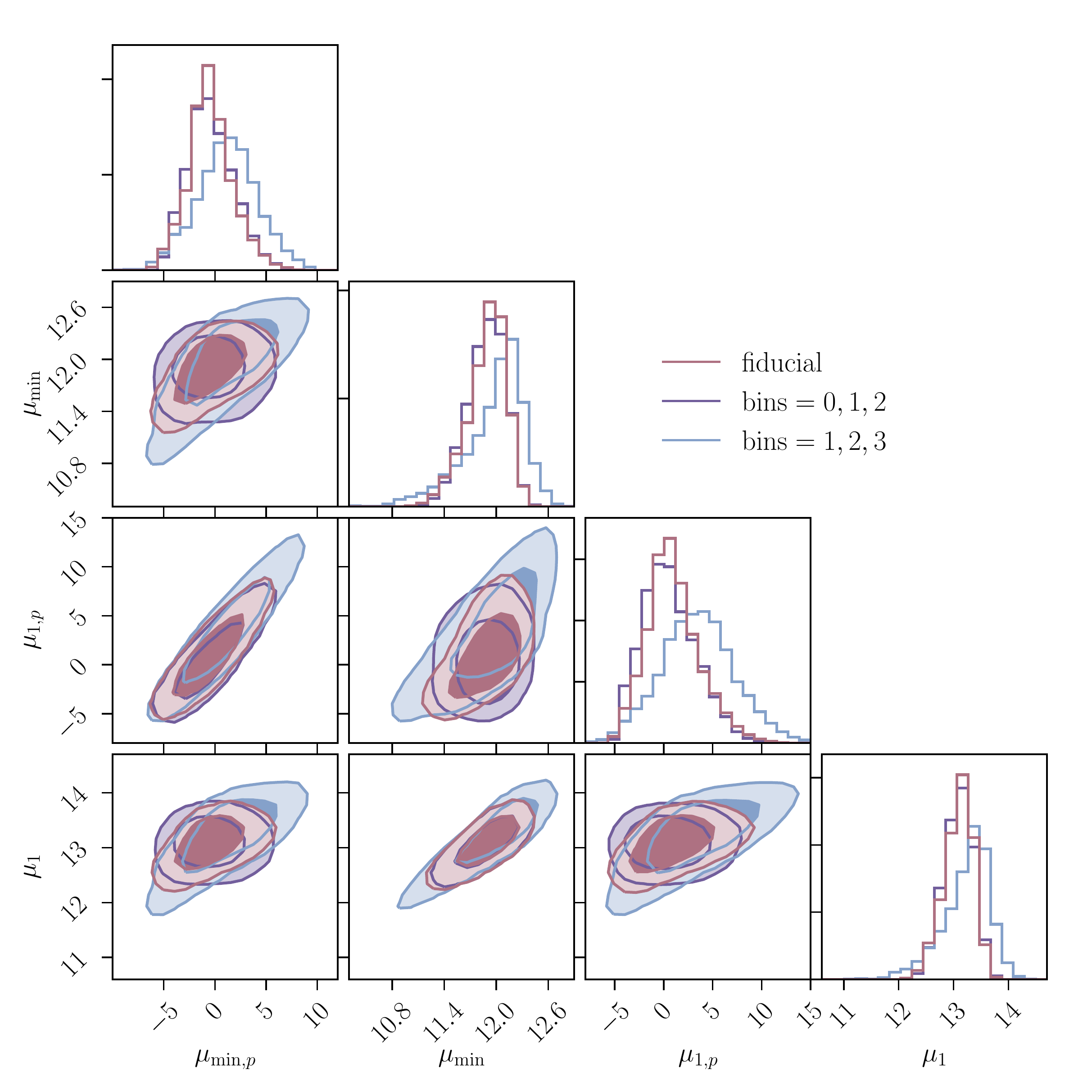}
          \caption{Comparison of our fiducial constraints on HOD parameters to those obtained from the three lowest redshift bins and their cross-correlations and those obtained from the three highest bins and their cross-correlations. The inner (outer) contour shows the $68 \%$ c.l. ($95 \%$ c.l.).}
          \label{fig:constraints-fit-bins=0+1+2+3-vs-fit-bins=0+1+2-vs-fit-bins=1+2+3}
        \end{center}
      \end{figure}

      As described in Sec.~\ref{ssec:methods.nz}, our fiducial constraints use the redshift distributions derived using COSMOS 30-band data \cite{2016ApJS..224...24L}. There are several potential caveats associated with this approach. First, the   photometric redshift accuracy decreases significantly for objects in the COSMOS 30-band catalog with $z > 1.4$ and second, the fraction of catastrophic outliers at faint magnitudes ($23 \leq$ \texttt{i+} $\leq 25$) is estimated to be around $6-10 \%$ \cite{2016ApJS..224...24L}. We expect the highest redshift bin to be mostly affected by decreasing photometric redshift accuracy. However, as noted in \cite{Joudaki:2019}, catastrophic outliers that are erroneously assigned to too low redshifts are more likely to fall into our sample than outliers assigned to too high redshifts. Therefore, the lowest redshift bin in particular might also be affected by photometric redshift errors. In order to test the robustness of our results to photometric redshift uncertainties at low and high redshifts, we compute parameter constraints separately neglecting the high- and low-redshift bin and all their cross-correlations. The results of these two analyses are shown alongside our fiducial constraints in Fig.~\ref{fig:constraints-fit-bins=0+1+2+3-vs-fit-bins=0+1+2-vs-fit-bins=1+2+3}. As can be seen, we find all constraints to agree well with each other. This suggests that our analysis is robust against photometric redshift uncertainties in the COSMOS 30-band catalog affecting the lowest or highest of our redshift bins. 

      We also investigate the impact of a secondary mode in the redshift distribution of the first tomographic bin. This secondary peak contains $\sim$5\% of all galaxies and peaks at around $z\sim 3$ (and hence cannot be seen in Fig.~\ref{fig:nzs}). To investigate the impact of this mode, we manually excise it from the redshift distribution and calculate the resulting change in $\chi^2$, finding it to be $\Delta \chi^2\sim 0.05$. This indicates that the effect of this peak is too small to be statistically significant. We therefore conclude that we do not need to worry whether it is real or an artifact of photometric redshift inference. 

\begin{figure}
        \begin{center}
          \includegraphics[width=0.95\textwidth]{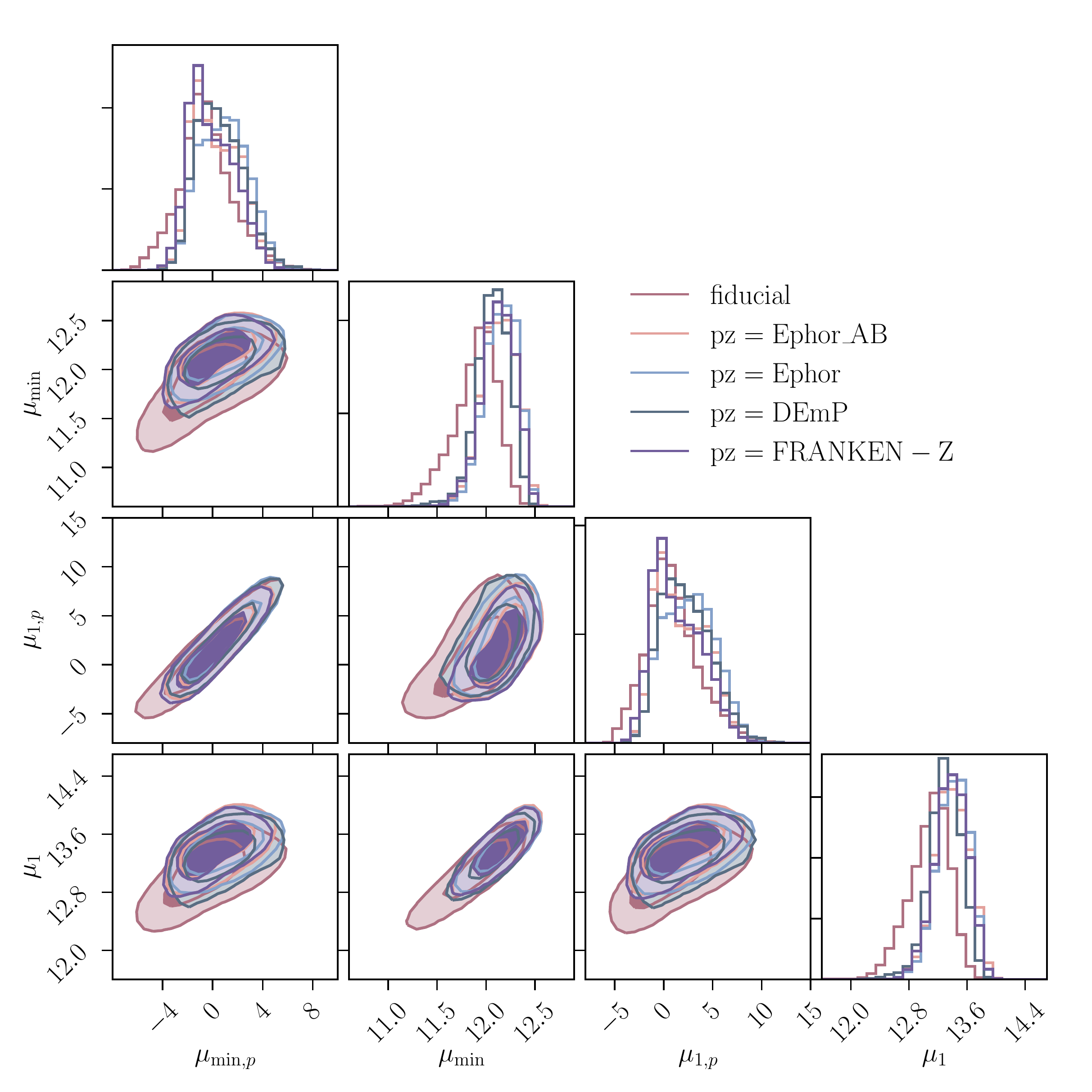}
          \caption{Comparison of our fiducial HOD constraints to those obtained using the redshift distributions derived from \texttt{Ephor\_AB}, \texttt{Ephor}, \texttt{DEmP} and \texttt{FRANKEN-Z} marginalized over both shifts in the means and changes in the widths of the distributions. The inner (outer) contour shows the $68 \%$ c.l. ($95 \%$ c.l.).}
          \label{fig:constraints-HOD-fit-pz-shifts+pz-widths-pz-methods}
        \end{center}
      \end{figure}

\begin{figure}
        \begin{center}
          \includegraphics[width=0.95\textwidth]{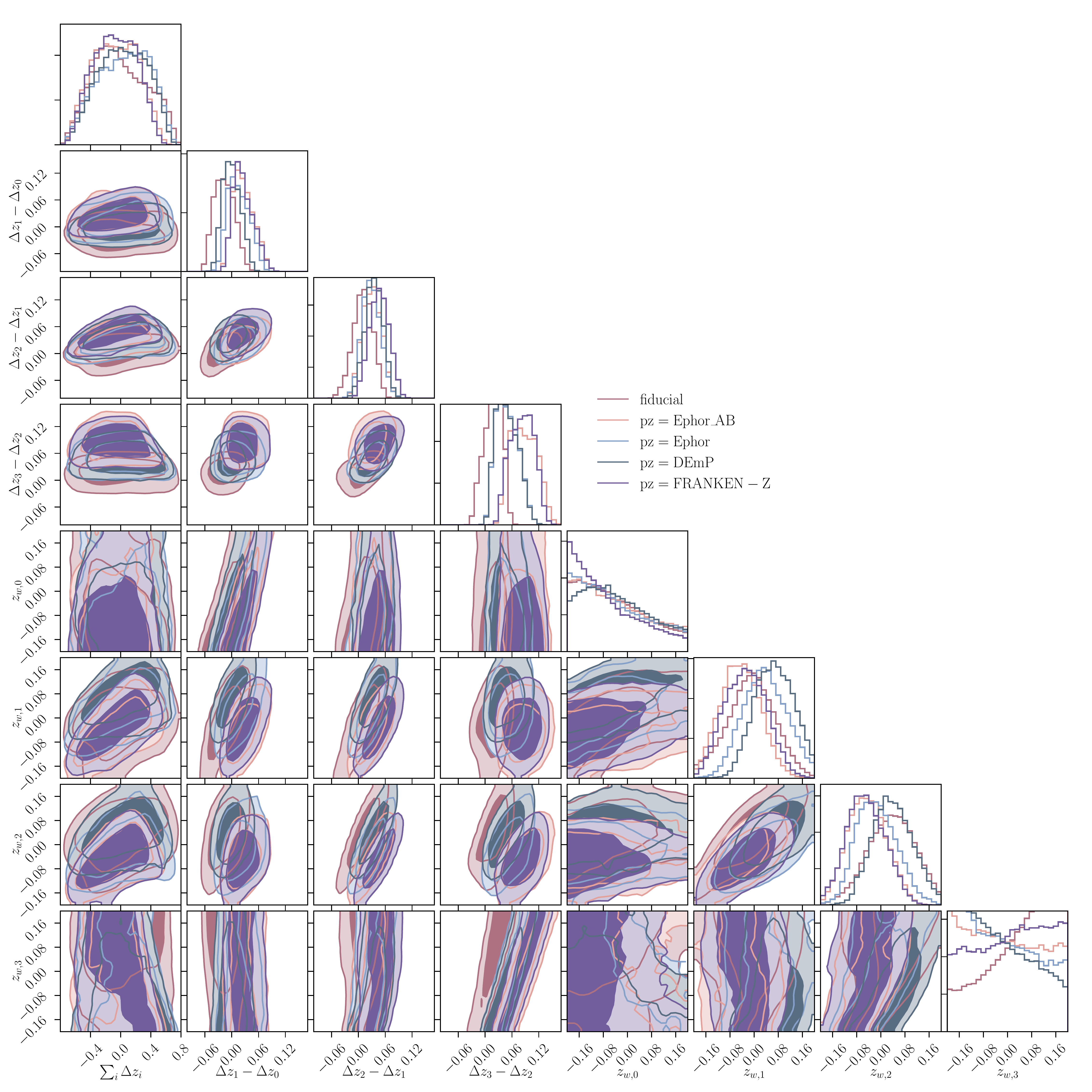}
          \caption{Comparison of our fiducial constraints on photo-$z$ systematics parameters to those obtained using the redshift distributions derived from \texttt{Ephor\_AB}, \texttt{Ephor}, \texttt{DEmP} and \texttt{FRANKEN-Z}. The inner (outer) contour shows the $68 \%$ c.l. ($95 \%$ c.l.).}
          \label{fig:constraints-pz-syst-fit-pz-shifts+pz-widths-pz-methods}
        \end{center}
      \end{figure}

      As a last test of robustness against photometric redshift errors, we compare our fiducial constraints to those obtained using the stacked pdfs from the photo-$z$ codes \texttt{Ephor\_AB}, \texttt{Ephor}, \texttt{DEmP} and \texttt{FRANKEN-Z}, as provided by the HSC Collaboration \cite{2018PASJ...70S...9T}. In analogy to our fiducial analysis, we use the photometric redshift error model given in Eq.~\ref{eq:photo-z-model} for each photo-$z$ code. The comparison of the constraints on HOD and photo-$z$ systematics parameters obtained for the five different methods are shown in Figures \ref{fig:constraints-HOD-fit-pz-shifts+pz-widths-pz-methods} and \ref{fig:constraints-pz-syst-fit-pz-shifts+pz-widths-pz-methods}. As can be seen, the constraints on HOD parameters from all codes are in very good agreement, while we see some discrepancies between the constraints on photometric redshift systematics parameters, especially between $z_{w, 3}$ and $\Delta z_{3} - \Delta z_{2}$. However, these differences are not unexpected as these redshift distributions differ in means and shapes and are therefore not required to give consistent constraints on photo-$z$ systematics parameters. 

      Even though the photo-$z$ error model given in Eq.~\ref{eq:photo-z-model} leads to consistent constraints on HOD parameters from all photo-$z$ methods considered, there is no guarantee that it is flexible enough to capture all redshift distribution differences important for galaxy clustering statistics. We investigate this by comparing the auto-power spectra for a fixed HOD model obtained when forcing the five redshift distributions to have approximately the same means and widths. We find that the resulting power spectra exhibit significant differences, which suggests that features in the redshift distribution beyond mean and width can significantly impact observed galaxy clustering power spectra. In our case, accounting for uncertainties in means and widths gives consistent constraints on HOD parameters and we therefore conclude that this error model is flexible enough to characterize photo-$z$ uncertainties in the present case. However, this will probably cease to be true for future photometric clustering analyses (e.g. using LSST data) and suggests that these data sets will have to be analyzed accounting for the full uncertainty on the shape of the photometric redshift distributions.

      In order to further illustrate the effects of an incomplete photo-$z$ error model, we perform two additional analyses in which we compare the constraints obtained from the five photo-$z$ methods setting $z_{w, i} = 0$ and $z_{w, i} = 0, \Delta z_{i} = 0$ respectively. In the former case we find that the values for the photo-$z$ systematics parameters $\Sigma_{i=0}^{3} \Delta z_{i}, \Delta z_{i} - \Delta z_{j}$ required by the five methods differ significantly, which leads to differences in both constraining power and constraints on HOD parameters. We especially find significant shifts in $\mu_{\mathrm{min}}$ and $\mu_{1}$ along their degeneracy direction. Investigating this further, we find that the observed differences in $\Delta z_{i}$ cannot be explained with the differences in mean redshift between the distributions derived using the five separate methods. In addition, we find that for the redshift distributions considered in this analysis, mean shifts mainly affect low redshift auto-power spectra but have a significantly smaller impact on high redshift bins. Variations in the widths of the photo-$z$ distributions on the other hand, have a similar impact at low and high redshifts. This suggests that a photo-$z$ error model accounting only for mean shifts is too simplistic to account for the redshift distribution differences observed in this analysis. As borne out by the observed discrepancies in HOD constraints, our analysis is sensitive to these differences and we therefore need to extend the photo-$z$ error model. From the discussion above, it follows that additionally accounting for variations in the widths of the distributions is sufficient, as this error model yields HOD constraints that are robust to changes in the photo-$z$ estimation method.

      The discrepancies between HOD parameters derived from the five different methods become even stronger when we do not account for photo-$z$ uncertainties, as can be seen from Fig.~\ref{fig:constraints-HOD-no-pz-shifts-pz-methods}. In contrast to the results obtained using our fiducial COSMOS30 photo-$z$ distributions, we find that we cannot obtain an acceptable fit to the data for the alternative four methods when not accounting for photometric redshift uncertainties. The corresponding reduced $\chi^{2}$s range between $\chi^{2}_{\mathrm{red}} = 1.97$ and $\chi^{2}_{\mathrm{red}} = 5.13$. This suggests that the data are able to detect the presence of significant photometric redshift errors in the stacked photo-$z$ distributions derived using \texttt{Ephor\_AB}, \texttt{Ephor}, \texttt{DEmP} and \texttt{FRANKEN-Z} as opposed to the distributions derived using COSMOS 30-band data.

      The positive side of the enhanced sensitivity of galaxy clustering to photometric redshift uncertainties compared to cosmic shear is the possibility of using clustering measurements to self-calibrate some of these uncertainties. As an example, our analysis hints towards a consistently positive value of the relative shift parameters $\Delta z_i-\Delta z_j,\,i>j$ (see e.g. Figs. \ref{fig:constraints-pz-syst-auto+cross-vs-auto} and \ref{fig:constraints-pz-syst-fit-pz-shifts+pz-widths-pz-methods}), i.e. the different bins are more widely separated than their fiducial distributions assume. Since the amplitude of the cosmic shear signal increases towards higher redshifts, a weak lensing analysis based on the fiducial redshift distributions inferred from the COSMOS 30-band catalog that does not account for these shifts would overestimate the amplitude of density perturbations $\sigma_8$ in order to match the comparatively high weak lensing signal (see \cite{2019arXiv190609262J} for a similar discussion with regards to COSMOS-based redshift distributions). Although the data analysed here are not yet sensitive enough to provide conclusive evidence of these types of systematics, larger samples will be able to provide useful insight on the presence of photometric redshift systematics.
 
    \subsubsection{Interpretation of HOD results}\label{sssec:results.hod-constraints.constraints-interpretation}
      As described in Sec.~\ref{sssec:results.hod-constraints.fiducial}, our best-fit HOD constraints predict both the minimal mass to host a central galaxy $M_{\mathrm{min}}(z)$ and the mass scale for satellites $M_{1}(z)$ to be redshift independent. As noted above, this is somewhat counter-intuitive, as we would expect at least $M_{\mathrm{min}}(z)$ to increase with redshift. In order to understand this trend, it would be ideal to look at the rest-frame properties of the galaxies in our sample and how these change with redshift. However, computing k-corrections (see e.g. \cite{Hogg:2002, Blanton:2007}) and absolute magnitudes for broad-band photometric data can be challenging and we therefore choose an alternative approach. We cross-match galaxies passing our selection criteria within the HSC COSMOS field to galaxies also included in the COSMOS 30-band photometric catalog of \cite{2016ApJS..224...24L}. The COSMOS 30-band catalog contains, amongst others, absolute magnitudes in Subaru \texttt{B}-band, $M_{B}$, and Subaru \texttt{r+}-band, $M_{R}$. Using these quantities, we construct color-magnitude diagrams for all cross-matched galaxies. Fig.~\ref{fig:color-mag} shows the $M_{B}-M_{R}$ color as a function of $M_{R}$ magnitude for 10 equally spaced redshift bins in $z \in [0.15, 1.5]$\footnote{The galaxies are split into bins according to COSMOS 30-band photometric redshifts. We note that we have repeated the analysis splitting the sample according to HSC \texttt{Ephor\_AB} photometric redshifts, finding consistent results.}. In this diagram, red galaxies populate the high $M_{B}-M_{R}$ color, high $M_{R}$ magnitude plane, while blue galaxies tend to have lower $M_{B}-M_{R}$ and $M_{R}$. As can be seen from the figure, red galaxies increasingly drop-out of our sample at high redshift. We can quantify this effect by computing the fraction of red galaxies in our sample as a function of redshift: we empirically determine the separation between the blue and red clouds (denoted by the solid lines in the figure) and compute the fraction of galaxies in the red cloud as $f_{R} = \sfrac{N_{R}}{(N_{R}+N_{B})}$, where $N_{R}$ denotes the number of galaxies above the separation line, while $N_{B}$ is the number of galaxies below the line. Consistent with the figure, we find a decreasing $f_{R}$ as a function of redshift.
      
\begin{figure}
        \begin{center}
          \begin{subfigure}{0.32\textwidth}
          \includegraphics[width=\textwidth]{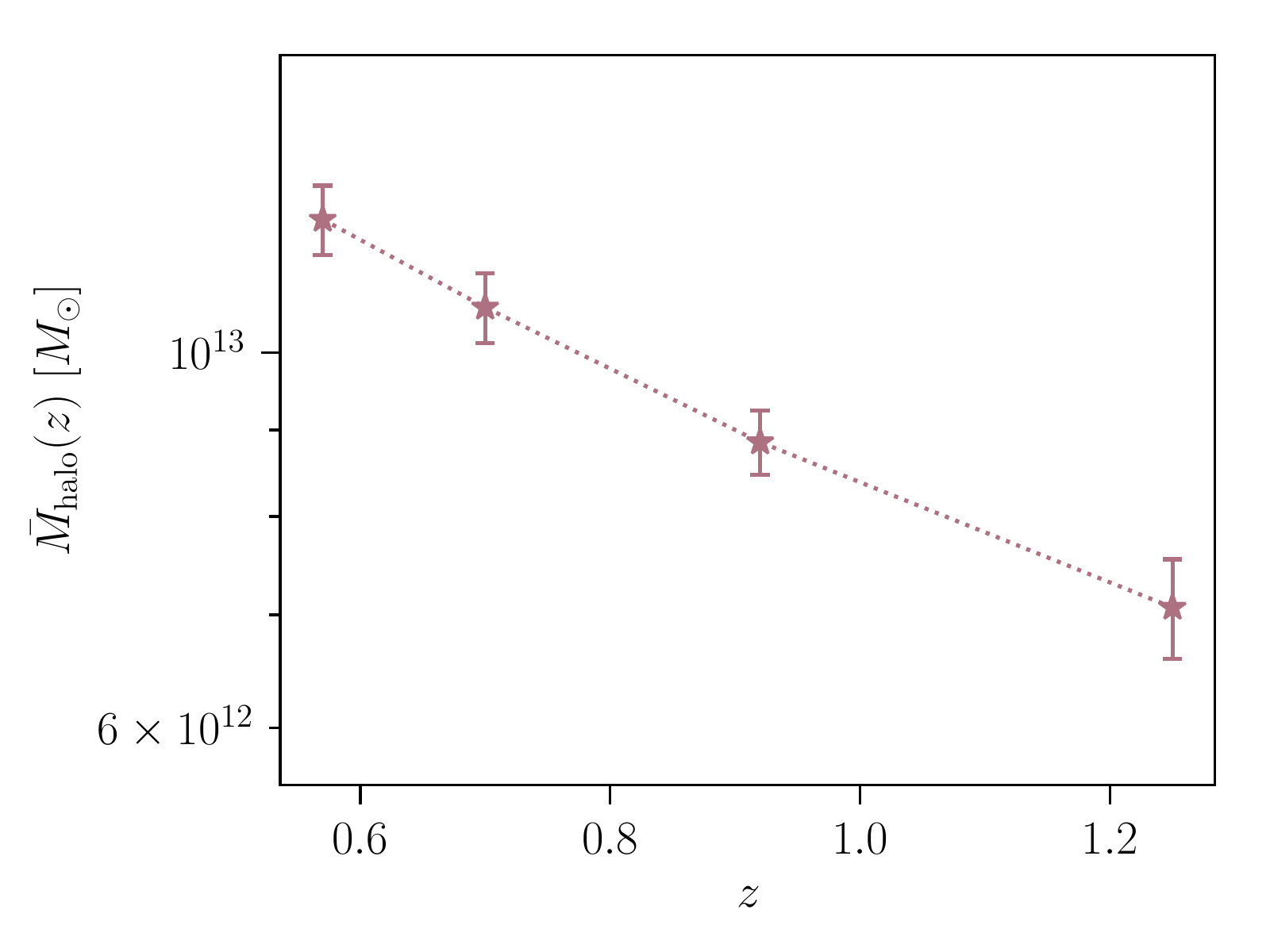}
          \end{subfigure}
          \begin{subfigure}{0.32\textwidth}
          \includegraphics[width=\textwidth]{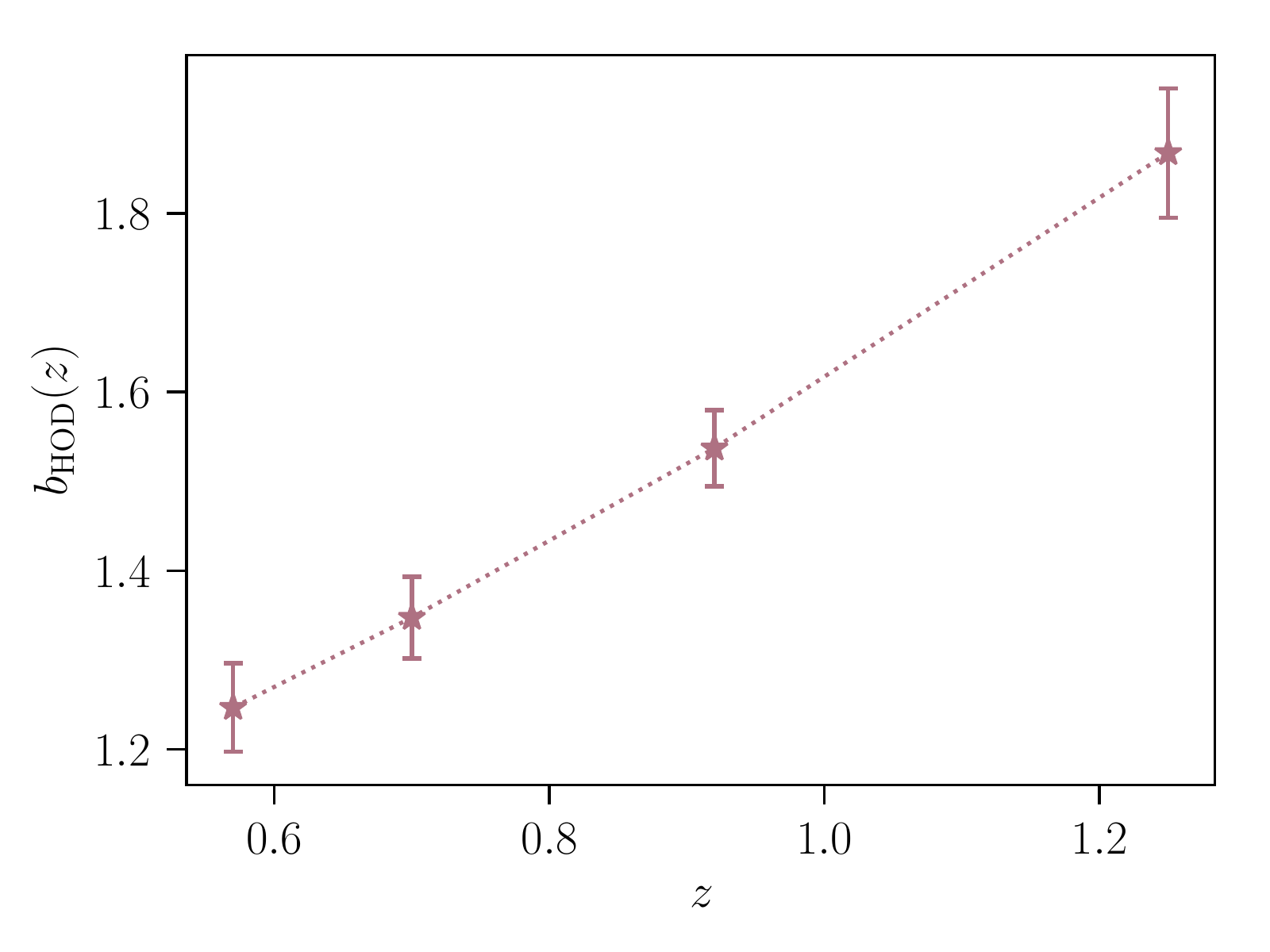}
          \end{subfigure}
          \begin{subfigure}{0.32\textwidth}
          \includegraphics[width=\textwidth]{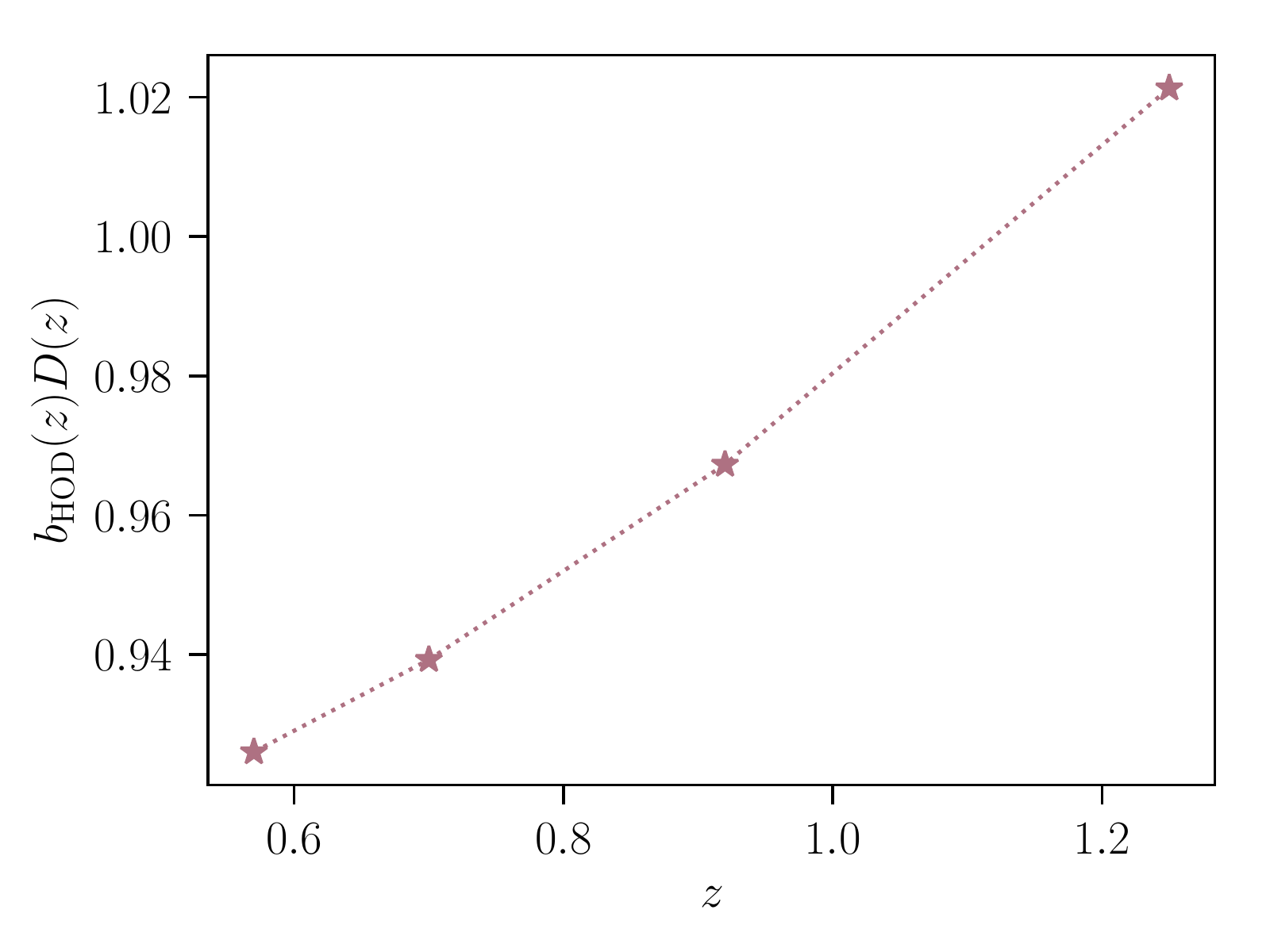}
          \end{subfigure}
          \caption{Redshift dependence of the mean halo mass $\bar{M}_{\mathrm{halo}}$ and the large-scale galaxy bias $b_{\mathrm{HOD}}$ as derived from our best-fit HOD model. The third panel illustrates the empirical scaling $b(z) \propto \sfrac{1}{D(z)}$ and shows that $b(z)$ is consistent with $\sfrac{1}{D(z)}$ to within approximately $3-7 \%$.}
          \label{fig:HOD-redshift}
        \end{center}
      \end{figure}

\begin{sidewaysfigure}
        \begin{center}
          \includegraphics[width=0.95\textwidth]{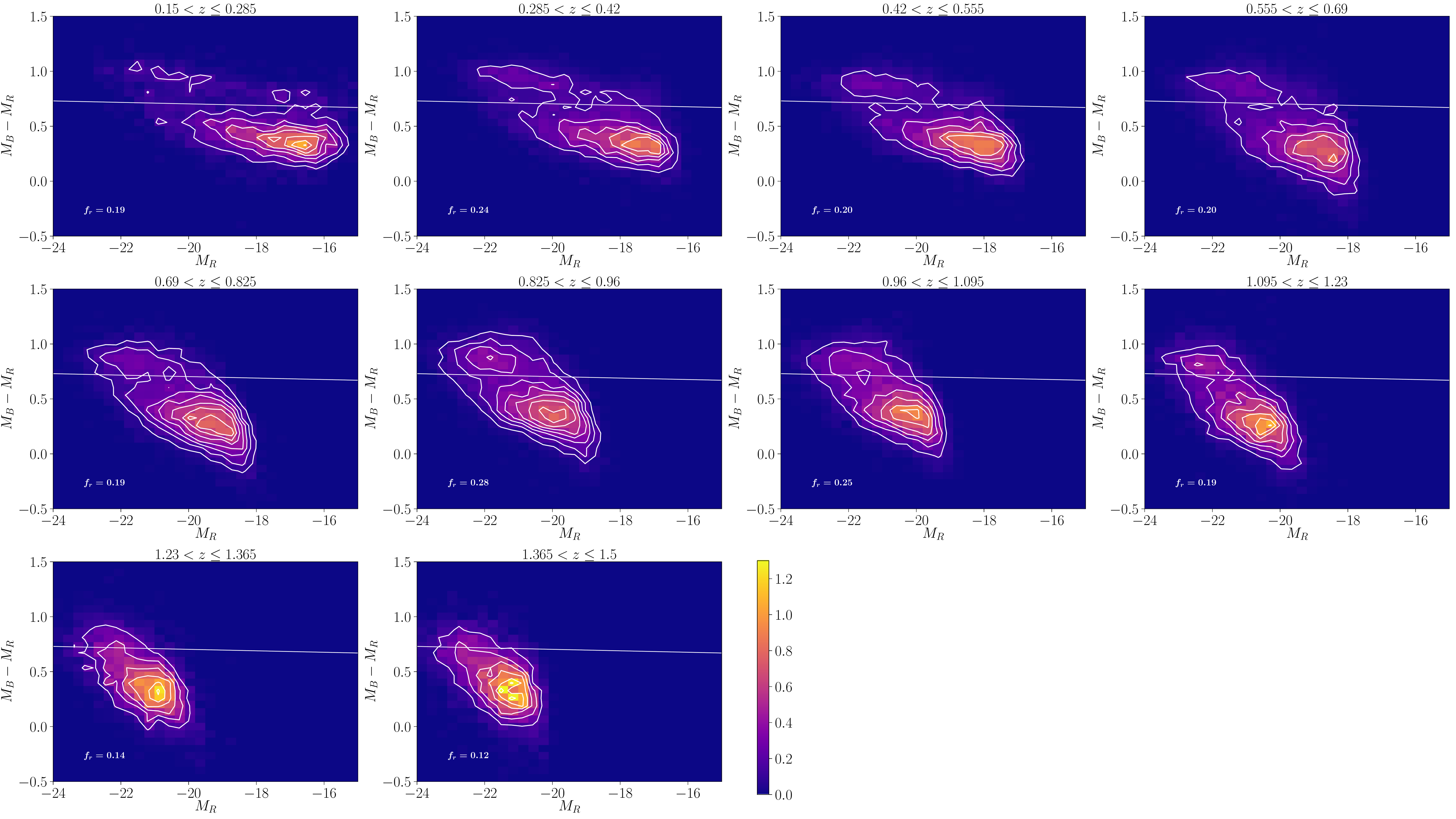}
          \caption{Normalized color-magnitude diagrams of COSMOS 30-band galaxies cross-matched to HSC galaxies passing our selection cuts for 10 tomographic redshift bins with $z \in [0.15, 1.5]$.}
          \label{fig:color-mag}
        \end{center}
      \end{sidewaysfigure}

      These results suggest that red galaxies are increasingly underrepresented with respect to blue galaxies in our sample at higher redshifts. This is due to the applied magnitude limit of $i < 24.5$: the $i$-band filter translates to rest-frame wavelengths smaller than $4000$ \AA \, at redshifts $z \gtrsim 0.9$. Red galaxies with a strong $4000$ \AA \, break therefore become very faint at these wavelengths and drop out of our selection, thus explaining the observed constancy of $M_{\mathrm{min}}(z)$ and $M_{1}(z)$ with redshift. This interpretation is consistent with the results from the HOD fits that predict a decreasing mean halo mass as a function of redshift for our sample (see Fig.~\ref{fig:HOD-redshift}).

      In Fig.~\ref{fig:HOD-redshift}, we additionally show the large-scale galaxy bias as a function of redshift derived from our best-fit HOD model. As can be seen, the large-scale bias scales approximately as $b(z) \propto \sfrac{1}{D(z)}$ (to within approximately $5 \%$), where $D(z)$ denotes the linear growth factor. This behavior is consistent with e.g. studies of galaxies in the DEEP2 survey, which found constant clustering strengths as a function of redshift (and thus $b(z) \propto \sfrac{1}{D(z)}$) for magnitude-limited samples \cite{Coil:2004}.
      
Finally we note that in our analysis, we assume the observed fraction of central galaxies $f_{c}$ to be unity and constant with redshift. In light of the observed drop-out of red galaxies from our sample at high redshift, this may not be true, as $f_{c}$ could potentially decrease with redshift. In order to test our assumption, we perform separate HOD fits to the four redshift bins, allowing for variations in $f_{c}$ on top of $\mu_{\mathrm{min}}, \mu_{1}$ and $\mu_{0}$\footnote{As we fit each redshift bin separately, we do not account for any redshift-dependence in the HOD parameters.}. We find the data to not have the statistical power to constrain $f_{c}$ and the derived values to be consistent with redshift-independent. Furthermore, our constraints on $\mu_{\mathrm{min}}, \mu_{1}$ and $\mu_{0}$ are robust to variations in $f_{c}$. We therefore conclude that our results are not significantly affected by the choice of $f_{c}=1$.  

  \subsection{Magnification}\label{ssec:results.magnification}
    In addition to intrinsic correlations between galaxy positions, there exist several additional effects impacting the observed clustering of galaxies. The most relevant one for projected clustering studies is the magnification of distant sources due to gravitational lensing by the intervening LSS, an effect usually called magnification bias (e.g. \cite{Schneider:1989, Narayan:1989}). The main effect of magnification on the statistical distribution of galaxies is to increase observed galaxy fluxes, thus allowing dim galaxies to pass survey selection thresholds, and to alter their observed angular positions. The combined effect is a position-dependent modulation of the number density that distorts the clustering pattern. It has been shown that the clustering signal due to magnification can be comparable to the intrinsic clustering signal for cross-correlations between widely-separated redshift bins (see e.g. \cite{Challinor:2011}).

    In the presence of magnification, the equations presented in Section \ref{ssec:methods.theory} acquire additional terms. The projected galaxy overdensity becomes:
    \begin{equation}
      \delta^i_g(\nv)=\int \mathrm{d}z\,\left[p^i(z)\,\Delta_g+\frac{W_\mu^i(z)}{H(z)}\nabla_\theta^2\nabla^{-2}\Delta_m\right],
    \end{equation}
    where $\Delta_m$ is the 3D matter overdensity (we have omitted the dependence on $t$ and $\chi\nv$ for brevity) and $W_\mu$ is the magnification kernel
    \begin{equation}
      W_\mu(z)=\frac{3H_0^2\Omega_{m}(1+z)}{2}\int_z^\infty \mathrm{d}z'\,p^i(z)\,\left(5s(z',m_{\rm lim})-2\right)\,\frac{\chi(z')-\chi(z)}{\chi(z)\chi(z')}.
    \end{equation}
    Here, $H_0\equiv H(z=0)$, $\Omega_{m}$ is the fractional matter density today, and $s$ is the logarithmic slope of the cumulative apparent magnitude distribution $N(<m, z)$:
    \begin{equation}
      \left. s(z,m_{\rm lim}) \equiv \frac{\partial \log_{10}N(<m, z)}{\partial m}\right \rvert_{m=m_{\mathrm{lim}}},
      \label{eq:s-func}
    \end{equation}
    where $m$ denotes the observed magnitude and $m_{\mathrm{lim}}$ is the magnitude limit of the galaxy sample considered\footnote{In the following, we omit the dependence of $s(z)$ on $m_{\mathrm{lim}}$ for brevity.}. In short, the observed density contrast does not depend solely on the distribution of galaxies in the tomographic redshift bin of interest, but also on the distribution of matter along the line of sight to these galaxies. Overdensities along the line of sight cause the local angular separation between galaxies to increase (lowering the observed number density), while lensing magnification causes fainter galaxies to enter the observed sample (increasing the observed number density). The two effects cancel at leading order when $s=\sfrac{2}{5}$. 

    The angular power spectrum is then given by
    \begin{equation}\label{eq:cell_gg_wmag}
      C^{ij}_\ell = \int \mathrm{d}z\,\frac{H(z)}{\chi^2(z)}\left[p^ip^j\,P_{gg}+\left(p^iW_\mu^j+p^jW_\mu^i\right)\frac{\ell(\ell+1)}{Hk^2}P_{gm}+W_\mu^iW_\mu^j\left(\frac{\ell(\ell+1)}{Hk^2}\right)^2P_{mm}\right],
    \end{equation}
    where $P_{mm}$ is the 3D power spectrum of matter fluctuations, $P_{gm}$ is the galaxy-matter cross-spectrum and we have omitted the dependence of all quantities on $z$ or $k=\sfrac{(\ell+\sfrac{1}{2})}{\chi}$.

\begin{figure}
      \begin{center}
        \begin{subfigure}[t]{0.47\textwidth}
        \centering
        \includegraphics[width=\textwidth]{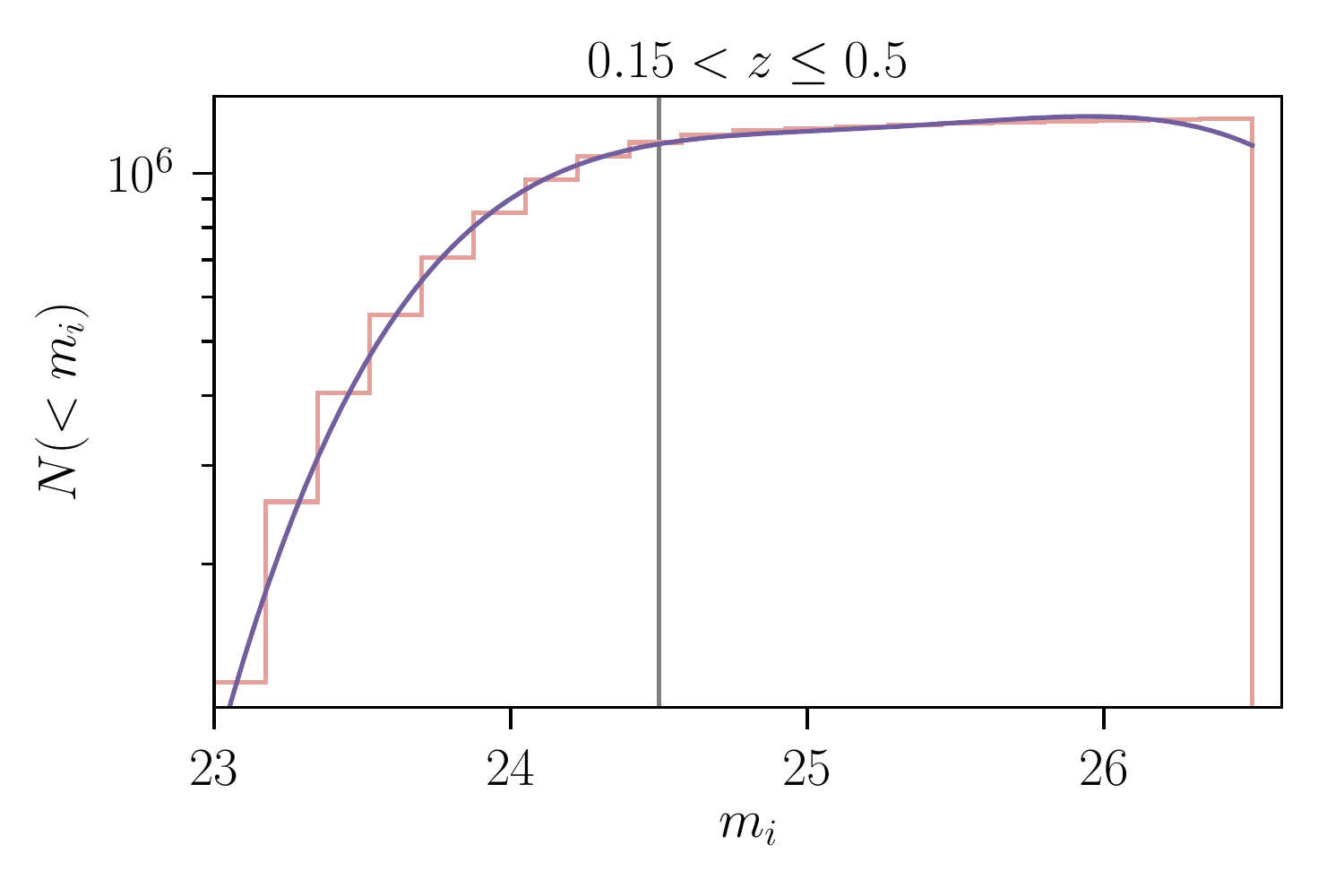}
        \caption{Cumulative galaxy number counts as a function of $i$-band magnitude for the lowest tomographic redshift bin used in this analysis.}
        \end{subfigure}\hfill
        \begin{subfigure}[t]{0.47\textwidth}
        \centering
        \includegraphics[width=\textwidth]{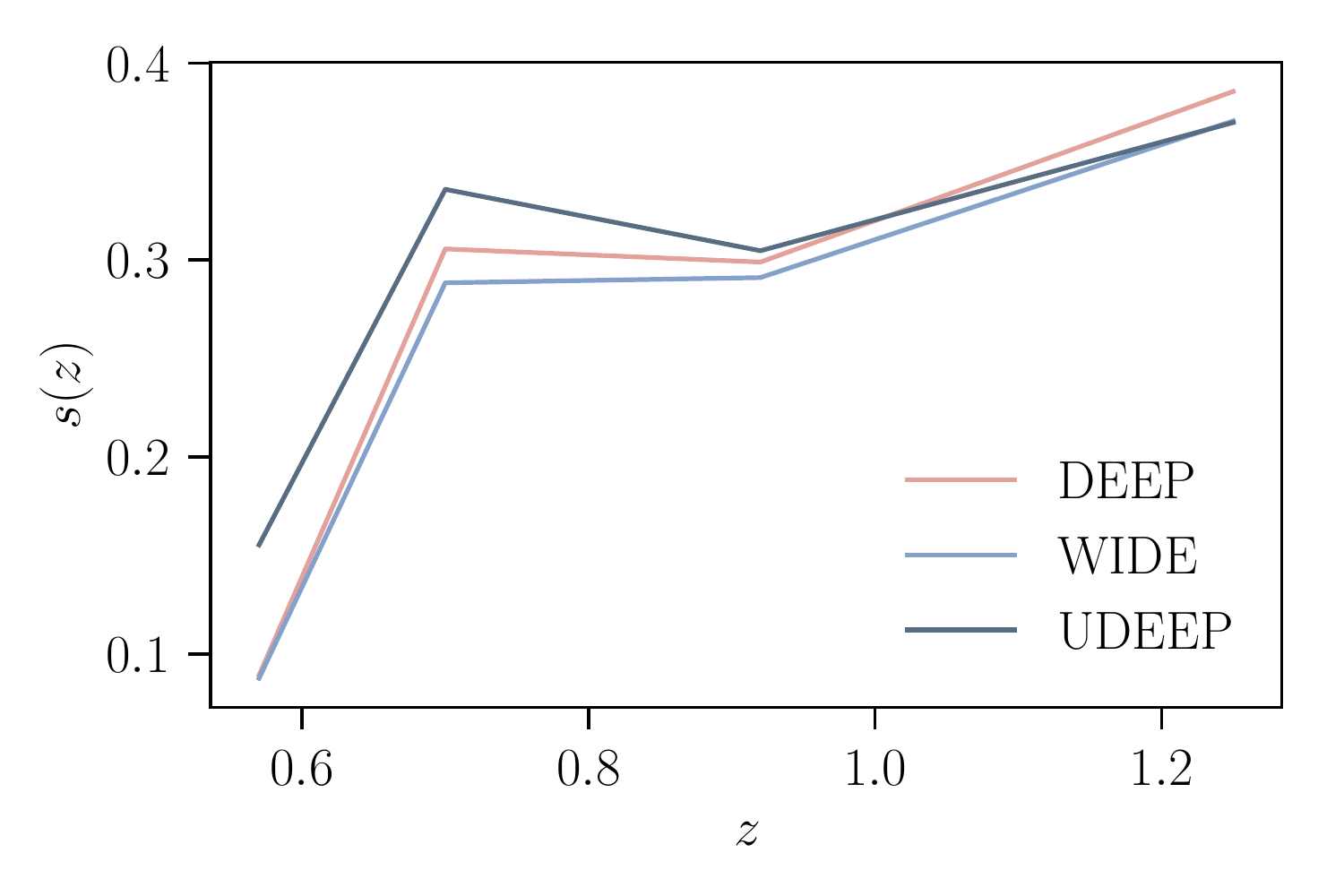}
        \caption{Comparison of $s(z)$ (see Eq.~\ref{eq:s-func}) derived using the WIDE, DEEP and UDEEP HSC galaxy samples.}
        \end{subfigure}
        \caption{Illustration of data-driven estimation of $s(z)$ for the galaxy sample used in our analysis.} 
        \label{fig:s-func-estimation}
      \end{center}
    \end{figure}

    The magnitude of the magnification signal crucially depends on the derivative of the cumulative galaxy number counts with respect to observed magnitude, $s(z)$. In order to obtain a data-driven model for magnification in our particular sample, we therefore estimate $s(z)$ from the observed number counts. Using linear least squares, we fit a fourth-order polynomial to the logarithm of the observed cumulative apparent $\tt{i}$-band magnitude distribution $N(<m, z)$ in each of the four tomographic redshift bins considered in our analysis. We then determine $s(z)$ at the effective redshift of each bin\footnote{We define the effective redshift for each tomographic bin as the mean redshift of the galaxy distribution.} by taking the derivative of the best-fit function at the magnitude limit of our sample, i.e. $m_{i, \mathrm{lim}} = 24.5$. In addition to the WIDE HSC galaxy data used in this work, HSC DR1 also contains data from two deep patches, called DEEP and UDEEP. We repeat the above analysis with these galaxy samples to ensure the stability of our results. As an example, Fig.~\ref{fig:s-func-estimation} shows the observed number counts as a function of $i$-band magnitude for the lowest redshift bin in the WIDE sample alongside the derived best-fit function. In Fig.~\ref{fig:s-func-estimation}, we also show the $s(z)$ functions derived as outlined above for HSC WIDE, DEEP and UDEEP. As can be seen, the three estimates agree well with each other and we use the $s(z)$ function derived from our fiducial WIDE sample in the following. Our uncertainty on the estimation of $s(z)$ is subdominant given the statistical precision of our measurement, and we therefore ignore it.

    Using the estimated $s(z)$, we can derive theoretical predictions for the angular galaxy power spectra in presence of magnification. In order to investigate the sensitivity of the data to magnification, we run two different MCMC analyses: in the first case, we fix the magnification amplitude to the predictions from the empirically determined $s(z)$ function. In the second case, we include an additional magnification parameter $A_{\mu}$ that scales the amplitude of the magnification kernel $W_{\mu}$ and we fit it alongside our fiducial parameter set.

\begin{figure}
      \begin{center}
        \includegraphics[width=0.3\textwidth]{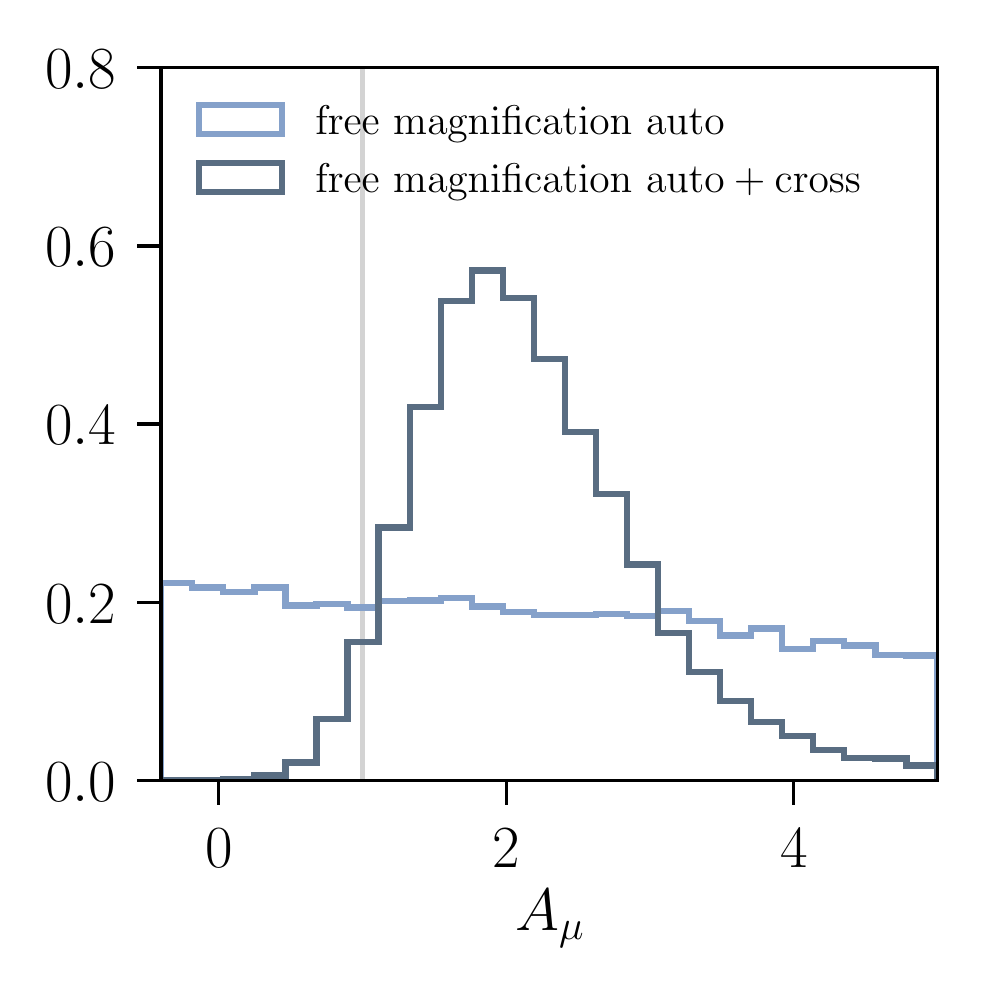}
        \caption{Comparison of the constraints on magnification amplitude obtained from auto-power spectra alone to those obtained from auto- and cross-power spectra. The solid line denotes the expectation value of the magnification amplitude determined from our data-driven fiducial model, described in Sec.~\ref{ssec:results.magnification}.}
        \label{fig:constraints-fit=mag-bias-ampl_fit=auto+cross-vs-fit=mag-bias-ampl_fit=auto}
      \end{center}
    \end{figure}

\begin{figure}
      \begin{center}
        \includegraphics[width=0.95\textwidth]{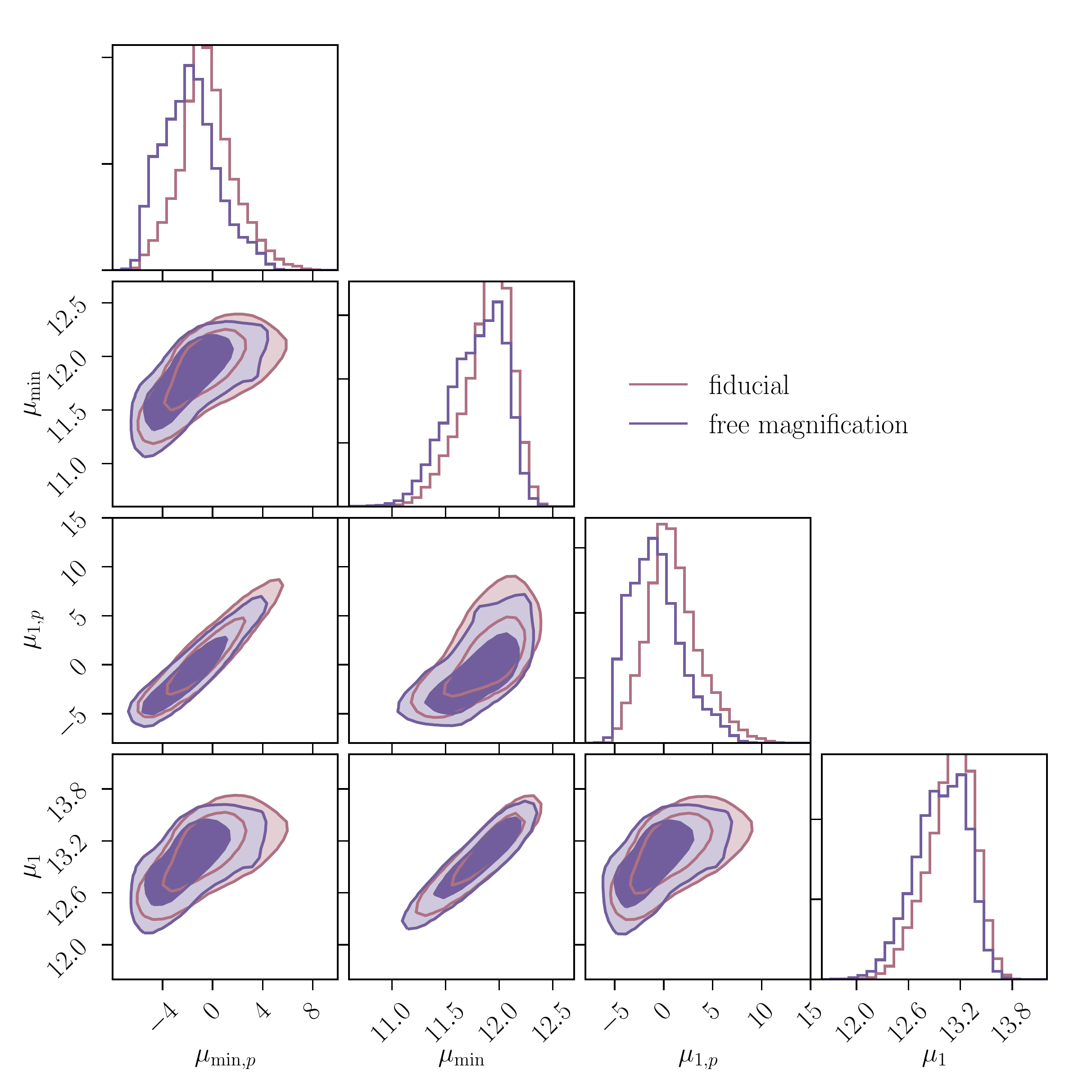}
        \caption{Comparison of our fiducial HOD constraints to those obtained when varying the magnification amplitude. The inner (outer) contour shows the $68 \%$ c.l. ($95 \%$ c.l.).}
        \label{fig:constraints-no-mag-bias-vs-fit=mag-bias-ampl}
      \end{center}
    \end{figure}

    From a joint fit to all auto- and cross-power spectra we obtain $A_{\mu} = 2.18 \pm 0.74$ and the constraint is shown in Fig.~\ref{fig:constraints-fit=mag-bias-ampl_fit=auto+cross-vs-fit=mag-bias-ampl_fit=auto}. This constitutes a $\sim 2.9 \sigma$ detection of magnification from our HSC sample. Furthermore, this constraint is consistent with our fiducial model for $s(z)$, which corresponds to $A_{\mu} = 1$, at the $\sim 1.6 \sigma$ level. This marginal inconsistency is probably due to uncertainties in the data-driven estimation of $s(z)$. The best-fit $\chi^{2}$ including magnification is $\chi^{2} = 69.0$, which, computing the degrees of freedom as $\nu = N_{\mathrm{data}} - N_{\mathrm{param}} = 94 - 15 = 79$, leads to $\chi^{2}_{\mathrm{red}} = \sfrac{\chi^{2}}{\nu} = 0.87$ ($p$-value $= 0.78$). This corresponds to an improvement in $\chi^{2}$ compared to the fiducial analysis of $\Delta \chi^{2} = 86.2 - 69.0 = 17.2$, roughly consistent with the $\sim 3 \sigma$ detection. Fig.~\ref{fig:cls-best-fit} shows the corresponding theoretical predictions derived from the maximum likelihood parameters alongside the observed data. The constraints on HOD parameters obtained when including magnification agree very well with our fiducial constraints, as can be seen from Fig.~\ref{fig:constraints-no-mag-bias-vs-fit=mag-bias-ampl}. In addition, we find the significance of the magnification detection to be largely insensitive to photo-$z$ systematics modeling choices, as it is not significantly affected by relaxing our photo-$z$ systematics model by fixing $z_{w, i} = 0$. 

    These results are consistent with those obtained when we fix $A_{\mu} = 1$, as can be seen from Tab.~\ref{tab:constraints_robustness}. In the latter case, we obtain a best-fit $\chi^{2}$ of $\chi^{2} = 72.8$, which corresponds to $\chi^{2}_{\mathrm{red}} = \sfrac{\chi^{2}}{\nu} = 0.91$ ($p$-value $= 0.70$) for $\nu = N_{\mathrm{data}} - N_{\mathrm{param}} = 94 - 14 = 80$. The improvement in $\chi^{2}$ with respect to the fiducial analysis amounts to $\Delta \chi^{2} = 86.2 - 72.8 = 13.4$, which is lower than the value obtained with free magnification amplitude, but nevertheless constitutes a sizable $\chi^{2}$ improvement.

    In order to identify the part of the data vector driving the constraints on $A_{\mu}$, we repeat this analysis only including the auto-spectra. The comparison between the constraints on $A_{\mu}$ from auto- and cross-power spectra to those obtained from auto-spectra alone is shown in Fig.~\ref{fig:constraints-fit=mag-bias-ampl_fit=auto+cross-vs-fit=mag-bias-ampl_fit=auto}. As can be seen, the auto-power spectra alone do not constrain $A_{\mu}$, which implies that the constraints on the magnification amplitude are solely driven by the cross-power spectra. This sensitivity of the cross-power spectra to magnification mainly stems from the fact that the cross-correlations between widely separated bins caused by intrinsic clustering are small, which leads to an increased sensitivity to any non-intrinsic source of cross-correlation, such as magnification. In analogy to the results for the photo-$z$ systematics parameters, we thus find that cross-power spectra carry important information and it will thus be beneficial to include these cross-correlations in current and future photometric clustering analyses\footnote{We note that we do not find the constraints on $A_{\mu}$ to be strongly degenerate with photo-$z$ systematics parameters. In fact, the detection significance of magnification is almost unchanged if we include shift and width parameters in our analysis as opposed to only accounting for shifts.}. 

  \subsection{Joint cosmology and HOD constraints}\label{ssec:results.cosmo}
    Additionally we also consider jointly fitting cosmological, HOD and photo-$z$ systematics parameters. This analysis should be regarded as exploratory and only provide a means to illustrate the potential of photometric galaxy clustering analyses to jointly constrain these parameters. An in-depth cosmological analysis would require validation of the analysis and HOD modeling framework on simulations, which is beyond the scope of this work.

    To illustrate the cosmological constraining power of the data considered in our analysis, we allow for variations in the r.m.s. of linear matter fluctuations in spheres of comoving radius 8 $h^{-1}$ Mpc, $\sigma_{8}$, and the fractional cold matter density today, $\Omega_{c}$, in addition to our baseline set of parameters (c.f. Sec.~\ref{ssec:methods.constr}). All other cosmological parameters remain fixed to their fiducial Planck 2018 values. From a joint fit to all auto- and cross-power spectra we obtain $\sigma_{8} = 0.807\substack{+0.149 \\ -0.143}$ and $\Omega_{m} = 0.286 \pm 0.025$\footnote{We have recast our constraints on $\Omega_{c}$ in terms of $\Omega_{m}$ for ease of comparison with other analyses.}, consistent with our assumed fiducial cosmology.  In particular, this is consistent with the constraints obtained from combining Planck 2018 and BAO measurements, which give $\Omega_m=0.3110\pm 0.006$ and $\sigma_8=0.8099 \pm 0.0071$ \cite{Planck:2018} albeit with a different parameterization. The weak lensing analysis of the full first year HSC data (similar in size, but somewhat larger than the public release) gives $\Omega_m=0.346^{+0.052}_{-0.100}$ \cite{1906.06041}, showing an excellent internal consistency. We note that our analysis makes considerably stronger modeling assumptions and thus gives error bars that are likely under-estimated. In order to illustrate this, we perform an additional analysis additionally allowing for variations in $n_{s}$, which, like $\Omega_{m}$, affects the shape of the matter power spectrum. The uncertainty on $\Omega_{m}$ in this case increases significantly to $\sigma(\Omega_{m}) \sim 0.15$, thus showing the impact of our strong modeling assumptions.

\begin{figure}
      \begin{center}
        \includegraphics[width=0.49\textwidth]{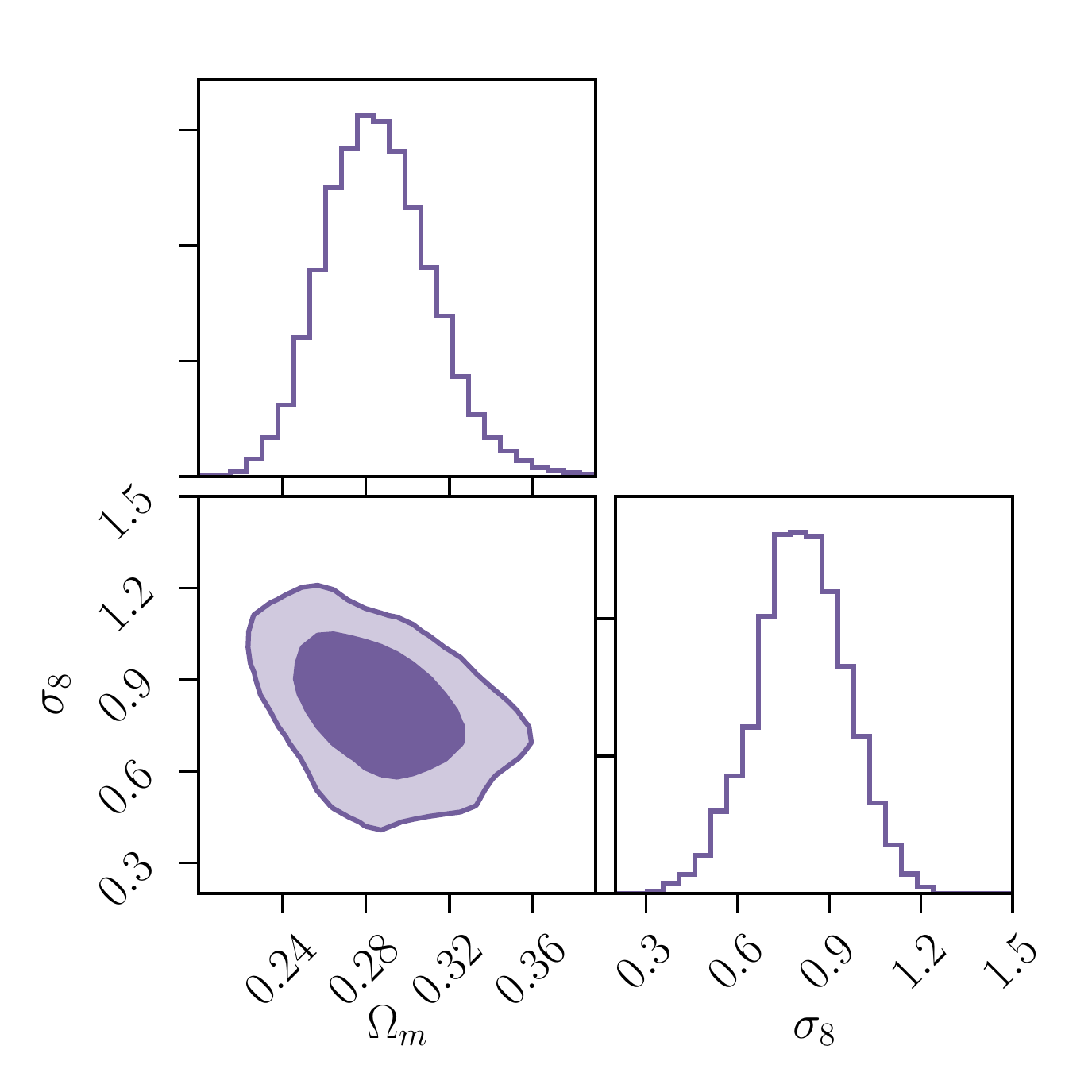}
        \caption{Constraints on $\Omega_{m}$ and $\sigma_{8}$ obtained from a joint fit to all auto- and cross-power spectra. The inner (outer) contour shows the $68 \%$ c.l. ($95 \%$ c.l.).}
        \label{fig:constraints-fit=Oc+s8}
      \end{center}
    \end{figure}

    The constraints obtained when fixing $n_{s}$ are shown in Fig.~\ref{fig:constraints-fit=Oc+s8}. As can be seen from Tab.~\ref{tab:constraints_robustness}, the derived constraints on HOD parameters agree well with our fiducial results, which shows that our HOD constraints are robust to variations in $\Omega_{c}$ and $\sigma_{8}$ as allowed by the data. This exploratory cosmological analysis with strong modeling assumptions further shows that the data are able to constrain $\Omega_{c}$ (and hence $\Omega_{m}$) while $\sigma_{8}$ is largely unconstrained, showing that the halo model is sufficiently flexible to allow effective marginalization over linear galaxy bias. Therefore, current and future photometric galaxy clustering analyses combined with external data constitute a promising way to jointly constrain HOD and cosmological parameters, which we leave to future work. 
 
  \subsection{Constraints on linear galaxy bias for different magnitude-limited samples}\label{ssec:results.mag_cuts}
    Finally we use our analysis pipeline described in Sec.~\ref{sec:methods} to investigate the dependence of galaxy clustering on limiting magnitude $m_{\mathrm{lim}}$ of the sample and derive an approximate fitting function for the linear galaxy bias as a function of redshift and $m_{\mathrm{lim}}$. We only consider magnitude cuts in the $i$-band and we do not investigate the dependence of our results on the photometric band used to select a magnitude-limited sample.
    
\begin{figure}
      \begin{center}
        \begin{subfigure}{0.48\textwidth}
        \includegraphics[width=0.9\textwidth]{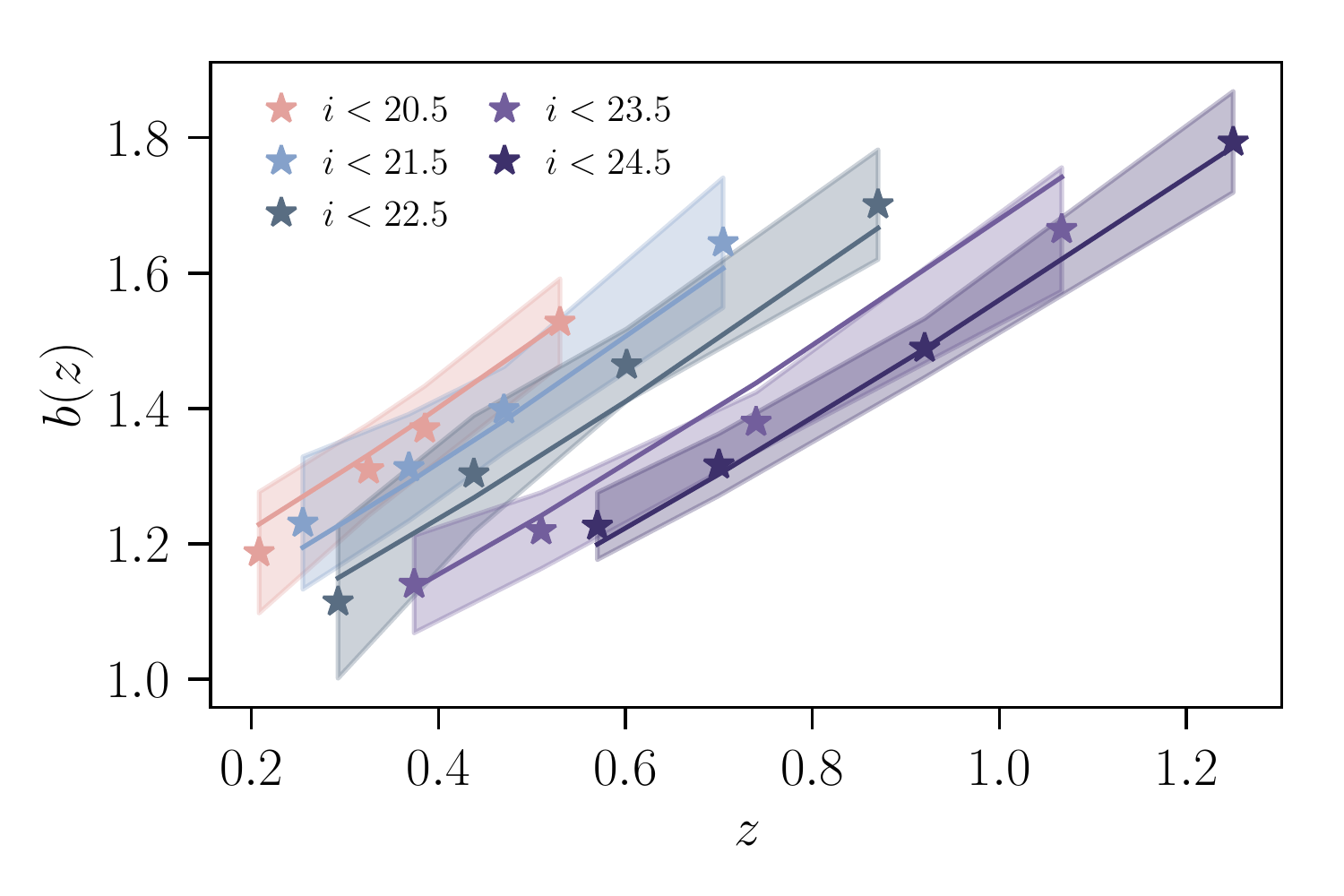}
        \end{subfigure}
        \begin{subfigure}{0.48\textwidth}
        \includegraphics[width=0.9\textwidth]{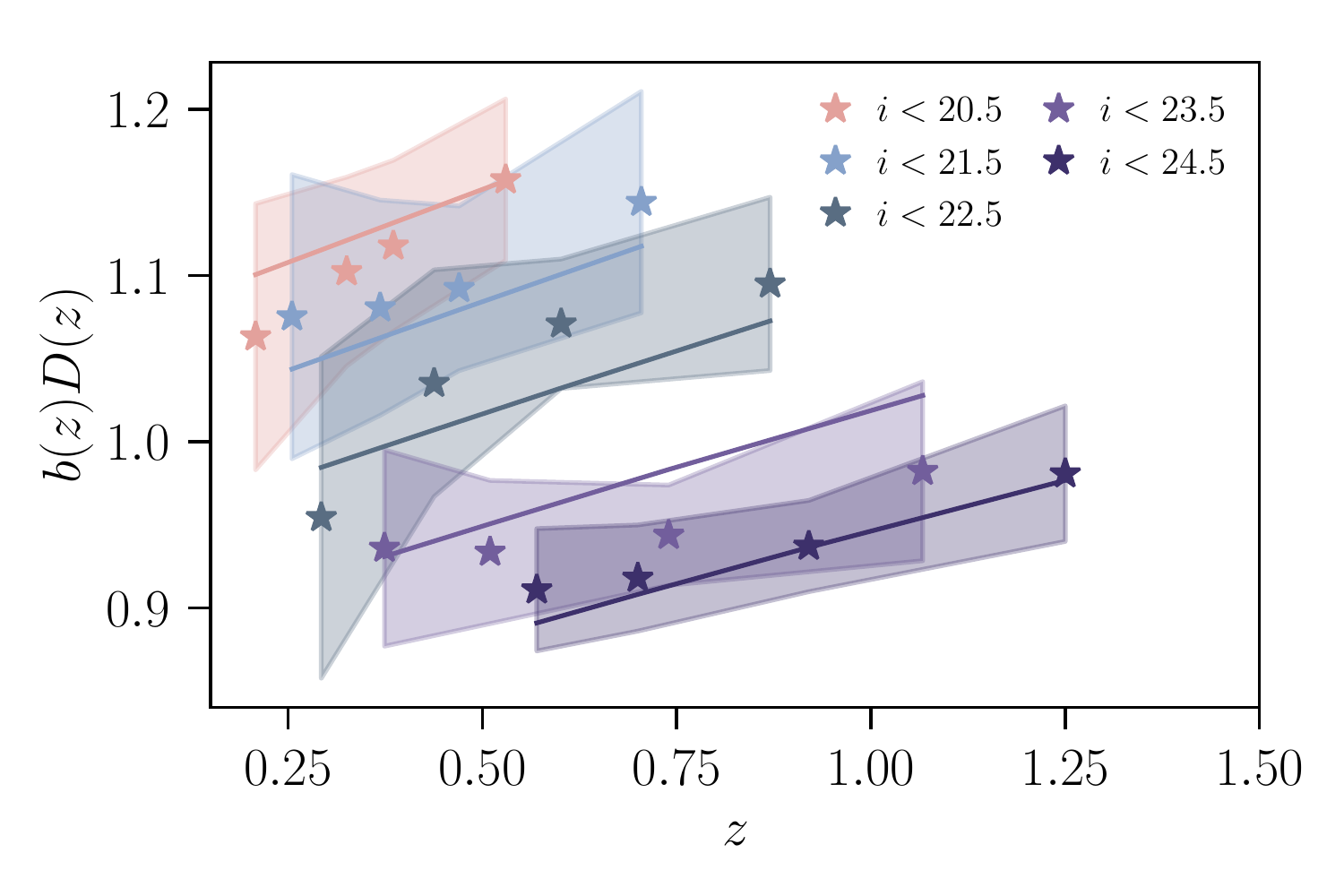}
        \end{subfigure}
       \begin{subfigure}{0.48\textwidth}
       \includegraphics[width=0.9\textwidth]{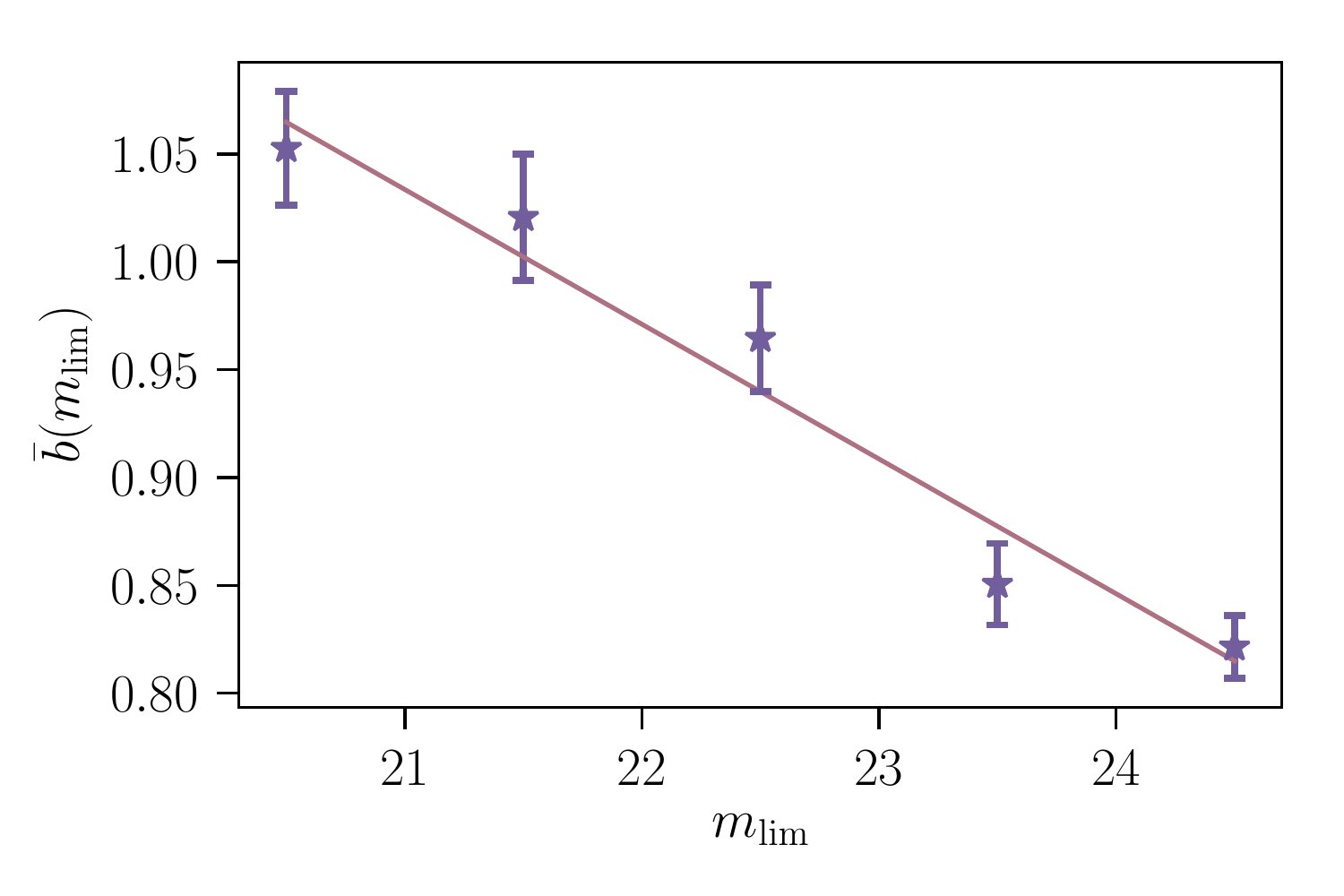}
       \end{subfigure}
          \begin{subfigure}{0.48\textwidth}
          \includegraphics[width=0.9\textwidth]{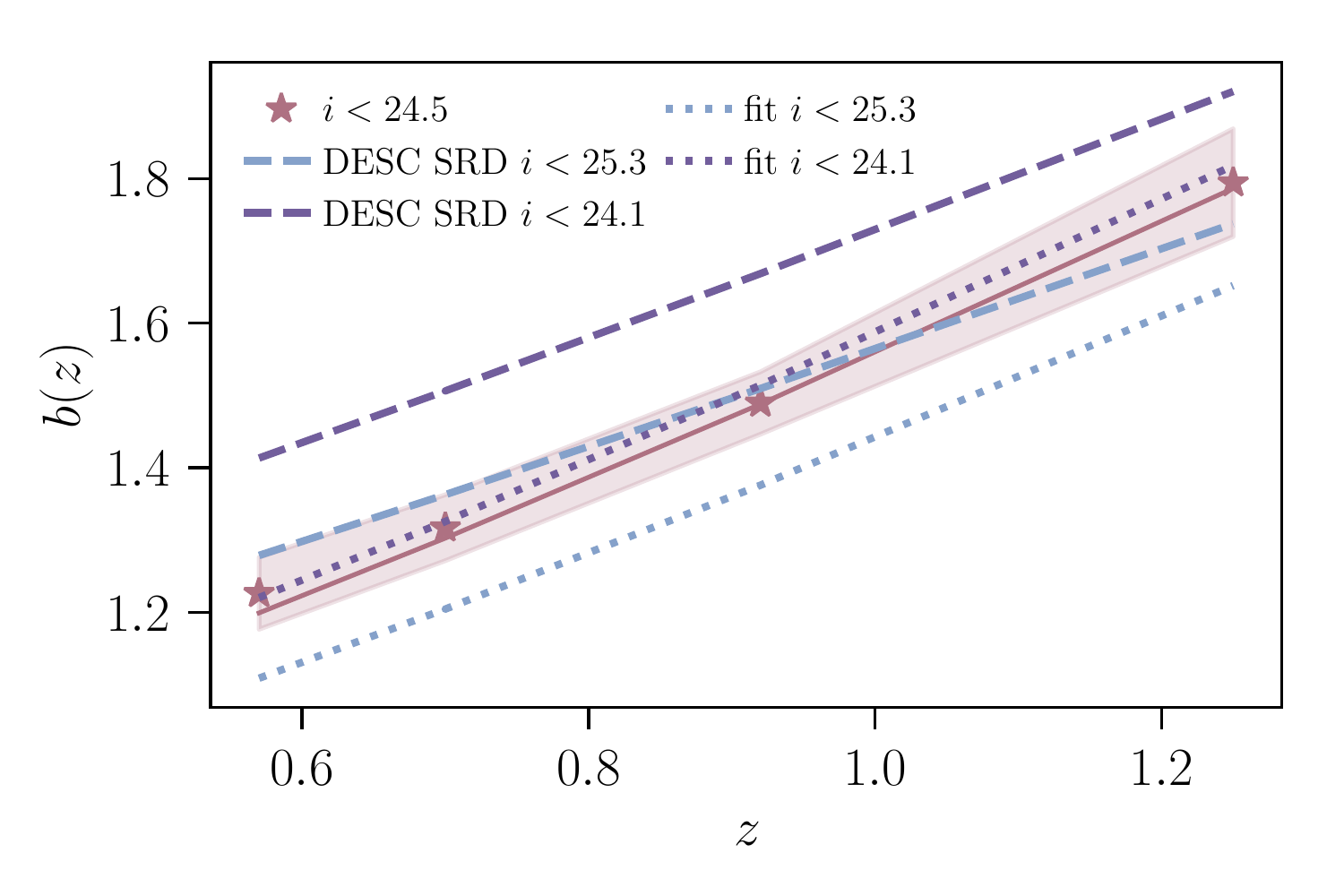}
           \end{subfigure}
        \caption{\textit{Upper panels:} Constraints on the large-scale galaxy bias $b(z)$ and $b(z)D(z)$ as a function of redshift for the different magnitude-limited galaxy samples considered in this analysis. For all plots, the points denote the values derived from the mean of the posterior distribution of the parameters and the solid lines show the predictions from the fitting function given in Equations \ref{eq:bias-fit-func} and \ref{eq:bias-fit-func-result}. The shaded regions denote the $68 \%$ c.l.. \textit{Lower left panel:} Constraints on $\bar{b}(m_{\mathrm{lim}})$, as defined in Eq.~\ref{eq:bias-fit-func}. The solid line illustrates the best-fit relation given in Eq.~\ref{eq:bias-fit-func-result}. \textit{Lower right panel:} Comparison of the results obtained in this analysis to the bias values assumed for LSST forecasts in the DESC SRD.}
        \label{fig:magnitude-cuts-bias}
      \end{center}
    \end{figure}

    To this end, we split our fiducial galaxy sample into four brighter subsamples with limiting magnitudes $i_{\mathrm{corr}} < 20.5$, $i_{\mathrm{corr}} < 21.5$, $i_{\mathrm{corr}} < 22.5$ and $i_{\mathrm{corr}} < 23.5$, respectively. We then subdivide each of these samples into four redshift bins of approximately equal galaxy number and use our pipeline to compute constraints on HOD parameters for our fiducial analysis variant. We find all subsamples to be well-fit by our fiducial model described in Sec.~\ref{ssec:methods.theory}. In fact, the reduced $\chi^{2}$s for some subsamples are quite low and thus lead to correspondingly high $p$-values, which seem to indicate the presence of overfitting and/or covariance overestimation in our brighter magnitude samples. However, for all subsamples, the distribution of the fit residuals is consistent with a Gaussian and we do not find any significant evidence that our covariance is overestimated by comparing the consistency of power spectra across different HSC fields. We therefore use the results of these HOD fits to derive constraints on the large-scale galaxy bias as a function of redshift for each sample. We determine the bias from the mean of the posterior distribution, as we find it to yield more stable constraints than the best-fit value. The results for the bias evaluated at the effective redshift of each tomographic bin are shown in the upper left panel of Fig.~\ref{fig:magnitude-cuts-bias}, alongside the results for our fiducial sample. As can be seen, we find the linear bias to increase for decreasing limiting magnitude, as expected. Furthermore, the redshift-dependence of the bias can be roughly approximated as inversely proportional to the growth factor, $b(z) \propto \sfrac{1}{D(z)}$, for all samples, as illustrated in the upper right panel of Fig.~\ref{fig:magnitude-cuts-bias}. Motivated by these results, we make the following Ansatz to derive a fitting function for the large-scale galaxy bias as a function of redshift and limiting magnitude:
    \begin{equation}
      b(z, m_{\mathrm{lim}}) = \bar{b}(m_{\mathrm{lim}}) D(z)^{\alpha},
      \label{eq:bias-fit-func}
    \end{equation}
    where we have assumed the bias to be separable in $z$ and $m_{\mathrm{lim}}$, and the parameter $\alpha$ accounts for deviations from the simple inverse growth function proportionality. We determine the values of $\alpha$ and $\bar{b}(m_{\mathrm{lim}})$ in a multi-step weighted, linear least squares fit: we first fit Eq.~\ref{eq:bias-fit-func} to all five samples separately and determine our fiducial $\alpha$ as the median of these values\footnote{We note that we find slightly better results using the median, which is the reason for preferring it over the mean.}. We then fit Eq.~\ref{eq:bias-fit-func} to all samples again, obtaining constraints on $\bar{b}(m_{\mathrm{lim}})$ for $\alpha$ fixed to our fiducial value. In a last step, we fit a linear function to $\bar{b}(m_{\mathrm{lim}})$, finally yielding
    \begin{equation}
      \begin{aligned}
        \alpha &= -1.30 \pm 0.19,\\
        \bar{b}(m_{\mathrm{lim}}) &= b_{1}(m_{\mathrm{lim}}-24) + b_{0},
        \label{eq:bias-fit-func-result}
      \end{aligned}
    \end{equation}
    where $b_{1} = -0.0624 \pm 0.0070$ and $b_{0} = 0.8346 \pm 0.161$\footnote{We note that to compute the uncertainty on $\alpha$, we use $\sigma(\tilde{x}) = 1.253 \; \sigma(\bar{x})$, where $\tilde{x}$ denotes the median and $\bar{x}$ is the mean of a given sample.}. The results from this fitting function are shown as solid lines in Fig.~\ref{fig:magnitude-cuts-bias}, alongside the values obtained from the HOD fit.

    This analysis yields a few interesting results. First, we observe a deviation from the scaling $b(z) \propto \sfrac{1}{D(z)}$ (i.e $\alpha=-1$), which can be expected from objects forming in rare peaks at early times with a high bias \cite{1986ApJ...304...15B,1996MNRAS.282..347M,1998ApJ...500L..79T,Coil:2004}. This is not unexpected: even in the case where this model is valid for objects of a given type, our cut on observed magnitude implies that the samples observed and high and low redshifts are different populations. We find that the bias increases with redshift somewhat faster than the standard empirical assumption of $\sfrac{1}{D(z)}$.

    We anticipate that these results will be useful for forecasting the constraining power of magnitude-limited samples, e.g. from LSST. In the lower right panel of Fig.~\ref{fig:magnitude-cuts-bias}, we show a comparison between the biases derived using our fitting function to those assumed in the DESC Science Requirements Document (SRD) \cite{LSST:SRD:2018} analysis. As can be seen, we find these two to differ at approximately $2 \sigma$, using the uncertainties on our bias measurement. An alternative use might be galaxy evolution or simulation studies, where these results provide a convenient cross-check formula. Our parameter chains for halo model and systematics parameters for the different limiting magnitudes are available upon request. 

\section{Discussion and Conclusions}\label{sec:discussion}
  In this work, we have presented a comprehensive analysis of photometric galaxy clustering in the public HSC-SSP DR1 data using a magnitude-limited sample with $i<24.5$. We have split this sample into four tomographic redshift bins and computed all possible auto- and cross-power spectra, accounting for observational systematics in the number density fluctuations through map-level deprojection. We have then fitted the data using a halo model coupled with a halo occupation distribution, marginalizing over a flexible model for photometric redshift uncertainties. Our results are well fitted by our theoretical model including scales up to $k_{\mathrm{max}} = 1$ Mpc$^{-1}$. We find monotonically decreasing average halo masses as a function of redshift, which can be interpreted in terms of the drop-out of red galaxies at high redshifts for a magnitude-limited sample, consistent with previous analyses. In order to test the stability of our results, we performed a suite of alternative analyses, finding our results to be robust to photometric redshift uncertainties. We extended our analysis to include the effects of lensing magnification, which results in a $\sim 3 \sigma$ detection of magnification from these data alone. In addition to this, we also allowed for variations in $\Omega_{c}$ and $\sigma_{8}$, finding that, within our strong model priors, the data are able to constrain $\Omega_{c}$, while $\sigma_{8}$ remains unconstrained. Finally, we used these data to derive a simple fitting function for the linear, large-scale galaxy bias of magnitude-limited samples as a function of redshift and limiting $i$-band magnitude (Eq. \ref{eq:bias-fit-func-result}).

  Our analysis hence shows that the quality of the HSC data allows scientific analyses leading to new insights into both physics of galaxy formation as well as analysis strategies for galaxy clustering. In terms of physics, we show that a simple magnitude cut leading to a sufficiently homogeneous sample can be used to carry out robust galaxy clustering analyses, provided that photo-$z$ systematic uncertainties can be kept under control. This contrasts with other approaches that focus on red galaxies to create significantly smaller samples with more precise redshifts (see e.g. \cite{1507.05460,1807.10163}). Using magnitude-limited samples, on the other hand, results in larger galaxy samples with smaller shot noise and therefore improved total statistical sensitivity of the survey, despite their less pronounced clustering amplitude due to lower galaxy bias and larger systematics deprojection effects. In addition we find that the cross-correlations between redshift bins carry important information, as they drive the constraints on both parameters accounting for photometric redshift uncertainties as well as lensing magnification.

  Despite the small fraction of sky covered by the data used in this work, HSC DR1 constitutes an ideal proxy for the data expected for LSST. Our analysis demonstrates the feasibility of photometric galaxy clustering studies using such high signal-to-noise datasets. In particular, this work shows the level of readiness of the LSST DESC analysis pipelines, and showcases the application of power spectrum methods for galaxy clustering and the use of template deprojection techniques to handle contamination from sky systematics. We anticipate that future development of this pipeline will include improvements to photo-$z$ marginalization and non-linear bias modeling, thus permitting its application to magnitude-limited galaxy samples with even higher signal-to-noise. These results therefore bode well for future cosmological constraints using galaxy clustering data from LSST.

\acknowledgments
  This paper has undergone internal review in the LSST Dark Energy Science Collaboration. 
  We would like to thank the internal reviewers Elisabeth Krause, Jonathan Loveday, Thomas McClintock and Masahiro Takada for their comments, which helped improve this paper. We would also like to thank David Spergel for numerous helpful discussions and suggestions.

  DA acknowledges support from the Beecroft trust and from the Science and Technology Facilities Council through an Ernest Rutherford Fellowship, grant reference ST/P004474/1.
  HA has been supported by the Rutgers Discovery Informatics Institute Fellowship of Excellence in Computational and Data Science (AY 2017-2020) and Rutgers University \& Bevier Dissertation Completion Fellowship (AY 2019-2020); HA also thanks the LSSTC Data Science Fellowship Program, which is funded by LSSTC, NSF Cybertraining Grant No. 1829740, the Brinson Foundation, and the Moore Foundation, as participation in the program has benefited this work. HA, AB, and EG were supported by the Department of Energy (grants DE-SC0011636 and DE-SC0010008).
  ZG is supported by a Rhodes Scholarship granted by the Rhodes Trust.
  RM is supported by the Department of Energy Cosmic Frontier program, grant DE-SC0010118.
  HM has been supported by the Jet Propulsion Laboratory, California Institute of Technology, under a contract with the National Aeronautics and Space Administration and Grant-in-Aid for Scientific Research from the JSPS Promotion of Science (No. 18H04350). 
  JAN us supported by grant DOE DE-SC0007914 from the U.S. Department of Energy Office of Science (Office of High Energy Physics).
  JD and AN acknowledge support from National Science Foundation Grant No. 1814971.
  This manuscript has been authored by Fermi Research Alliance, LLC under Contract No. DE-AC02-07CH11359 with the U.S. Department of Energy, Office of Science, Office of High Energy Physics.
  The work by JS was supported by the U.S. Department of Energy award DE-SC0009920.
  This work was supported by the grant MDM-2015-0509 from the Spanish Ministry of Science and Innovation.
  This project was supported in part by the U.S. Department of Energy, Office of Science, Office of Workforce Development for Teachers and Scientists (WDTS) under the Science Undergraduate Laboratory Internships Program (SULI).

  The contributions from the authors are listed below:
  AN: co-led project, led signal modeling, photo-$z$ systematics characterization, likelihood analysis and interpretation of results.
  DA: co-led project, led development of the data analysis pipeline, contributed to likelihood and modeling efforts.
  JS: development of data analysis pipeline, comparisons with simulated data, checked impact of observing conditions (systematics).
  AS: co-led project, advised student, cross checked MCMC chains, calculated $N(z)$ sample variance uncertainty, contributed to modeling efforts, wrote initial likelihood code.
  HA: carried out initial data exploration, analyzed HSC photo-$z$ pdfs, and determined tomographic bin selection.
  AB: shot noise estimation, small updates to HSC pipeline.
  JD: provided advice and guidance.
  ZG: calibrated redshift distributions from COSMOS 30-band data.
  EG: galaxy sample selection and shot noise estimation.
  RM: provided guidance on how to use the HSC data.
  HM: provided guidance on how to use the HSC data.
  JN: assistance in interpretation of galaxy clustering results.
  SS: model parametrization, testing MCMC chains, studied uncertainties in the first bin's $N(z)$.
  EW: produced observing condition maps.

  The DESC acknowledges ongoing support from the Institut National de Physique Nucl\'eaire et de Physique des Particules in France; the Science \& Technology Facilities Council in the United Kingdom; and the Department of Energy, the National Science Foundation, and the LSST Corporation in the United States.  DESC uses resources of the IN2P3 Computing Center (CC-IN2P3--Lyon/Villeurbanne - France) funded by the Centre National de la Recherche Scientifique; the National Energy Research Scientific Computing Center, a DOE Office of Science User Facility supported by the Office of Science of the U.S.\ Department of Energy under Contract No.\ DE-AC02-05CH11231; STFC DiRAC HPC Facilities, funded by UK BIS National E-infrastructure capital grants; and the UK particle physics grid, supported by the GridPP Collaboration.  This work was performed in part under DOE Contract DE-AC02-76SF00515.

  The Hyper Suprime-Cam (HSC) collaboration includes the astronomical communities of Japan and Taiwan, and Princeton University. The HSC instrumentation and software were developed by the National Astronomical Observatory of Japan (NAOJ), the Kavli Institute for the Physics and Mathematics of the Universe (Kavli IPMU), the University of Tokyo, the High Energy Accelerator Research Organization (KEK), the Academia Sinica Institute for Astronomy and Astrophysics in Taiwan (ASIAA), and Princeton University. Funding was contributed by the FIRST program from Japanese Cabinet Office, the Ministry of Education, Culture, Sports, Science and Technology (MEXT), the Japan Society for the Promotion of Science (JSPS), Japan Science and Technology Agency (JST), the Toray Science Foundation, NAOJ, Kavli IPMU, KEK, ASIAA, and Princeton University. 

  This paper makes use of software developed for the Large Synoptic Survey Telescope. We thank the LSST Project for making their code available as free software at \url{http://dm.lsst.org}.

  This paper is based on data collected at the Subaru Telescope and retrieved from the HSC data archive system, which is operated by Subaru Telescope and Astronomy Data Center at National Astronomical Observatory of Japan. Data analysis was in part carried out with the cooperation of Center for Computational Astrophysics, National Astronomical Observatory of Japan.

\appendix
\counterwithin{figure}{section}

\section{Reconstructing depth from discrete sources}\label{app:depth}
  The depth map used in our analysis (see Section \ref{ssec:methods.syst}) is constructed from the signal-to-noise ratios of sources classified as stars. The use of all sources, including galaxies, to generate this map, would result in spurious correlations with the true galaxy distribution that would cause a bias in the power spectra after deprojecting it. We explore this effect further in this section.
  
  In a simplified model, let us assume that galaxies can be separated into discrete types labelled by an index $\alpha$. If we want to compute the value of a given quantity $Q$ in a pixel $p$ by averaging over all galaxies in that pixel, the result is:
  \begin{equation}
    \bar{Q}_p = \frac{\sum_\alpha N_{\alpha,p}Q_{\alpha,p}}{\sum_\alpha N_{\alpha,p}},
  \end{equation}
  where $N_{\alpha,p}$ is the number of galaxies of type $\alpha$ in pixel $p$, and $Q_{\alpha,p}$ is the value of $Q$ for those galaxies. Expanding $N_{\alpha,p}=\bar{N}_\alpha(1+\delta_{\alpha,p})$, where $\bar{N}_\alpha$ is the average number of galaxies of type $\alpha$ per pixel, and $\delta_{\alpha,p}$ is their overdensity, we obtain:
  \begin{align}\nonumber
    \bar{Q}_p&=\frac{\langle Q\rangle_p+\langle Q\delta\rangle_p}{1+\langle \delta\rangle_p}\\
             &\simeq\langle Q\rangle_p+\langle Q\delta\rangle_p-\langle Q\rangle_p\langle\delta\rangle_p+\mathcal{O}(\delta^2),
  \end{align}
  where we have defined $\langle f\rangle_p\equiv \sum_\alpha \bar{N}_\alpha f_{\alpha,p}/\sum_\alpha \bar{N}_\alpha$, and we have expanded to first order in $\delta$ in the second line. If $Q$ is constant across all galaxies, the second and third term in the second line cancel, and the resulting map would receive no noise from the inhomogeneous galaxy distribution ($\bar{Q}_p=\langle Q\rangle_p$) up to second order. This is however not the case for the flux errors used to estimate the depth map. Brighter sources have lower number densities (and therefore stronger shot-noise fluctuations), and also larger flux errors. Therefore, although the depth map estimated from mean galaxy flux errors is dominated by the large-scale depth fluctuations, the small-scale fluctuations trace the spatial distribution of the brightest galaxies, and thus correlate with the true galaxy distribution. This violates the core assumption of template deprojection (that the deprojected contaminant templates do not correlate with the true map fluctuations), and causes a bias in the estimated power spectrum.
  
  Our depth map is computed from star-classified objects and therefore, although it will be subject to the same small-scale noise due to this discrete sampling, these noise fluctuations do not correlate with the galaxy distribution, and therefore do not bias our results. A second-order correlation could arise from the small fraction of galaxies mis-classified as stars, but the associated effect is negligible given our uncertainties. Due to these complications, in the future it will be desirable to estimate the survey depth using image properties directly (e.g. by injecting fake galaxies, or using the background noise integrated over PSF-size apertures).
  
\begin{figure}
\begin{center}
\includegraphics[width=0.95\textwidth]{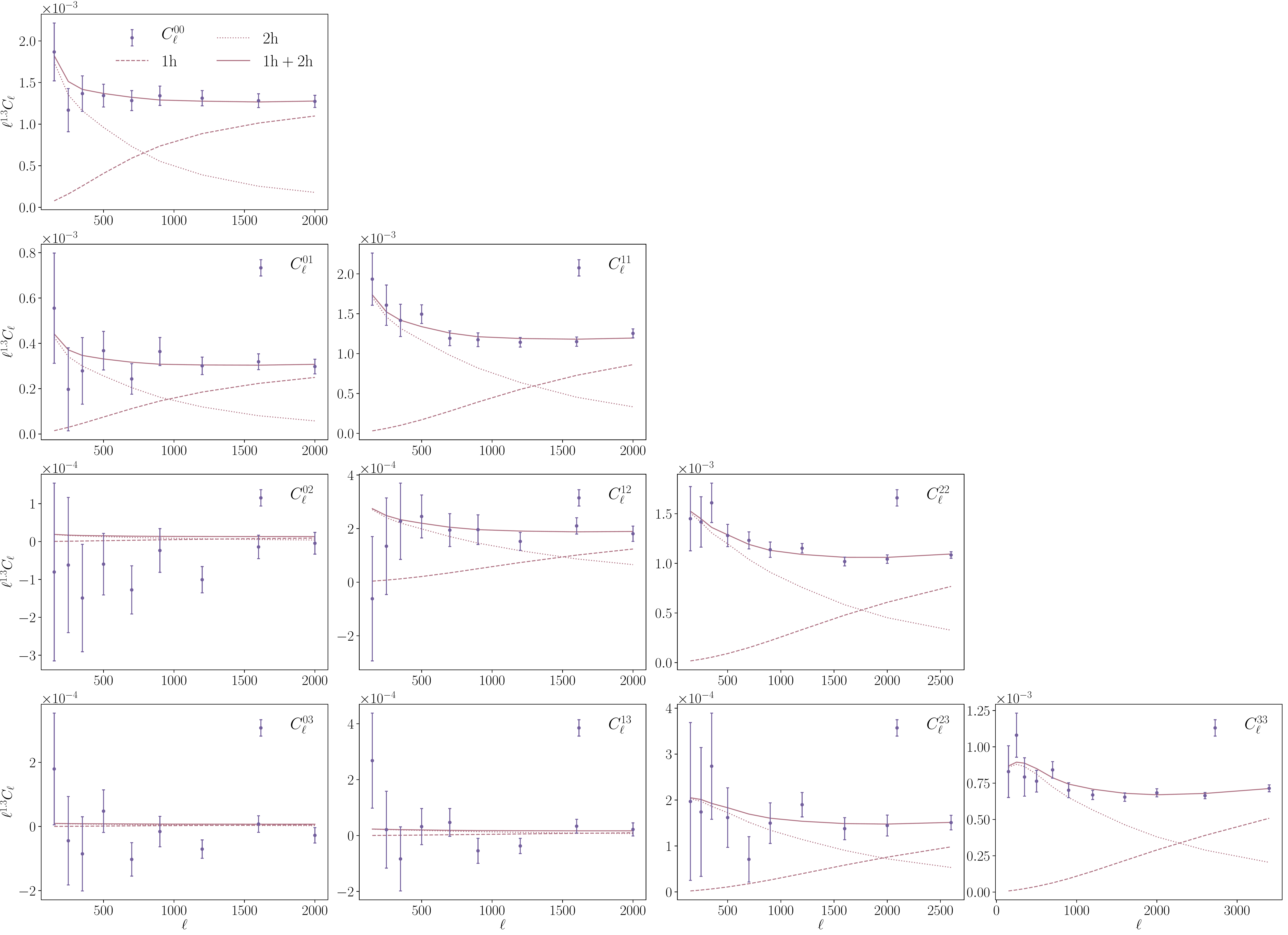}
\caption{Measured auto- and cross-power spectra obtained in our analysis. The lines show the theoretical predictions derived from our fiducial best-fit model parameters, additionally split into the 1- and 2-halo contributions to the total power spectrum.}
\label{fig:cls-best-fit-1h-2h}
\end{center}
\end{figure}  
  
\begin{figure}
  \begin{center}
    \includegraphics[width=0.95\textwidth]{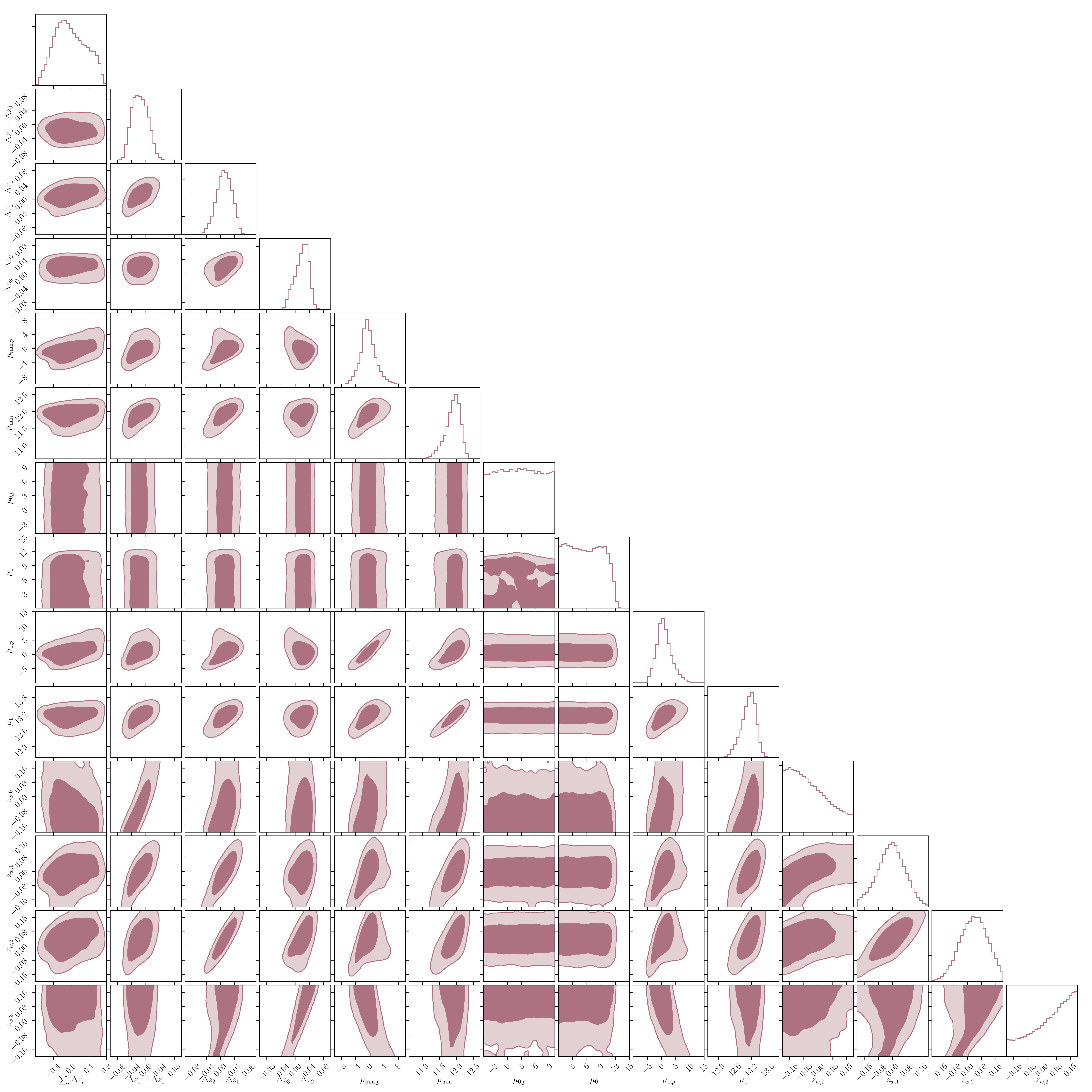}
    \caption{Fiducial constraints on HOD and systematics parameters obtained in this work. The inner (outer) contour shows the $68 \%$ c.l. ($95 \%$ c.l.).}
    \label{fig:constraints-fid-full}
  \end{center}
\end{figure}

\begin{figure}
  \begin{center}
    \includegraphics[width=0.95\textwidth]{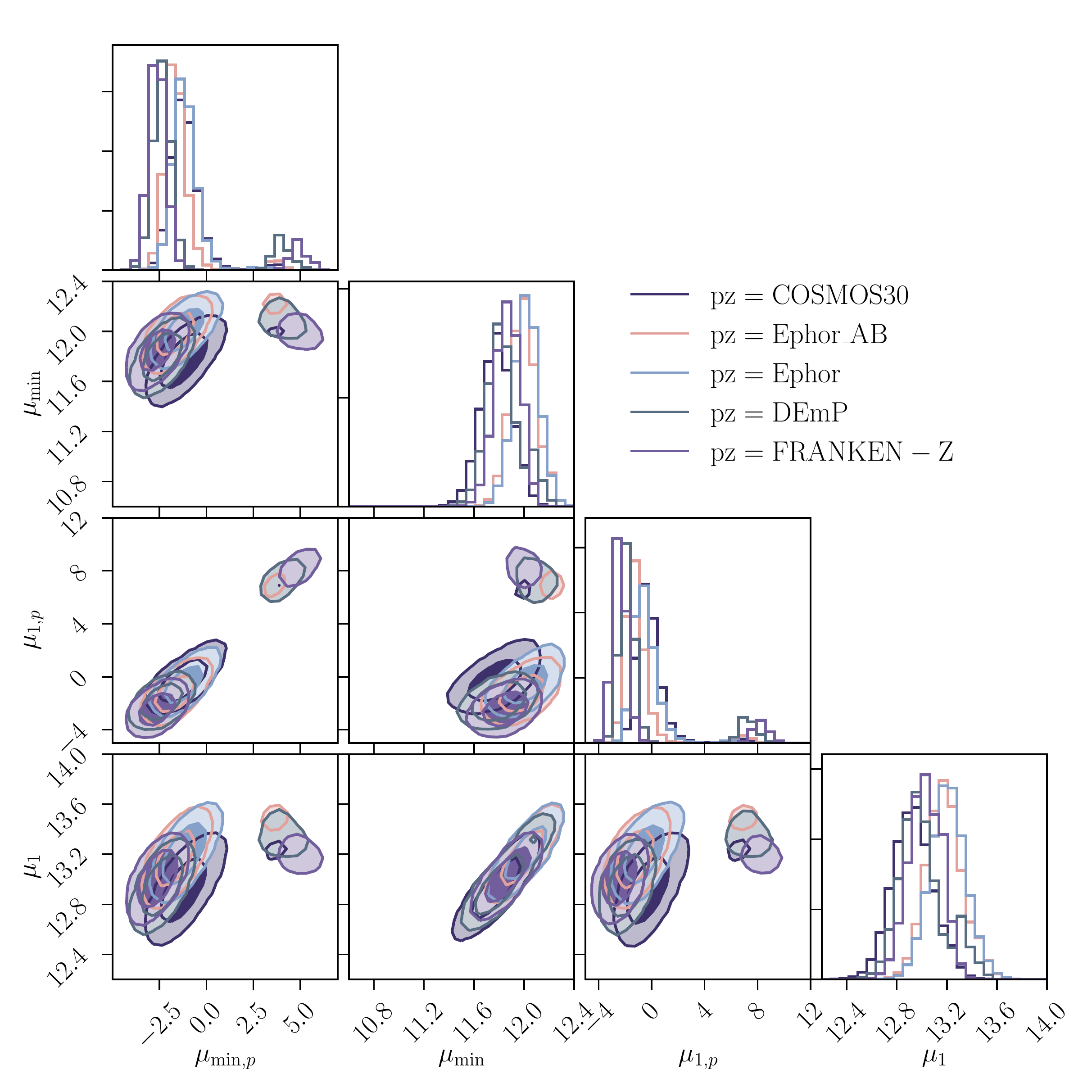}
    \caption{Comparison of the HOD constraints obtained using the redshift distributions derived from COSMOS30, \texttt{Ephor\_AB}, \texttt{Ephor}, \texttt{DEmP} and \texttt{FRANKEN-Z} setting $z_{w, i} = 0, \Delta z_{i} = 0$. The inner (outer) contour shows the $68 \%$ c.l. ($95 \%$ c.l.).}
    \label{fig:constraints-HOD-no-pz-shifts-pz-methods}
  \end{center}
\end{figure}  

\bibliography{bibliography}

\end{document}